\documentclass[aps,prc,showpacs,showkeys,nofootinbib]{revtex4}

\usepackage{color} 

\usepackage{graphicx}
\usepackage{dcolumn}
\usepackage{amsthm}
\usepackage{bm}

\usepackage{isotope}

\newcommand{\sigmabf}{\mbox{\boldmath $\sigma$}}
\newcommand{\xibf}{\mbox{\boldmath $\xi$}}
\newcommand{\rhobf}{\mbox{\boldmath $\rho$}}

\begin{document}

\title{What can we know about hypernuclei via analysis of bremsstrahlung photons?
}

\author{Xin~Liu$^{(1,2)}$}%
\email{liuxin@impcas.ac.cn} %
\author{Sergei~P.~Maydanyuk$^{(2,3)}$}%
\email{maidan@kinr.kiev.ua}%
\author{Peng-Ming~Zhang$^{2}$}%
\email{zhpm@impcas.ac.cn} %
\author{Ling~Liu$^{(1)}$}%
\email{lling216@163.com}%
\affiliation{$(1)$College of Physical Science and Technology, Shenyang Normal University, Shenyang, 110034, China}
\affiliation{$(2)$Institute of Modern Physics, Chinese Academy of Sciences, Lanzhou, 730000, China}
\affiliation{$(3)$Institute for Nuclear Research, National Academy of Sciences of Ukraine, Kiev, 03680, Ukraine}


\date{\small\today}


\begin{abstract}
We investigate possibility of emission of the bremsstrahlung photons in nuclear reactions with hypernuclei for the first time.
A new model of the bremsstrahlung emission which accompanies interactions between $\alpha$ particles and hypernuclei is constructed,
where a new formalism for the magnetic momenta of nucleons and hyperon inside hypernucleus is added.
For first calculations, we choose $\alpha$ decay of the normal nucleus \isotope[210]{Po} and the hypernucleus \isotope[211][\Lambda]{Po}.
We find that
(1) emission for the hypernucleus \isotope[211][\Lambda]{Po} is larger than for normal nucleus \isotope[210]{Po},
(2) difference between these spectra is small.
We propose a way how to find hypernuclei, where role of hyperon is the most essential in emission of bremsstrahlung photons during $\alpha$ decay.
As demonstration of such a property, we show that the spectra for the hypernuclei \isotope[107][\Lambda]{Te} and \isotope[109][\Lambda]{Te} are essentially larger than the spectra for the normal nuclei \isotope[106]{Te} and \isotope[108]{Te}.
Such a difference is explained by additional contribution of emission to the full bremsstrahlung, which is formed by magnetic moment of hyperon inside hypernucleus.
The bremsstrahlung emission formed by such a mechanism, is of the magnetic type.
A new formula for fast estimations of bremsstrahlung spectra for even-even hypernuclei is proposed,
where role of magnetic moment of hyperon of hypernucleus in formation of the bremsstrahlung emission is shown explicitly.
Such an analysis 
opens possibility of new experimental study of properties of hypernuclei via bremsstrahlung study.
\end{abstract}




\maketitle

\section{Introduction
\label{sec.introduction}}

Physics of hypernuclei 
is an important branch of nuclear physics at
low 
as well as at intermediate energies 
\cite{Danysz.1953.PhilMag,Gal.2016.RMP,Lenske.2018.LNP,Botta.2012.EPJA,Davis.1986.ContPhys,Hashimoto.2006.PPNP,Hiyama.2009.PPNP}.
Hypernucleus is a kind of nucleus with atomic weight $A$ and atomic number $Z$, contained at least one hyperon ($\Lambda$, $\Sigma$, $\Theta $, and perhaps $\Omega$) except protons and neutrons.
A hypernucleus is characterized by its spin, isospin, in the case of $\Lambda$ hypernuclei, a strangeness of $-1$~\cite{Gal.2016.RMP},
and in a case of double-$\Lambda$ hypernuclei, a
strangeness of $-2$~\cite{Tang.1965.PRL,Nakazawa.2010.NPA,Hiyama.2002.PRC,Gaitanos.2012.NPA,Ahn.2001.PRL,Hiyama.2010.PRL,Ahn.2013.PRC,Morita.2015.PRC,Lanskoy.1998.PRC,Vidana.2004.PRC}.
During last decades, many hypernuclei have been produced experimentally (for example, see Refs.~\cite{Saito.2018.LNP,Bando.1990.IJMPA,Pile.1991.PRL,Hasegawa.1996.PRC}).
Among all hypernuclei, $\Lambda$ hypernuclei have been studied the most deeply \cite{Hu.2016.PRC}.

An important question is how nuclear interactions will be changed, if to include strange baryon to the normal nucleus.
Many types of potentials~\cite{James.1970.NPA,Dalitz.1972.NPA,Motoba.1983.PTP,Kurihara.1985.PRC,Motoba.1988.PRC,Millener.1988.PRC,Lalazissis.1988.PRC,Siegel.1990.AmJP,Koutroulos.1991.JPG}
were investigated to study interactions between $\Lambda$-hyperons and nuclei, and reactions with hypernuclei also.
In last case, some of interest is given for more heavy hypernuclei, which have possibility to emit $\alpha$ particles.
$\alpha$-Nucleus interactions are understood deeply and determined well
(see investigations \cite{Denisov.2009.ADNDT,Denisov.2015.PRC,Denisov.2005.PHRVA,Denisov.2009.PRC.v79,Denisov.2009.PRC.v80}
providing the accurate potential of interactions between the $\alpha$ particles and nuclei basing on existed experimental information of $\alpha$ decay and $\alpha$ capture,
reviews and databases~\cite{Buck.1993.ADNDT,Akovali.1998.NDS,Duarte.2002.ADNDT,Audi.2003.NPA,Dasgupta-Schubert.2007.ADNDT,Silisteanu.2012.ADNDT,Lovas.1998.PRep,Sobiczewski.2007.PPNP,www_library},
other approaches~\cite{Stewart.1996.NPA,Xu.2006.PRC,Nazarewicz.2012.PRC,Delion.2013.PRC,%
Silisteanu.2015.RJP,Silisteanu.2015.AIP,Silisteanu.2016.EPJWebConf,Silisteanu.2017.PRC,AOSilisteanu.2017.AIP,Silisteanu.2017.AIP},
approaches of sharp angular momentum cut-off in $\alpha$-capture~\cite{Glas.1975.NPA,Glas.1974.PRC},
quantum mechanical calculations of fusion~\cite{Maydanyuk_Zhang_Zou.2017.PRC,Maydanyuk.2015.NPA},
experimental data for $\alpha$-capture~\cite{Eberhard.1979.PRL,DAuria.1968.PR,Barnett.2000.PRC},
evaluations of the $\alpha$-particle capture rates in stars~\cite{Mohr.2000.PRC,Demetriou.2002.NPA,Rauscher.2000.NPA} 
).
By such a reason, $\alpha$ decay can be considered for proper test of new calculations with hypernuclei.
Note that emission of $\alpha$-particles from \isotope[10][\Lambda]{Be} and \isotope[10][\Lambda]{B} was studied experimentally in Nuclotron accelerator at JINR,
Dubna~\cite{Majling.2004.PAN,Majling.2005.NPA,Batusov.2005.PAN}.

Unfortunately, opportunities to study hypernuclei experimentally are restricted.
In such a situation, we put attention to bremsstrahlung emission of photons which accompany nuclear reactions.
Such a topic is traditional in nuclear physics which has been causing much interest for a long time
(see reviews~\cite{Pluiko.1987.PEPAN,Kamanin.1989.PEPAN}). 
This is because of bremsstrahlung photons provide rich independent information about the studied nuclear process.
Dynamics of the nuclear process, interactions between nucleons, types of nuclear forces, structure of nuclei, quantum effects, anisotropy (deformations) can be included in the model describing the bremsstrahlung emission.
At the same time, measurements of such photons and their analysis provide information about all these aspects, and verify suitability of the models.
So, bremsstrahlung photons is the independent tool to obtain experimental information for all above questions.
So, in this paper we investigate idea if is it possible to use bremsstrahlung photons, which can be emitted in reactions with participation of hypernuclei, in order to obtain a new information about properties of these hypernuclei.
From analysis of literature we conclude, that this question has not been studied yet.
On the other side, we can base our new investigations on our current formalism of emission of bremsstrahlung photons in
nuclear reactions~\cite{Maydanyuk.2003.PTP,Maydanyuk.2009.NPA,Maydanyuk.2009.TONPPJ,Maydanyuk.2009.JPS,Maydanyuk.2010.PRC,Maydanyuk.2011.JPCS,Maydanyuk.2011.JPG,
Maydanyuk.2006.EPJA,Maydanyuk.2008.EPJA,Giardina.2008.MPLA,Maydanyuk.2012.PRC,Maydanyuk_Zhang.2015.PRC,Maydanyuk_Zhang_Zou.2016.PRC,Maydanyuk_Zhang_Zou.2018.PRC}.

At present, we do not know anything about emission of bremsstrahlung photons during reactions with hypernuclei.
By such a reason, in order to perform the first estimation of bremsstrahlung spectra,
we choose reaction with nuclei, where our bremsstrahlung formalism was the most successful
(in description of existed experimental data of bremsstrahlung).
This is bremsstrahlung in $\alpha$ decay of nuclei
(the experimental bremsstrahlung data obtained with the highest accuracy are data~\cite{Boie.2007.PRL,Boie.2009.PhD} for $\alpha$ decay of nucleus \isotope[211]{Po},
the most accurate description of these data were obtained
in Refs.~\cite{Maydanyuk.2006.EPJA,Maydanyuk.2008.EPJA}; also see calculations~\cite{Papenbrock.1998.PRLTA,Tkalya.1999.PHRVA,Jentschura.2008.PRC} and reference therein).
So, in current research we will study bremsstrahlung in $\alpha$ decay of hypernuclei.
Note that some investigations of properties of hypernuclei in $\alpha$ decay have been already performed~\cite{Santhosh.2018.Pramana}.
This reinforces our motivation for current research
(one can use parameters of potentials between $\alpha$-particles and hypernuclei obtained in Ref.~\cite{Santhosh.2018.Pramana} for new calculations of the bremsstrahlung spectra).

Hyperon (with zero electric charge) has anomalous magnetic moment, which is essentially different from anomalous magnetic moments of neutrons and protons.
So, one can suppose that hyperon inside nucleus should form emission of photons with another intensity (and another type of emission as possible) of the bremsstrahlung photons, in comparison with neutrons inside the same nucleus (protons also has non-zero electric charge, so by such a reason they are principally different from hyperons in formation of bremsstrahlung).
In order to clarify this question, we need in bremsstrahlung model which takes into account magnetic momenta of nucleons and hyperon.
Note that this formalism have not been constructed yet.
Importance of investigations of magnetic emission in nuclear reactions in stars (see Ref.~\cite{Maydanyuk_Zhang_Zou.2016.PRC})
reinforces motivation to create this bremsstrahlung formalism.
Another point of application is corrections of incoherent emission (after inclusion of anomalous magnetic moments of nucleons to model) which can be not small in nuclear reactions~\cite{Maydanyuk_Zhang.2015.PRC}.

In this paper we focus on realization of ideas above.
The paper is organized in the following way.
In Sec.~\ref{sec.model} a new bremsstrahlung model of bremsstrahlung emission during $\alpha$-decay of normal nuclei and hypernuclei is presented.
Here, we include a new formalism of emission of photons due to magnetic momenta of nucleons and hyperons inside nucleus.
In Sec.~\ref{sec.results} we study bremsstrahlung emitted during $\alpha$ decay of normal nuclei \isotope[210]{Po}, \isotope[106]{Te}, \isotope[108]{Te}
and hypernuclei \isotope[211][\Lambda]{Po}, \isotope[107][\Lambda]{Te}, \isotope[109][\Lambda]{Te}.
We summarize conclusions in Sect.~\ref{sec.conclusions}.
Details of calculations of
relative coordinates and corresponding momenta,
operator of emission in relative coordinates,
electric and magnetic form-factors are given in Appendixes~\ref{sec.app.1}--\ref{sec.app.3}.

\section{Model
\label{sec.model}}

\subsection{Generalized Pauli equation for nucleons in the $\alpha$--nucleus system and operator of emission of photons
\label{sec.2.1}}


Let us consider $\alpha$-particle interacting with nucleus (which can be hypernucleus).
In order to describe evolution of nucleons of such a complicated system  in the laboratory frame
(we have $A+4$ nucleons of the system of nucleus and $\alpha$-particle),
we shall use many-nucleon generalization of Pauli equation
(obtained starting from Eq.~(1.3.6) in Ref.~\cite{Ahiezer.1981}, p.~33;
this formalism if along Refs.~\cite{Maydanyuk.2012.PRC,Maydanyuk_Zhang.2015.PRC,Maydanyuk_Zhang_Zou.2016.PRC}, and reference therein)
\begin{equation}
\begin{array}{lcl}
  i \hbar \displaystyle\frac{\partial \Psi}{\partial t} = \hat{H} \Psi,
\end{array}
\label{eq.pauli.1}
\end{equation}
where hamiltonian is
\begin{equation}
\begin{array}{lcl}
\vspace{1mm}
  \hat{H} & = &
  \displaystyle\sum_{i=1}^{4}
  \biggl\{
    \displaystyle\frac{1}{2m_{i}} \Bigl( \mathbf{p}_{i} - \displaystyle\frac{z_{i}e}{c} \mathbf{A}_{i} \Bigr)^{2} +
    z_{i}e\, A_{i,0} - \displaystyle\frac{z_{i}e \hbar}{2m_{i}c}\; \sigmabf \cdot \mathbf{rot\, A}_{i}
  \biggr\}\; + \\

  & + &
  \displaystyle\sum_{j=1}^{A}
  \biggl\{
    \displaystyle\frac{1}{2m_{j}} \Bigl( \mathbf{p}_{j} - \displaystyle\frac{z_{j}e}{c} \mathbf{A}_{j} \Bigr)^{2} +
    z_{j}e\, A_{j,0} - \displaystyle\frac{z_{j}e \hbar}{2m_{j}c}\; \sigmabf \cdot \mathbf{rot\, A}_{j}
  \biggr\} +

  V(\mathbf{r}_{1} \ldots \mathbf{r}_{A+4}).
\end{array}
\label{eq.pauli.2}
\end{equation}
Here, $m_{i}$ and $z_{i}$ are mass and electromagnetic charge of nucleon with number $i$,
$\mathbf{p}_{i} = -i\hbar\, \mathbf{d}/\mathbf{dr}_{i} $ is momentum operator for nucleon with number $i$,
$V(\mathbf{r}_{1} \ldots \mathbf{r}_{A+1})$ is general form of the potential of interactions between nucleons,
$\sigmabf$ are Pauli matrixes,
$A_{i} = (\mathbf{A}_{i}, A_{i,0})$ is potential of electromagnetic field formed by moving nucleon with number $i$,
$A$ in summation is mass number of a daughter nucleus%
\footnote{%
Note that Eqs.~(\ref{eq.pauli.1})--(\ref{eq.pauli.2}) is modification of Pauli equation, which is obtained as the first approximation of Dirac equation.
Wave function for Pauli equation is spinor $\Psi$ (it has two components), while wave function of Dirac equation is bi-spinor $\Psi^{\rm (Dir)} = (\chi, \Psi)$ (it has four components).
In case of one-nucleon problem, another spinor component $\chi$ of bi-spinor wave function of Dirac equation has form
[see Eq.~(2) in Ref.~\cite{Maydanyuk.2012.PRC}; also Eq.~(1.3.4) in Ref.~\cite{Ahiezer.1981}]:
%
$\chi = \displaystyle\frac{1}{2mc}\; \sigmabf\, \Bigl( \mathbf{p} - \displaystyle\frac{ze}{c} \mathbf{A} \Bigr)\, \Psi$.
}.


We rewrite hamiltonian (\ref{eq.pauli.2}) as
\begin{equation}
\begin{array}{lcl}
  \hat{H} = \hat{H}_{0} + \hat{H}_{\gamma},
\end{array}
\label{eq.pauli.4}
\end{equation}
where
\begin{equation}
\begin{array}{lcl}
\vspace{1mm}
  \hat{H}_{0} & = &
  \displaystyle\sum_{i=1}^{4} \displaystyle\frac{1}{2m_{i}}\, \mathbf{p}_{i}^{2} +
  \displaystyle\sum_{j=1}^{A} \displaystyle\frac{1}{2m_{j}}\, \mathbf{p}_{j}^{2} +
  V(\mathbf{r}_{1} \ldots \mathbf{r}_{A+4}), \\

\vspace{1mm}
  \hat{H}_{\gamma} & = &

  \displaystyle\sum_{i=1}^{4}
  \biggl\{
    - \displaystyle\frac{z_{i} e}{m_{i}c}\; \mathbf{p}_{i} \cdot \mathbf{A}_{i} +
    \displaystyle\frac{z_{i}^{2}e^{2}}{2m_{i}c^{2}} \mathbf{A}_{i}^{2} -
    \displaystyle\frac{z_{i}e\hbar}{2m_{i}c}\, \sigmabf \cdot \mathbf{rot A}_{i} +
    z_{i}e\, A_{i,0}
  \biggr\}\; + \\

  & + &
  \displaystyle\sum_{j=1}^{A}
  \biggl\{
    - \displaystyle\frac{z_{j} e}{m_{j}c}\; \mathbf{p}_{j} \cdot \mathbf{A}_{j} +
    \displaystyle\frac{z_{j}^{2}e^{2}}{2m_{j}c^{2}} \mathbf{A}_{j}^{2} -
    \displaystyle\frac{z_{j}e\hbar}{2m_{j}c}\, \sigmabf \cdot \mathbf{rot A}_{j} +
    z_{j}e\, A_{j,0}
  \biggr\}.
\end{array}
\label{eq.pauli.5}
\end{equation}
Here,
$\hat{H}_{0}$ is hamiltonian describing evolution of nucleons of the $\alpha$-particle and nucleus in the studied reaction (without photons),
$\hat{H}_{\gamma}$ is operator describing emission of bremsstrahlung photons in the $\alpha$-nucleus reaction.


We introduce magnetic momentum of particle with number $i$ (which is given by Dirac's theory for proton,
see Eq.~(1.3.8) in page 33 in Ref.~\cite{Ahiezer.1981}),
defining it as
$\mu_{i}^{\rm (Dir)} = z_{i} \cdot e\hbar\, /\, 2m_{i}c$.%
%
%
\footnote{This magnetic moment represents a potential energy of magnetic dipole inside external magnetic field $\mathbf{H}$.
But in this definition we include also the electric charge $z_{i}$, and do not use Pauli matrixes, in contrast to Ref.~\cite{Ahiezer.1981}.}
In order to go to anomalous magnetic momenta of particle $\mu_{i}^{\rm (an)}$, we use the following change:
\begin{equation}
\begin{array}{lcl}
  \mu_{i}^{\rm (Dir)} \to \mu_{i}^{\rm (an)}.
\end{array}
\label{eq.2.2.2}
\end{equation}
We have the following anomalous magnetic momenta for proton, neutron and $\Lambda$-hyperon \cite{RewPartPhys_PDG.2018}:
\begin{equation}
\begin{array}{llll}
\vspace{1mm}
  \mu_{\rm p}^{\rm (an)} = 2.79284734462\: \mu_{N}, & 
  \mu_{\rm n}^{\rm (an)} = -1.91304273\: \mu_{N},  & 
  \mu_{\rm \Lambda}^{\rm (an)} = -0.613\: \mu_{N}, 
\end{array}
\label{eq.2.2.3}
\end{equation}
where
$\mu_{N} = e\hbar / (2m_{\rm p}c) = 3.152 451 2550\; 10^{-14}$~MeV T$^{-1}$ is nuclear magneton.
Using change (\ref{eq.2.2.2}),
neglecting terms at $\mathbf{A}_{j}^{2}$ and $A_{j,0}$,
using Coulomb gauge,
operator of emission~(\ref{eq.pauli.5}) is transformed to
\begin{equation}
\begin{array}{lcl}
  \hat{H}_{\gamma} & = &
  \displaystyle\sum_{i=1}^{4}
  \biggl\{
    - \displaystyle\frac{z_{i} e}{m_{i}c}\; \mathbf{A}_{i}\, \mathbf{p}_{i} -
    \mu_{i}^{\rm (an)}\, \sigmabf \cdot \mathbf{H}_{i}
  \biggr\} +
  \displaystyle\sum_{j=1}^{A}
  \biggl\{
    - \displaystyle\frac{z_{j} e}{m_{j}c}\; \mathbf{A}_{j}\, \mathbf{p}_{j} -
    \mu_{j}^{\rm (an)}\, \sigmabf \cdot \mathbf{H}_{j}
  \biggr\},
\end{array}
\label{eq.2.2.5}
\end{equation}
where
\begin{equation}
  \mathbf{H} = \mathbf{rot\: A} = \bigl[ \mathbf{\nabla} \times \mathbf{A} \bigr].
\label{eq.2.2.6}
\end{equation}
This expression is many-nucleon generalization of operator of emission $\hat{W}$ in Eq.~(4) in Ref.~\cite{Maydanyuk.2012.PRC} with included anomalous magnetic momenta of nucleons.

\subsection{Formalism in space representation
\label{sec.2.3}}

Principle of uncertainty forms grounds of quantum mechanics.
This gives us relations between space coordinates and corresponding momenta.
By such a reason, we need to obtain full formalism in space variables or momenta.
For further convenience, we will rewrite the operator of emission (and perform all further calculations) in the space representation.

Substituting the following definition for the potential of electromagnetic field:
\begin{equation}
\begin{array}{lcl}
  \mathbf{A} & = &
  \displaystyle\sum\limits_{\alpha=1,2}
    \sqrt{\displaystyle\frac{2\pi\hbar c^{2}}{w_{\rm ph}}}\; \mathbf{e}^{(\alpha),\,*}
    e^{-i\, \mathbf{k_{\rm ph}r}},
\end{array}
\label{eq.2.3.1}
\end{equation}
we obtain:
\begin{equation}
\begin{array}{lcl}
  \mathbf{H} & = &
  \mathbf{rot\: A} = \bigl[ \mathbf{\nabla} \times \mathbf{A} \bigr] =
  \sqrt{\displaystyle\frac{2\pi\hbar c^{2}}{w_{\rm ph}}}\,
    \displaystyle\sum\limits_{\alpha=1,2}
    \Bigl\{ -i\, e^{-i\, \mathbf{k_{\rm ph}r}}\, \bigl[ \mathbf{k_{\rm ph}} \times \mathbf{e}^{(\alpha),\,*} \bigr] +
      e^{-i\, \mathbf{k_{\rm ph}r}}\, \bigl[ \mathbf{\nabla} \times \mathbf{e}^{(\alpha),\,*} \bigr] \Bigr\}.
\end{array}
\label{eq.2.3.2}
\end{equation}
Here, $\mathbf{e}^{(\alpha)}$ are unit vectors of polarization of the photon emitted [$\mathbf{e}^{(\alpha), *} = \mathbf{e}^{(\alpha)}$], $\mathbf{k}_{\rm ph}$ is wave vector of the photon and $w_{\rm ph} = k_{\rm ph} c = \bigl| \mathbf{k}_{\rm ph}\bigr|c$. Vectors $\mathbf{e}^{(\alpha)}$ are perpendicular to $\mathbf{k}_{\rm ph}$ in Coulomb calibration. We have two independent polarizations $\mathbf{e}^{(1)}$ and $\mathbf{e}^{(2)}$ for the photon with impulse $\mathbf{k}_{\rm ph}$ ($\alpha=1,2$). One can develop formalism simpler in the system of units where $\hbar = 1$ and $c = 1$, but we shall write constants $\hbar$ and $c$ explicitly.
Also we have properties:
\begin{equation}
\begin{array}{lclc}
  \Bigl[ \mathbf{k}_{\rm ph} \times \mathbf{e}^{(1)} \Bigr] = k_{\rm ph}\, \mathbf{e}^{(2)}, &
  \Bigl[ \mathbf{k}_{\rm ph} \times \mathbf{e}^{(2)} \Bigr] = -\, k_{\rm ph}\, \mathbf{e}^{(1)}, &
  \Bigl[ \mathbf{k}_{\rm ph} \times \mathbf{e}^{(3)} \Bigr] = 0, &
  \displaystyle\sum\limits_{\alpha=1,2,3} \Bigl[ \mathbf{k}_{\rm ph} \times \mathbf{e}^{(\alpha)} \Bigr] = k_{\rm ph}\, (\mathbf{e}^{(2)} - \mathbf{e}^{(1)}).
\end{array}
\label{eq.2.3.3}
\end{equation}

We substitute formulas (\ref{eq.2.3.1}) and (\ref{eq.2.3.2}) to formula~(\ref{eq.2.2.5}) for operator of emission and obtain:
\begin{equation}
\begin{array}{lcl}
  \hat{H}_{\gamma} & = &
  \sqrt{\displaystyle\frac{2\pi\hbar c^{2}}{w_{\rm ph}}}\;
  \displaystyle\sum_{i=1}^{4}
  \displaystyle\sum\limits_{\alpha=1,2}
    e^{-i\, \mathbf{k_{\rm ph}r}_{i}}\,
  \biggl\{
    i\, \mu_{N}\, \displaystyle\frac{2 z_{i} m_{\rm p}}{m_{\alpha i}}\:
    \mathbf{e}^{(\alpha)} \cdot \nabla_{i} +
    \mu_{i}^{\rm (an)}\, \sigmabf \cdot \Bigl( i\, \bigl[ \mathbf{k_{\rm ph}} \times \mathbf{e}^{(\alpha)} \bigr] - \bigl[ \mathbf{\nabla}_{i} \times \mathbf{e}^{(\alpha)} \bigr] \Bigr)
  \biggr\}\; + \\

  & + &
  \sqrt{\displaystyle\frac{2\pi\hbar c^{2}}{w_{\rm ph}}}\;
  \displaystyle\sum_{j=1}^{A}
  \displaystyle\sum\limits_{\alpha=1,2}
    e^{-i\, \mathbf{k_{\rm ph}r}_{j}}\;
    \biggl\{
      i\, \mu_{N}\, \displaystyle\frac{2 z_{j} m_{\rm p}}{m_{Aj}}\:
      \mathbf{e}^{(\alpha)} \cdot \nabla_{j} +
      \mu_{j}^{\rm (an)}\, \sigmabf \cdot \Bigl( i\, \bigl[ \mathbf{k_{\rm ph}} \times \mathbf{e}^{(\alpha)} \bigr] - \bigl[ \mathbf{\nabla}_{j} \times \mathbf{e}^{(\alpha)} \bigr] \Bigr)
    \biggr\},
\end{array}
\label{eq.2.3.4}
\end{equation}
where $\mu_{N}$ is nuclear magneton defined after Eqs.~(\ref{eq.2.2.3}).
This expression coincides with operator of emission $\hat{W}$ in form (6) in Ref.~\cite{Maydanyuk.2012.PRC} in limit case of problem of one nucleon with charge $Z_{\rm eff}$ in the external field
[taking Eqs.~(\ref{eq.2.2.1}), (\ref{eq.2.2.2}), $\mathbf{e}^{(\alpha), *} = \mathbf{e}^{(\alpha)}$ and $\hbar = 1$ into account].

\subsection{Transition to coordinates of relative distances
\label{sec.2.4}}

Let us rewrite formalism above via coordinates of relative distances.
We define coordinate of centers of masses for the $\alpha$ particle  as $\mathbf{r}_{\alpha}$, for the daughter nucleus as $\mathbf{R}_{A}$, and
for the complete system as $\mathbf{R}$:
%
%
\begin{equation}
\begin{array}{lll}
   \mathbf{r}_{\alpha} = \displaystyle\frac{1}{m_{\alpha}} \displaystyle\sum_{i=1}^{4} m_{i}\, \mathbf{r}_{\alpha i}, &
   \mathbf{R}_{A}      = \displaystyle\frac{1}{m_{A}} \displaystyle\sum_{j=1}^{A} m_{j}\, \mathbf{r}_{A j}, &
   \mathbf{R}          = \displaystyle\frac{m_{A}\mathbf{R}_{A} + m_{\alpha}\mathbf{r}_{\alpha}}{m_{A}+m_{\alpha}} =
     c_{A}\, \mathbf{R}_{A} + c_{\alpha}\, \mathbf{R}_{\alpha},
\end{array}
\label{eq.2.4.1}
\end{equation}
where $m_{\alpha}$ and $m_{A}$ are masses of the $\alpha$ particle and daughter nucleus, and
we introduced new coefficients $c_{A} = \frac{m_{A}}{m_{A}+m_{\alpha}}$ and $c_{\alpha} = \frac{m_{\alpha}}{m_{A}+m_{\alpha}}$.
Introducing new relative coordinate $\mathbf{r}$,
new relative coordinates $\rhobf_{\alpha i}$ for nucleons of the $\alpha$-particle,
new relative coordinates $\rhobf_{A j}$ for nucleons (with possible hyperon) for the daughter nucleus
as
\begin{equation}
\begin{array}{lll}
   \mathbf{r} = \mathbf{r}_{\alpha} - \mathbf{R}_{A}, &
   \rhobf_{\alpha i} = \mathbf{r}_{\alpha i} - \mathbf{r}_{\alpha}, &
   \rhobf_{A j} = \mathbf{r}_{j} - \mathbf{R}_{A},
\end{array}
\label{eq.2.4.2}
\end{equation}
we obtain new independent variables $\mathbf{R}$, $\mathbf{r}$,
$\rhobf_{\alpha j}$ ($i=1, 2, 3$) and
$\rhobf_{Aj}$ ($j=1 \ldots A-1$).
%
We rewrite old coordinates $\mathbf{r}_{\alpha i}$, $\mathbf{r}_{Aj}$ of nucleons via new coordinates $\rhobf_{\alpha i}$
(see calculations in Appendix~\ref{sec.app.1}):
\begin{equation}
\begin{array}{lll}
  \mathbf{r}_{\alpha i} = \rhobf_{\alpha i} + \mathbf{R} + c_{A}\, \mathbf{r}, &
  \mathbf{r}_{Aj} = \rhobf_{A j} + \mathbf{R} - c_{\alpha}\, \mathbf{r}.
\end{array}
\label{eq.2.4.6}
\end{equation}
For numbers $i=n$ and $j=A$ it is more convenient to use
\begin{equation}
\begin{array}{ll}
  \mathbf{r}_{\alpha n} = \mathbf{R} + c_{A} \mathbf{r} -
    \displaystyle\frac{1}{m_{n}} \displaystyle\sum_{k=1}^{n-1} m_{k}\, \rhobf_{\alpha k}, &
\hspace{5mm}
  \mathbf{r}_{AA} = \mathbf{R} - c_{\alpha} \mathbf{r} -
    \displaystyle\frac{1}{m_{AA}} \displaystyle\sum_{k=1}^{A-1} m_{k}\, \rhobf_{A k}.
\end{array}
\label{eq.2.4.7}
\end{equation}

We calculate momenta connected with new independent variables $\mathbf{R}$, $\mathbf{r}$, $\rhobf_{\alpha i}$, $\rhobf_{A j}$ (at $j = 1 \ldots A-1$, $i = 1 \ldots n-1$, $n=4$ for the $\alpha$ particle).
We obtain (see Appendix~\ref{sec.app.1} for details):
\begin{equation}
\begin{array}{lclll}
  \vspace{1mm}
  \mathbf{p}_{\alpha i} =
  \displaystyle\frac{m_{\alpha i}}{m_{A} + m_{\alpha}}\, \mathbf{P} +
    \displaystyle\frac{m_{\alpha i}}{m_{\alpha}}\,\mathbf{p} +
    \displaystyle\frac{m_{\alpha} - m_{\alpha i}}{m_{\alpha}}\, \mathbf{\tilde{p}}_{\alpha i} -
    \displaystyle\frac{m_{\alpha i}}{m_{\alpha}}\,
      \displaystyle\sum_{k=1, k \ne i}^{n-1} \mathbf{\tilde{p}}_{\alpha k} \quad
      {\rm at }\; i = 1 \ldots n-1, \\

  \vspace{1mm}
  \mathbf{p}_{\alpha n} =
  \displaystyle\frac{m_{\alpha n}}{m_{A} + m_{\alpha}}\, \mathbf{P} +
    \displaystyle\frac{m_{\alpha n}}{m_{\alpha}}\,\mathbf{p} -
    \displaystyle\frac{m_{\alpha n}}{m_{\alpha}}\,
      \displaystyle\sum_{k=1}^{n-1} \mathbf{\tilde{p}}_{\alpha k}, \\

  \vspace{1mm}
  \mathbf{p}_{Aj} =
  \displaystyle\frac{m_{Aj}}{m_{A} + m_{\alpha}}\, \mathbf{P} -
    \displaystyle\frac{m_{Aj}}{m_{A}}\,\mathbf{p} +
    \displaystyle\frac{m_{A} - m_{Aj}}{m_{A}}\, \mathbf{\tilde{p}}_{Aj} -
    \displaystyle\frac{m_{Aj}}{m_{A}}\,
      \displaystyle\sum_{k=1, k \ne j}^{A-1} \mathbf{\tilde{p}}_{Ak} \quad {\rm at }\; j = 1 \ldots A-1, \\

  \vspace{1mm}
  \mathbf{p}_{AA} =
  \displaystyle\frac{m_{AA}}{m_{A} + m_{\alpha}}\, \mathbf{P} -
    \displaystyle\frac{m_{AA}}{m_{A}}\,\mathbf{p} -
    \displaystyle\frac{m_{AA}}{m_{A}}\,
      \displaystyle\sum_{k=1}^{A-1} \mathbf{\tilde{p}}_{Ak}.
\end{array}
\label{eq.2.4.13}
\end{equation}

\subsection{Operator of emission with relative coordinates
\label{sec.2.5}}

Now we will find operator of emission in new relative coordinates.
For this, we start from (\ref{eq.2.3.4}), rewriting this expression via relative momenta:
\begin{equation}
\begin{array}{lcl}
  \hat{H}_{\gamma} & = &
  -\, \sqrt{\displaystyle\frac{2\pi c^{2}}{\hbar w_{\rm ph}}}\;
  \displaystyle\sum_{i=1}^{4}
  \displaystyle\sum\limits_{\alpha=1,2}
    e^{-i\, \mathbf{k_{\rm ph}r}_{i}}\,
  \biggl\{
    \mu_{N}\, \displaystyle\frac{2 z_{i} m_{\rm p}}{m_{\alpha i}}\: \mathbf{e}^{(\alpha)} \cdot \mathbf{p}_{\alpha i} +
    i\, \mu_{i}^{\rm (an)}\, \sigmabf \cdot \Bigl( - \hbar \bigl[ \mathbf{k_{\rm ph}} \times \mathbf{e}^{(\alpha)} \bigr] + \bigl[ \mathbf{p}_{\alpha i} \times \mathbf{e}^{(\alpha)} \bigr] \Bigr)
  \biggr\}\; - \\

  & - &
  \sqrt{\displaystyle\frac{2\pi c^{2}}{\hbar w_{\rm ph}}}\;
  \displaystyle\sum_{j=1}^{A}
  \displaystyle\sum\limits_{\alpha=1,2}
    e^{-i\, \mathbf{k_{\rm ph}r}_{j}}\;
    \biggl\{
      \mu_{N}\, \displaystyle\frac{2 z_{j} m_{\rm p}}{m_{Aj}}\: \mathbf{e}^{(\alpha)} \cdot \mathbf{p}_{Aj} +
      i\, \mu_{j}^{\rm (an)}\, \sigmabf \cdot \Bigl( - \hbar \bigl[ \mathbf{k_{\rm ph}} \times \mathbf{e}^{(\alpha)} \bigr] + \bigl[ \mathbf{p}_{Aj} \times \mathbf{e}^{(\alpha)} \bigr] \Bigr)
    \biggr\}.
\end{array}
\label{eq.2.5.1}
\end{equation}
Substituting formulas (\ref{eq.2.4.13}) to these expressions, we find (see calculations in Appendix~\ref{sec.app.2}):
\begin{equation}
  \hat{H}_{\gamma} = \hat{H}_{P} + \hat{H}_{p} + \Delta\,\hat{H}_{\gamma} + \hat{H}_{k},
\label{eq.2.5.2}
\end{equation}
where
\begin{equation}
\begin{array}{lll}
\vspace{-0.1mm}
  & & \hat{H}_{P} =
  -\, \sqrt{\displaystyle\frac{2\pi c^{2}}{\hbar w_{\rm ph}}}\;
  \mu_{N}\, \displaystyle\frac{2 m_{\rm p}}{m_{A} + m_{\alpha}}\;
  e^{-i\, \mathbf{k_{\rm ph}} \mathbf{R}}
  \displaystyle\sum\limits_{\alpha=1,2}
  \biggl\{
    e^{-i\, c_{A}\, \mathbf{k_{\rm ph}} \mathbf{r}}
      \displaystyle\sum_{i=1}^{4} z_{i}\, e^{-i\, \mathbf{k_{\rm ph}} \rhobf_{\alpha i}} +
    e^{i\, c_{\alpha}\, \mathbf{k_{\rm ph}} \mathbf{r} }
      \displaystyle\sum_{j=1}^{A} z_{j}\, e^{-i\, \mathbf{k_{\rm ph}} \rhobf_{Aj}}
  \biggr\}\, \mathbf{e}^{(\alpha)} \cdot \mathbf{P} + \\
\vspace{0.5mm}
  & - &
  \sqrt{\displaystyle\frac{2\pi c^{2}}{\hbar w_{\rm ph}}}\;
    \displaystyle\frac{i}{m_{A} + m_{\alpha}}\;
    e^{-i\, \mathbf{k_{\rm ph}} \mathbf{R}}\,
    \displaystyle\sum\limits_{\alpha=1,2}
    \biggl\{
      e^{-i\, c_{A}\, \mathbf{k_{\rm ph}} \mathbf{r}}\,
        \displaystyle\sum_{i=1}^{4} \mu_{i}^{\rm (an)}\, m_{\alpha i}\, e^{-i\, \mathbf{k_{\rm ph}} \rhobf_{\alpha i}}\, \sigmabf  +
      e^{i\, c_{\alpha}\, \mathbf{k_{\rm ph}} \mathbf{r}}\,
        \displaystyle\sum_{j=1}^{A} \mu_{j}^{\rm (an)}\, m_{Aj}\, e^{-i\, \mathbf{k_{\rm ph}} \rhobf_{Aj}}\, \sigmabf
  \biggr\}\, \times \\
  & \times & \cdot \bigl[ \mathbf{P} \times \mathbf{e}^{(\alpha)} \bigr],
\end{array}
\label{eq.2.5.3}
\end{equation}
\begin{equation}
\begin{array}{lll}
\vspace{-0.1mm}
  \hat{H}_{p} & = &
  -\, \sqrt{\displaystyle\frac{2\pi c^{2}}{\hbar w_{\rm ph}}}\;
  2\, \mu_{N}\,  m_{\rm p}\,
  e^{-i\, \mathbf{k_{\rm ph}} \mathbf{R}}
  \displaystyle\sum\limits_{\alpha=1,2}
  \biggl\{
    e^{-i\, c_{A} \mathbf{k_{\rm ph}} \mathbf{r}}\, \displaystyle\frac{1}{m_{\alpha}}\,
      \displaystyle\sum_{i=1}^{4} z_{i}\, e^{-i\, \mathbf{k_{\rm ph}} \rhobf_{\alpha i}} -
    e^{i\, c_{\alpha} \mathbf{k_{\rm ph}} \mathbf{r}}\,  \displaystyle\frac{1}{m_{A}}\,
      \displaystyle\sum_{j=1}^{A} z_{j}\, e^{-i\, \mathbf{k_{\rm ph}} \rhobf_{Aj}}
  \biggr\}\; \mathbf{e}^{(\alpha)} \cdot \mathbf{p}\; - \\
\vspace{0.3mm}
  & - &
  i\, \sqrt{\displaystyle\frac{2\pi c^{2}}{\hbar w_{\rm ph}}}\;
  e^{-i\, \mathbf{k_{\rm ph}} \mathbf{R}}
  \displaystyle\sum\limits_{\alpha=1,2}
  \biggl\{
    e^{-i\, c_{A} \mathbf{k_{\rm ph}} \mathbf{r}} \displaystyle\frac{1}{m_{\alpha}}\,
    \displaystyle\sum_{i=1}^{4}
      \mu_{i}^{\rm (an)}\, m_{\alpha i}\;
      e^{-i\, \mathbf{k_{\rm ph}} \rhobf_{\alpha i}}\, \sigmabf - \\
  & - &
    e^{i\, c_{\alpha} \mathbf{k_{\rm ph}} \mathbf{r}} \displaystyle\frac{1}{m_{A}}
    \displaystyle\sum_{j=1}^{A}
      \mu_{j}^{\rm (an)}\, m_{Aj}\;
      e^{-i\, \mathbf{k_{\rm ph}} \rhobf_{Aj}}\, \sigmabf \biggr\}
  \cdot \bigl[ \mathbf{p} \times \mathbf{e}^{(\alpha)} \bigr],
\end{array}
\label{eq.2.5.4}
\end{equation}
\begin{equation}
\begin{array}{lcl}
  \hat{H}_{k} & = &
  i\, \hbar\,
  \sqrt{\displaystyle\frac{2\pi c^{2}}{\hbar w_{\rm ph}}}\:
  e^{-i\, \mathbf{k_{\rm ph}} \mathbf{R}}\,
  \displaystyle\sum\limits_{\alpha=1,2}
  \biggl\{
    e^{-i\, c_{A}\, \mathbf{k_{\rm ph}} \mathbf{r}}\,
    \displaystyle\sum_{i=1}^{4}
      \mu_{i}^{\rm (an)}\, e^{-i\, \mathbf{k_{\rm ph}} \rhobf_{\alpha i}}\, \sigmabf +
    e^{i\, c_{\alpha}\, \mathbf{k_{\rm ph}} \mathbf{r}}\,
    \displaystyle\sum_{j=1}^{A}
      \mu_{j}^{\rm (an)}\, e^{-i\, \mathbf{k_{\rm ph}} \rhobf_{Aj}}\, \sigmabf
  \biggr\}
  \cdot \bigl[ \mathbf{k_{\rm ph}} \times \mathbf{e}^{(\alpha)} \bigr],
\end{array}
\label{eq.2.5.5}
\end{equation}
and $\Delta\,\hat{H}_{\gamma}$ is calculated in Appendix~\ref{sec.app.2} [see Eqs.~(\ref{eq.app.2.8}) and (\ref{eq.app.2.9}) in this Section].
A summation of expression (\ref{eq.2.5.4}) and $\hat{H}_{k}$ is many-nucleon generalization of operator of emission $\hat{W}$ in Eq.~(6) in Ref.~\cite{Maydanyuk.2012.PRC} with
included anomalous magnetic momenta for nucleons.


\subsection{Wave function of the $\alpha$-nucleus system
\label{sec.2.6}}

Emission of the bremsstrahlung photons is caused by the relative motion of nucleons of the full nuclear system. However, as the most intensive emission of photons is formed by relative motion of the $\alpha$ particle related to the nucleus, it is sensible to represent the total wave function via coordinates of relative motion of these complicated objects.
We follow the formalism given in \cite{Maydanyuk_Zhang.2015.PRC} for the proton-nucleus scattering, and we add description of many-nucleon structure of the $\alpha$-particle as in Ref.~\cite{Maydanyuk_Zhang_Zou.2016.PRC}.
Such a presentation of the wave function allows us to take into account the most accurately the leading contribution of the wave function of relative motion into the bremsstrahlung spectrum,
while the many nucleon structure of the $\alpha$ particle and nucleus should provide only minor corrections .
%
%
Before developing a detailed many-nucleon formalism for such a problem, we shall clarify first if the many-nucleon structure of the $\alpha$ nucleus system is visible in the experimental bremsstrahlung spectra.
In this regard, estimation of many-nucleon contribution in the full bremsstrahlung spectrum is well described task.
Thus, we define the wave function of the full nuclear system as
\begin{equation}
  \Psi =
  \Phi (\mathbf{R}) \cdot
  \Phi_{\rm \alpha - nucl} (\mathbf{r}) \cdot
  \psi_{\rm nucl} (\beta_{A}) \cdot
  \psi_{\alpha} (\beta_{\alpha}) +
  \Delta \Psi,
\label{eq.2.6.1}
\end{equation}
where
\begin{equation}
\begin{array}{lcl}
  \psi_{\rm nucl} (\beta_{A}) =
  \psi_{\rm nucl} (1 \cdots A ) =
  \displaystyle\frac{1}{\sqrt{A!}}
  \displaystyle\sum\limits_{p_{A}}
    (-1)^{\varepsilon_{p_{A}}}
    \psi_{\lambda_{1}}(1)
    \psi_{\lambda_{2}}(2) \ldots
    \psi_{\lambda_{A}}(A), \\

  \psi_{\alpha} (\beta_{\alpha}) =
  \psi_{\alpha} (1 \cdots 4) =
  \displaystyle\frac{1}{\sqrt{4!}}
  \displaystyle\sum\limits_{p_{\alpha}}
    (-1)^{\varepsilon_{p_{\alpha}}}
    \psi_{\lambda_{1}}(1)
    \psi_{\lambda_{2}}(2)
    \psi_{\lambda_{3}}(3)
    \psi_{\lambda_{4}}(4).
\end{array}
\label{eq.2.6.2}
\end{equation}
Here, $\beta_{\alpha}$ is the set of numbers $1 \cdots 4$ of nucleons of the $\alpha$ particle,
$\beta_{A}$ is the set of numbers $1 \cdots A$ of nucleons of the nucleus,
$\Phi (\mathbf{R})$ is the function describing motion of center-of-mass of the full nuclear system,
$\Phi_{\rm \alpha - nucl} (\mathbf{r})$ is the function describing relative motion of the $\alpha$ particle concerning to nucleus (without description of internal relative motions of nucleons in the $\alpha$ particle and nucleus),
$\psi_{\alpha} (\beta_{\alpha})$ is the many-nucleon function dependent on nucleons of the $\alpha$ particle (it determines space state on the basis of relative distances $\rhobf_{1} \cdots \rhobf_{4}$ of nucleons of the $\alpha$ particle concerning to its center-of-mass),
$\psi_{\rm nucl} (\beta_{A})$ is the many-nucleon function dependent on nucleons of the nucleus.
Summation in Eqs.~(\ref{eq.2.3.3}) is performed over all $A!$ permutations of coordinates or states of nucleons. One-nucleon functions $\psi_{\lambda_{s}}(s)$ represent the multiplication of space and spin-isospin
functions as
$\psi_{\lambda_{s}} (s) = \varphi_{n_{s}} (\mathbf{r}_{s})\, \bigl|\, \sigma^{(s)} \tau^{(s)} \bigr\rangle$,
%
%
where
$\varphi_{n_{s}}$ is the space function of the nucleon with number $s$,
$n_{s}$ is the number of state of the space function of the nucleon with number $s$,
$\bigl|\, \sigma^{(s)} \tau^{(s)} \bigr\rangle$ is the spin-isospin function of the nucleon with number $s$.
%

In definition (\ref{eq.2.6.1}) of the wave function we have also included the new term $\Delta \Psi$. It is correction, which should take into account fully anty-symmetric formulation of wave function for all nucleons.
However, in this work we shall neglect by this correction, supposing that it has minor influence on studied physical effects here.


\subsection{Matrix element of emission
\label{sec.2.7}}

We define matrix element of emission of the bremsstrahlung photons, using the wave functions $\Psi_{i}$ and $\Psi_{f}$ of
the full nuclear system in states before emission of photons ($i$-state) and after such emission ($f$-state),
as
%
$\langle \Psi_{f} |\, \hat{H}_{\gamma} |\, \Psi_{i} \rangle$.
%
In this matrix element we should integrate over all independent variables.
These variables are space variables $\mathbf{R}$, $\mathbf{r}$, $\rhobf_{\alpha n}$, $\rhobf_{Am}$.
Here, we should take into account space representation of all used momenta $\mathbf{P}$, $\mathbf{p}$, $\mathbf{\tilde{p}}_{\alpha n}$, $\mathbf{\tilde{p}}_{A m}$.
Substituting formulas (\ref{eq.2.5.1}) for operator of emission to matrix element, we obtain:
\begin{equation}
\begin{array}{lcl}
  \vspace{0mm}
  \langle \Psi_{f} |\, \hat{H}_{\gamma} |\, \Psi_{i} \rangle  \;\; = \;\;
    \sqrt{\displaystyle\frac{2\pi\, c^{2}}{\hbar w_{\rm ph}}}\,
  \Bigl\{ M_{1} + M_{2} + M_{3} \Bigr\},
\end{array}
\label{eq.2.7.2}
\end{equation}
where
\begin{equation}
\begin{array}{lcl}
\vspace{1mm}
  M_{1} & = &
  \sqrt{\displaystyle\frac{\hbar w_{\rm ph}} {2\pi c^{2}}}\;
  \biggl\langle \Psi_{f}\, \biggl|\, \hat{H}_{P} \biggr|\, \Psi_{i} \biggr\rangle\; = \\

  & = &
  -\, \displaystyle\frac{1}{m_{A} + m_{\alpha}}\,
  \displaystyle\sum\limits_{\alpha=1,2}
  \biggl\langle \Psi_{f}\, \biggl|\,
    2\, \mu_{N}\, m_{\rm p}\;
    e^{-i\, \mathbf{k_{\rm ph}} \mathbf{R}}
    \biggl\{
      e^{-i\, c_{A}\, \mathbf{k_{\rm ph}} \mathbf{r}}
        \displaystyle\sum_{i=1}^{4} z_{i}\, e^{-i\, \mathbf{k_{\rm ph}} \rhobf_{\alpha i}} +
      e^{i\, c_{\alpha}\, \mathbf{k_{\rm ph}} \mathbf{r} }
        \displaystyle\sum_{j=1}^{A} z_{j}\, e^{-i\, \mathbf{k_{\rm ph}} \rhobf_{Aj}}
    \biggr\}\, \mathbf{e}^{(\alpha)} \cdot \mathbf{P} + \\

  & + &
    i\:
    e^{-i\, \mathbf{k_{\rm ph}} \mathbf{R}}\,
    \biggl\{
      e^{-i\, c_{A}\, \mathbf{k_{\rm ph}} \mathbf{r}}\,
        \displaystyle\sum_{i=1}^{4} \mu_{i}^{\rm (an)}\, m_{\alpha i}\, e^{-i\, \mathbf{k_{\rm ph}} \rhobf_{\alpha i}}\, \sigmabf  +
      e^{i\, c_{\alpha}\, \mathbf{k_{\rm ph}} \mathbf{r}}\,
        \displaystyle\sum_{j=1}^{A} \mu_{j}^{\rm (an)}\, m_{Aj}\, e^{-i\, \mathbf{k_{\rm ph}} \rhobf_{Aj}}\, \sigmabf
  \biggr\}\, \cdot \bigl[ \mathbf{P} \times \mathbf{e}^{(\alpha)} \bigr]\,
  \biggr|\, \Psi_{i} \biggr\rangle,
\end{array}
\label{eq.2.7.3.a}
\end{equation}
\begin{equation}
\begin{array}{lcl}
\vspace{1mm}
  M_{2} & = &
  \sqrt{\displaystyle\frac{\hbar w_{\rm ph}} {2\pi c^{2}}}\;
  \biggl\langle \Psi_{f}\, \biggl|\, \hat{H}_{p} \biggr|\, \Psi_{i} \biggr\rangle\; = \\

  & = &
  - \displaystyle\sum\limits_{\alpha=1,2}
    \biggl\langle
      \Psi_{f}\,
    \biggl|\,
  2\, \mu_{N}\,  m_{\rm p}\,
  e^{-i\, \mathbf{k_{\rm ph}} \mathbf{R}}
  \biggl\{
    e^{-i\, c_{A} \mathbf{k_{\rm ph}} \mathbf{r}}\, \displaystyle\frac{1}{m_{\alpha}}\,
      \displaystyle\sum_{i=1}^{4} z_{i}\, e^{-i\, \mathbf{k_{\rm ph}} \rhobf_{\alpha i}} -
    e^{i\, c_{\alpha} \mathbf{k_{\rm ph}} \mathbf{r}}\,  \displaystyle\frac{1}{m_{A}}\,
      \displaystyle\sum_{j=1}^{A} z_{j}\, e^{-i\, \mathbf{k_{\rm ph}} \rhobf_{Aj}}
  \biggr\}\; \mathbf{e}^{(\alpha)} \cdot \mathbf{p}\; + \\
  & + &
  i\,
  e^{-i\, \mathbf{k_{\rm ph}} \mathbf{R}}
  \biggl\{
    e^{-i\, c_{A} \mathbf{k_{\rm ph}} \mathbf{r}} \displaystyle\frac{1}{m_{\alpha}}\,
    \displaystyle\sum_{i=1}^{4}
      \mu_{i}^{\rm (an)}\, m_{\alpha i}\;
      e^{-i\, \mathbf{k_{\rm ph}} \rhobf_{\alpha i}}\, \sigmabf -
    e^{i\, c_{\alpha} \mathbf{k_{\rm ph}} \mathbf{r}} \displaystyle\frac{1}{m_{A}}
    \displaystyle\sum_{j=1}^{A}
      \mu_{j}^{\rm (an)}\, m_{Aj}\;
      e^{-i\, \mathbf{k_{\rm ph}} \rhobf_{Aj}}\, \sigmabf \biggr\}
    \cdot \bigl[ \mathbf{p} \times \mathbf{e}^{(\alpha)} \bigr]
  \biggr|\, \Psi_{i}\, \biggr\rangle ,
\end{array}
\label{eq.2.7.3.b}
\end{equation}
\begin{equation}
\begin{array}{lcl}
  M_{3} & = &
    \biggl\langle
      \Psi_{f}
    \biggl|
      \Delta \hat{H}_{\gamma} + \hat{H}_{k}
    \biggr|
      \Psi_{i}
    \biggr\rangle.
\end{array}
\label{eq.2.7.3.c}
\end{equation}


\subsection{Integration over space variable $\mathbf{R}$
\label{sec.2.8}}

In Eq.~(\ref{eq.2.6.1}) we defined the wave function of the full nuclear system.
Here, $\Phi (\mathbf{R})$ is wave function describing evolution of center of mass of the full nuclear system.
We rewrite this wave function as
\begin{equation}
\begin{array}{lcl}
  \Psi = \Phi (\mathbf{R}) \cdot F (\mathbf{r}, \beta_{A}, \beta_{\alpha}), &

  F (\mathbf{r}, \beta_{A}, \beta_{\alpha}) =
  \Phi_{\rm \alpha - nucl} (\mathbf{r}) \cdot
  \psi_{\rm nucl} (\beta_{A}) \cdot
  \psi_{\alpha} (\beta_{\alpha}),
\end{array}
\label{eq.2.8.1}
\end{equation}
Now we assume approximated form for this wave function before and after emission of photons as
\begin{equation}
  \Phi_{\bar{s}} (\mathbf{R}) =  e^{-i\,\mathbf{K}_{\bar{s}}\cdot\mathbf{R}},
\label{eq.2.8.2}
\end{equation}
%
%
%
where $\bar{s} = i$ or $f$ (indexes $i$ and $f$ denote the initial state,
i.e. the state before emission of photon,
and the final state, i.e. the state after emission of photon),
$\mathbf{K}_{s}$ is momentum of the total system~\cite{Kopitin.1997.YF}.
%
Also we assume $\mathbf{K}_{i} = 0$
as the $\alpha$-decaying nuclear system before emission of photons is not moving in the laboratory frame.

Let us consider just contribution $M_{2}$ to the full matrix element.
We calculate
\begin{equation}
\begin{array}{lll}
\vspace{-0.1mm}
  & & M_{2} =
  -\, \displaystyle\int_{-\infty}^{+\infty}  e^{i\, (\mathbf{K}_{f} - \mathbf{k}_{\rm ph}) \cdot\mathbf{R}}\;  \mathbf{dR}\; \times \\
\vspace{-0.1mm}
  & \times &
  \displaystyle\sum\limits_{\alpha=1,2}
  \biggl\langle F_{f}\, \biggl|\,
  2\, \mu_{N}\,  m_{\rm p}\,
  \Bigl\{
    e^{-i\, c_{A} \mathbf{k_{\rm ph}} \mathbf{r}}\, \displaystyle\frac{1}{m_{\alpha}}\,
      \displaystyle\sum_{i=1}^{4} z_{i}\, e^{-i\, \mathbf{k_{\rm ph}} \rhobf_{\alpha i}} -
    e^{i\, c_{\alpha} \mathbf{k_{\rm ph}} \mathbf{r}}\,  \displaystyle\frac{1}{m_{A}}\,
      \displaystyle\sum_{j=1}^{A} z_{j}\, e^{-i\, \mathbf{k_{\rm ph}} \rhobf_{Aj}}
  \Bigr\}\; \mathbf{e}^{(\alpha)} \cdot \mathbf{p}\; + \\
  & + &
  i\,
  \Bigl\{
    e^{-i\, c_{A} \mathbf{k_{\rm ph}} \mathbf{r}} \displaystyle\frac{1}{m_{\alpha}}\,
    \displaystyle\sum_{i=1}^{4}
      \mu_{i}^{\rm (an)}\, m_{\alpha i}\;
      e^{-i\, \mathbf{k_{\rm ph}} \rhobf_{\alpha i}}\, \sigmabf -
    e^{i\, c_{\alpha} \mathbf{k_{\rm ph}} \mathbf{r}} \displaystyle\frac{1}{m_{A}}
    \displaystyle\sum_{j=1}^{A}
      \mu_{j}^{\rm (an)}\, m_{Aj}\;
      e^{-i\, \mathbf{k_{\rm ph}} \rhobf_{Aj}}\, \sigmabf \Bigr\}
    \cdot \bigl[ \mathbf{p} \times \mathbf{e}^{(\alpha)} \bigr]\,
  \biggr|\, F_{i}\, \biggr\rangle.
\end{array}
\label{eq.2.8.5}
\end{equation}
From definition of $\delta$-function we have:
\begin{equation}
\begin{array}{lll}
  \displaystyle\int_{-\infty}^{+\infty}  e^{i\, (\mathbf{K}_{f} - \mathbf{k}) \cdot\mathbf{R}}\;  \mathbf{dR}\; =
  (2\pi)^{3} \delta (\mathbf{K}_{f} - \mathbf{k}).
\end{array}
\label{eq.2.8.6}
\end{equation}
Then, from (\ref{eq.2.8.5}) we obtain:
\begin{equation}
\begin{array}{lll}
\vspace{-0.1mm}
  & & M_{2} = -\, (2\pi)^{3} \delta (\mathbf{K}_{f} - \mathbf{k}_{\rm ph})\; \times \\
\vspace{-0.1mm}
  & \times &
  \displaystyle\sum\limits_{\alpha=1,2}
  \biggl\langle F_{f}\, \biggl|\,
  2\, \mu_{N}\,  m_{\rm p}\,
  \Bigl\{
    e^{-i\, c_{A} \mathbf{k_{\rm ph}} \mathbf{r}}\, \displaystyle\frac{1}{m_{\alpha}}\,
      \displaystyle\sum_{i=1}^{4} z_{i}\, e^{-i\, \mathbf{k_{\rm ph}} \rhobf_{\alpha i}} -
    e^{i\, c_{\alpha} \mathbf{k_{\rm ph}} \mathbf{r}}\,  \displaystyle\frac{1}{m_{A}}\,
      \displaystyle\sum_{j=1}^{A} z_{j}\, e^{-i\, \mathbf{k_{\rm ph}} \rhobf_{Aj}}
  \Bigr\}\; \mathbf{e}^{(\alpha)} \cdot \mathbf{p}\; + \\
  & + &
  i\,
  \Bigl\{
    e^{-i\, c_{A} \mathbf{k_{\rm ph}} \mathbf{r}} \displaystyle\frac{1}{m_{\alpha}}\,
    \displaystyle\sum_{i=1}^{4}
      \mu_{i}^{\rm (an)}\, m_{\alpha i}\;
      e^{-i\, \mathbf{k_{\rm ph}} \rhobf_{\alpha i}}\, \sigmabf -
    e^{i\, c_{\alpha} \mathbf{k_{\rm ph}} \mathbf{r}} \displaystyle\frac{1}{m_{A}}
    \displaystyle\sum_{j=1}^{A}
      \mu_{j}^{\rm (an)}\, m_{Aj}\;
      e^{-i\, \mathbf{k_{\rm ph}} \rhobf_{Aj}}\, \sigmabf \Bigr\}
    \cdot \bigl[ \mathbf{p} \times \mathbf{e}^{(\alpha)} \bigr]\,
  \biggr|\, F_{i}\, \biggr\rangle.
\end{array}
\label{eq.2.8.7}
\end{equation}
In this formula, we have integration over space variables $\mathbf{r}$, $\rhobf_{\alpha n}$, $\rhobf_{Am}$.

\subsection{Electric and magnetic form-factors
\label{sec.2.9}}

We substitute explicit formulation (\ref{eq.2.8.1}) for wave function $F (\mathbf{r}, \beta_{A}, \beta_{\alpha})$ to the obtained matrix element (\ref{eq.2.8.7}).
We calculate this matrix element and obtain (see Appendix~\ref{sec.app.3} for details):
\begin{equation}
\begin{array}{lll}
\vspace{-0.1mm}
  M_{2} & = &
  i \hbar\, (2\pi)^{3} \delta (\mathbf{K}_{f} - \mathbf{k}_{\rm ph}) \cdot
  \displaystyle\sum\limits_{\alpha=1,2}
  \displaystyle\int\limits_{}^{}
    \Phi_{\rm \alpha - nucl, f}^{*} (\mathbf{r})\;
    e^{i\, \mathbf{k}_{\rm ph} \mathbf{r}}\; \times \\
\vspace{0.5mm}
  & \times &
  \biggl\{
  2\, \mu_{N}\,  m_{\rm p}\,
  \Bigl[
    e^{-i\, c_{A} \mathbf{k_{\rm ph}} \mathbf{r}}\, \displaystyle\frac{1}{m_{\alpha}}\, F_{\alpha,\, {\rm el}} -
    e^{i\, c_{\alpha} \mathbf{k_{\rm ph}} \mathbf{r}}\,  \displaystyle\frac{1}{m_{A}}\, F_{A,\, {\rm el}}
  \Bigr]\,
  e^{- i\, \mathbf{k}_{\rm ph} \mathbf{r}} \cdot
  \mathbf{e}^{(\alpha)}\, \mathbf{\displaystyle\frac{d}{dr}}\; + \\
  & + &
  i\,
  \Bigl[
    e^{-i\, c_{A} \mathbf{k_{\rm ph}} \mathbf{r}} \displaystyle\frac{1}{m_{\alpha}}\, \mathbf{F}_{\alpha,\, {\rm mag}} -
    e^{i\, c_{\alpha} \mathbf{k_{\rm ph}} \mathbf{r}} \displaystyle\frac{1}{m_{A}}\, \mathbf{F}_{A,\, {\rm mag}}
  \Bigr]\,
  e^{- i\, \mathbf{k}_{\rm ph} \mathbf{r}} \cdot
  \Bigl[ \mathbf{\displaystyle\frac{d}{dr}} \times \mathbf{e}^{(\alpha)} \Bigr]\,
  \biggr\} \cdot
  \Phi_{\rm \alpha - nucl, i} (\mathbf{r})\; \mathbf{dr}.
\end{array}
\label{eq.2.9.11}
\end{equation}
Here, we introduce new definitions of electric and magnetic form-factors of the $\alpha$-particle and nucleus as
\begin{equation}
\begin{array}{lll}
\vspace{1mm}
  F_{\alpha,\, {\rm el}} = &
    \displaystyle\sum\limits_{n=1}^{4}
    \biggl\langle \psi_{\alpha, f} (\beta_{\alpha})\, \biggl|\,
      z_{n}\, e^{-i \mathbf{k}_{\rm ph} \rhobf_{\alpha n} }
    \biggr|\,  \psi_{\alpha, i} (\beta_{\alpha}) \biggr\rangle , \\

\vspace{1mm}
  F_{A,\, {\rm el}} = &
    \displaystyle\sum\limits_{m=1}^{A}
    \biggl\langle \psi_{\rm nucl, f} (\beta_{A}) \biggl|\,
      z_{m}\, e^{-i \mathbf{k}_{\rm ph} \rhobf_{A m} }
    \biggr|\, \psi_{\rm nucl, i} (\beta_{A}) \biggr\rangle , \\

\vspace{1mm}
  \mathbf{F}_{\alpha,\, {\rm mag}} = &
    \displaystyle\sum_{i=1}^{4}
    \Bigl\langle \psi_{\alpha, f} (\beta_{\alpha})\, \Bigl|\,
      \mu_{i}^{\rm (an)}\, m_{\alpha i}\; e^{-i\, \mathbf{k_{\rm ph}} \rhobf_{\alpha i}}\, \sigmabf
    \Bigr| \psi_{\alpha, i} (\beta_{\alpha}) \Bigr\rangle, \\

  \mathbf{F}_{A,\, {\rm mag}} = &
    \displaystyle\sum_{j=1}^{A}
    \Bigl\langle \psi_{\rm nucl, f} (\beta_{A})\, \Bigl|\,
        \mu_{j}^{\rm (an)}\, m_{Aj}\; e^{-i\, \mathbf{k_{\rm ph}} \rhobf_{Aj}}\, \sigmabf
    \Bigr| \psi_{\rm nucl, i} (\beta_{A}) \Bigr\rangle.
\end{array}
\label{eq.2.9.8}
\end{equation}

\subsection{Introduction of effective electric charge and magnetic moment of the full system
\label{sec.2.10}}

Let us consider the first two terms inside the first brackets in Eq.~(\ref{eq.2.9.11}).
In the first approximation, electrical form-factors tend to electric charges of $\alpha$-particle and the daughter nucleus.
So, we write:
\begin{equation}
\begin{array}{lll}
  e^{-i\, c_{A} \mathbf{k_{\rm ph}} \mathbf{r}}\, \displaystyle\frac{1}{m_{\alpha}}\, F_{\alpha,\, {\rm el}} -
  e^{i\, c_{\alpha} \mathbf{k_{\rm ph}} \mathbf{r}}\,  \displaystyle\frac{1}{m_{A}}\, F_{A,\, {\rm el}} =

  \displaystyle\frac{1}{m} \cdot
  \Bigl[
    e^{-i\, c_{A} \mathbf{k_{\rm ph}} \mathbf{r}}\, \displaystyle\frac{m_{A}}{m_{\alpha} + m_{A}}\, F_{\alpha,\, {\rm el}} -
    e^{i\, c_{\alpha} \mathbf{k_{\rm ph}} \mathbf{r}}\,  \displaystyle\frac{m_{\alpha}}{m_{\alpha} + m_{A}}\, F_{A,\, {\rm el}}
  \Bigr],
\end{array}
\label{eq.2.10.1}
\end{equation}
where
\begin{equation}
  m = \displaystyle\frac{m_{\alpha}\, m_{A}}{m_{\alpha} + m_{A}}.
\label{eq.2.10.2}
\end{equation}
Here, $m$ is reduced mass of system of $\alpha$-particle and the daughter nucleus.

We introduce a new definitions of \emph{effective electric charge} and \emph{effective magnetic moment} of the full system as
\begin{equation}
\begin{array}{lll}
  Z_{\rm eff} (\mathbf{k}_{\rm ph}, \mathbf{r}) =
  e^{i\, \mathbf{k_{\rm ph}} \mathbf{r}}\,
  \Bigl[
    e^{-i\, c_{A} \mathbf{k_{\rm ph}} \mathbf{r}}\, \displaystyle\frac{m_{A}}{m_{\alpha} + m_{A}}\, F_{\alpha,\, {\rm el}} -
    e^{i\, c_{\alpha} \mathbf{k_{\rm ph}} \mathbf{r}}\,  \displaystyle\frac{m_{\alpha}}{m_{\alpha} + m_{A}}\, F_{A,\, {\rm el}}
  \Bigr],
\end{array}
\label{eq.2.10.3}
\end{equation}
\begin{equation}
\begin{array}{lll}
  \textbf{M}_{\rm eff} (\mathbf{k}_{\rm ph}, \mathbf{r}) =
  e^{i\, \mathbf{k_{\rm ph}} \mathbf{r}}\,
  \Bigl[
    e^{-i\, c_{A} \mathbf{k_{\rm ph}} \mathbf{r}}\, \displaystyle\frac{m_{A}}{m_{\alpha} + m_{A}}\, \textbf{F}_{\alpha,\, {\rm mag}} -
    e^{i\, c_{\alpha} \mathbf{k_{\rm ph}} \mathbf{r}}\,  \displaystyle\frac{m_{\alpha}}{m_{\alpha} + m_{A}}\, \textbf{F}_{A,\, {\rm mag}}
  \Bigr].
\end{array}
\label{eq.2.10.4}
\end{equation}
So, from Eq.~(\ref{eq.2.10.1}) we obtain
\begin{equation}
\begin{array}{lll}
  e^{i\, \mathbf{k_{\rm ph}} \mathbf{r}}\,
  \Bigl[
    e^{-i\, c_{A} \mathbf{k_{\rm ph}} \mathbf{r}}\, \displaystyle\frac{1}{m_{\alpha}}\, F_{\alpha,\, {\rm el}} -
    e^{i\, c_{\alpha} \mathbf{k_{\rm ph}} \mathbf{r}}\,  \displaystyle\frac{1}{m_{A}}\, F_{A,\, {\rm el}}
  \Bigr] =
  \displaystyle\frac{1}{m} \cdot Z_{\rm eff} (\mathbf{k}_{\rm ph}, \mathbf{r}),
\end{array}
\label{eq.2.10.5}
\end{equation}
\begin{equation}
\begin{array}{lll}
  e^{i\, \mathbf{k_{\rm ph}} \mathbf{r}}\,
  \Bigl[
    e^{-i\, c_{A} \mathbf{k_{\rm ph}} \mathbf{r}}\, \displaystyle\frac{1}{m_{\alpha}}\, \textbf{F}_{\alpha,\, {\rm mag}} -
    e^{i\, c_{\alpha} \mathbf{k_{\rm ph}} \mathbf{r}}\,  \displaystyle\frac{1}{m_{A}}\, \textbf{F}_{A,\, {\rm mag}} =
  \Bigr] =
  \displaystyle\frac{1}{m} \cdot \textbf{M}_{\rm eff} (\mathbf{k}_{\rm ph}, \mathbf{r}).
\end{array}
\label{eq.2.10.6}
\end{equation}
Now we can rewrite expression (\ref{eq.2.9.11}) for $M_{2}$ via effective electric charge and magnetic moment in a compact form as
\begin{equation}
\begin{array}{lll}
\vspace{-0.2mm}
  M_{2} & =  &
  i \hbar\, (2\pi)^{3}\,
  \displaystyle\frac{1}{m}\,
  \delta (\mathbf{K}_{f} - \mathbf{k}_{\rm ph}) \cdot
  \displaystyle\sum\limits_{\alpha=1,2}
  \displaystyle\int\limits_{}^{}
    \Phi_{\rm \alpha - nucl, f}^{*} (\mathbf{r})\;
    e^{-i\, \mathbf{k_{\rm ph}} \mathbf{r}}\;
    \times \\
  & \times &
  \biggl\{
  2\, \mu_{N}\,  m_{\rm p} \cdot
  Z_{\rm eff} (\mathbf{k}_{\rm ph}, \mathbf{r}) \cdot
  \mathbf{e}^{(\alpha)}\, \mathbf{\displaystyle\frac{d}{dr}} +

  i\, \textbf{M}_{\rm eff} (\mathbf{k}_{\rm ph}, \mathbf{r}) \cdot
  \Bigl[ \mathbf{\displaystyle\frac{d}{dr}} \times \mathbf{e}^{(\alpha)} \Bigr]\,
  \biggr\}\;
  \Phi_{\rm \alpha - nucl, i} (\mathbf{r})\; \mathbf{dr}.
\end{array}
\label{eq.2.10.7}
\end{equation}
So, we have obtained the final formula for the matrix element, where we have our new introduced effective electric charge and magnetic momentum of the full nuclear system (of the $\alpha$ particle and nucleus).

In order to connect the found matrix element~(\ref{eq.2.10.7}) with our previous formalism, we use Eq.~(20) in Ref.~\cite{Maydanyuk_Zhang_Zou.2016.PRC} and
obtain (we use only term of $M_{2}$, $\hbar = 1$ and $c=1$):
\begin{equation}
  M_{2} = -\, \displaystyle\frac{e}{m_{\rm p}} \cdot p_{\rm full}\, \delta(\mathbf{K}_{f} - \mathbf{k}),
\label{eq.2.11.3}
\end{equation}
where
\begin{equation}
\begin{array}{lll}
\vspace{-0.2mm}
  p_{\rm full} & =  &
  -\, \displaystyle\frac{m_{\rm p}}{e}\,
  i \hbar\, (2\pi)^{3}\,
  \displaystyle\frac{1}{m} \cdot
  \displaystyle\sum\limits_{\alpha=1,2}
  \displaystyle\int\limits_{}^{}
    \Phi_{\rm \alpha - nucl, f}^{*} (\mathbf{r})\;
    e^{-i\, \mathbf{k_{\rm ph}} \mathbf{r}}\;
    \times \\
  & \times &
  \biggl\{
  2\, \mu_{N}\,  m_{\rm p} \cdot
  Z_{\rm eff} (\mathbf{k}_{\rm ph}, \mathbf{r}) \cdot
  \mathbf{e}^{(\alpha)}\, \mathbf{\displaystyle\frac{d}{dr}} +

  i\, \textbf{M}_{\rm eff} (\mathbf{k}_{\rm ph}, \mathbf{r}) \cdot
  \Bigl[ \mathbf{\displaystyle\frac{d}{dr}} \times \mathbf{e}^{(\alpha)} \Bigr]\,
  \biggr\}\;
  \Phi_{\rm \alpha - nucl, i} (\mathbf{r})\; \mathbf{dr}.
\end{array}
\label{eq.2.11.4}
\end{equation}

\subsection{Dipole approximation of the effective charge
\label{sec.2.12}}

For first estimations of the bremsstrahlung emission for hypernuclei,
one can consider the dipole approximation of effective charge (i.e. at $\mathbf{k}_{\rm ph} \mathbf{r} \to 0$).
In such an approximation, the effective electric charge of the full system is simplified as
\begin{equation}
\begin{array}{lll}
  Z_{\rm eff} (\mathbf{k}_{\rm ph}, \mathbf{r}) \to
  Z_{\rm eff}^{\rm (dip)} =
  \displaystyle\frac{m_{A}}{m_{\alpha} + m_{A}}\, Z_{\alpha} -
  \displaystyle\frac{m_{\alpha}}{m_{\alpha} + m_{A}}\, Z_{A} =
  \displaystyle\frac{m_{A}\, Z_{\alpha} - m_{\alpha}\, Z_{A}}{m_{\alpha} + m_{A}}.
\end{array}
\label{eq.2.12.1}
\end{equation}
For the effective magnetic moment, we obtain correspondingly
\begin{equation}
\begin{array}{lll}
  \textbf{M}_{\rm eff} (\mathbf{k}_{\rm ph}, \mathbf{r}) \to
  \textbf{M}_{\rm eff}^{\rm (dip)} (\mathbf{k}_{\rm ph}, \mathbf{r}) =
  \displaystyle\frac{m_{A}}{m_{\alpha} + m_{A}}\, \textbf{F}_{\alpha,\, {\rm mag}} -
  \displaystyle\frac{m_{\alpha}}{m_{\alpha} + m_{A}}\, \textbf{F}_{A,\, {\rm mag}}.
\end{array}
\label{eq.2.12.2}
\end{equation}
In particular, for even-even nuclei with additional hyperon (for example, for \isotope[211][\Lambda]{Po}), we obtain:
\begin{equation}
\begin{array}{lll}
  \mathbf{F}_{\alpha,\, {\rm mag}}^{\rm (dip)} = 0, &
  \mathbf{F}_{A,\, {\rm mag}}^{\rm (dip)} = \mu_{\Lambda}^{\rm (an)}\, m_{\Lambda}\; \sigmabf,
\end{array}
\label{eq.2.12.3}
\end{equation}
and from Eq.~(\ref{eq.2.12.2}) we obtain:
\begin{equation}
\begin{array}{lll}
  \textbf{M}_{\rm eff}^{\rm (dip)} (\mathbf{k}_{\rm ph}, \mathbf{r}) =
  \displaystyle\frac{m_{A}}{m_{\alpha} + m_{A}}\, \textbf{F}_{\alpha,\, {\rm mag}}^{\rm (dip)} -
  \displaystyle\frac{m_{\alpha}}{m_{\alpha} + m_{A}}\, \textbf{F}_{A,\, {\rm mag}}^{\rm (dip)} =

  - \displaystyle\frac{m_{\alpha}}{m_{\alpha} + m_{A}}\, \mu_{\Lambda}^{\rm (an)}\, m_{\Lambda}\; \sigmabf =
  - \mu_{\Lambda}^{\rm (an)}\, m_{\Lambda}\, \displaystyle\frac{\mu }{m_{A}}\; \sigmabf.
\end{array}
\label{eq.2.12.4}
\end{equation}
The matrix element in Eq.~(\ref{eq.2.10.7}) is transformed to the following:
\begin{equation}
\begin{array}{lll}
\vspace{-0.2mm}
  p_{\rm full}^{\rm (dip)} & =  &
  -\, Z_{\rm eff}^{\rm (dip)} \cdot
  i \hbar\, (2\pi)^{3}\,
  \displaystyle\frac{m_{\rm p}}{e}\,
  \displaystyle\frac{2\, \mu_{N}\,  m_{\rm p}}{m} \cdot
  \displaystyle\sum\limits_{\alpha=1,2}
  \displaystyle\int\limits_{}^{}
    \Phi_{\rm \alpha - nucl, f}^{*} (\mathbf{r})\;
    e^{-i\, \mathbf{k_{\rm ph}} \mathbf{r}}\;
    \times \\
  & \times &
  \biggl\{
    \mathbf{e}^{(\alpha)}\, \mathbf{\displaystyle\frac{d}{dr}} +
    i\, \displaystyle\frac{\textbf{M}_{\rm eff}^{\rm (dip)}} {2\, \mu_{N}\,  m_{\rm p}\, Z_{\rm eff}^{\rm (dip)}} \cdot
    \Bigl[ \mathbf{\displaystyle\frac{d}{dr}} \times \mathbf{e}^{(\alpha)} \Bigr]\,
  \biggr\}\;
  \Phi_{\rm \alpha - nucl, i} (\mathbf{r})\; \mathbf{dr}.
\end{array}
\label{eq.2.12.6}
\end{equation}
In particular, for even-even nuclei with additional hyperon (for example, for \isotope[211][\Lambda]{Po}), we obtain:
\begin{equation}
\begin{array}{lll}
\vspace{-0.2mm}
  p_{\rm full}^{\rm (dip)} & =  &
  -\, Z_{\rm eff}^{\rm (dip)} \cdot
  i \hbar\, (2\pi)^{3}\,
  \displaystyle\frac{m_{\rm p}}{e}\,
  \displaystyle\frac{2\, \mu_{N}\,  m_{\rm p}}{m}\; \times \\
  & \times &
  \displaystyle\sum\limits_{\alpha=1,2}
  \displaystyle\int\limits_{}^{}
    \Phi_{\rm \alpha - nucl, f}^{*} (\mathbf{r})\;
    e^{-i\, \mathbf{k_{\rm ph}} \mathbf{r}}\;
  \biggl\{
    \mathbf{e}^{(\alpha)}\, \mathbf{\displaystyle\frac{d}{dr}} -
    i\, c_{0}\, \sigmabf \cdot \Bigl[ \mathbf{\displaystyle\frac{d}{dr}} \times \mathbf{e}^{(\alpha)} \Bigr]\,
  \biggr\}\;
  \Phi_{\rm \alpha - nucl, i} (\mathbf{r})\; \mathbf{dr},
\end{array}
\label{eq.2.12.8}
\end{equation}
where we introduce a new factor:
\begin{equation}
\begin{array}{lll}
  c_{0} & =  &
  \displaystyle\frac{\mu_{\Lambda}^{\rm (an)}}{\mu_{N}}\,
  \displaystyle\frac{m_{\Lambda}}{2\, m_{\rm p}\, Z_{\rm eff}^{\rm (dip)}} \, \displaystyle\frac{m}{m_{A}}.
\end{array}
\label{eq.2.12.9}
\end{equation}
Substituting values 
for masses and magnetic moments for proton and hyperon~\cite{RewPartPhys_PDG.2018}, we calculate:
\begin{equation}
\begin{array}{lll}
  c_{0} & =  &
  -\, 0.364453 \cdot \displaystyle\frac{m_{\alpha}} {m_{A}\, Z_{\alpha} - m_{\alpha}\, Z_{A}}.
\end{array}
\label{eq.2.12.10}
\end{equation}

\subsection{Calculations of matrix elements of emission in multipolar expansion
\label{sec.2.13}}

We have to calculate the following matrix elements:
\begin{equation}
\begin{array}{ll}
  \biggl< k_{f} \biggl| \,  e^{-i\mathbf{kr}} \displaystyle\frac{\partial}{\partial \mathbf{r}} \,
  \biggr| \,k_{i} \biggr>_\mathbf{r} =
  \displaystyle\int
    \varphi^{*}_{f}(\mathbf{r}) \:
    e^{-i\mathbf{kr}} \displaystyle\frac{\partial}{\partial \mathbf{r}}\:
    \varphi_{i}(\mathbf{r}) \;
    \mathbf{dr}.
\end{array}
\label{eq.2.13.1}
\end{equation}
%
Such matrix elements were calculated in Ref.~\cite{Maydanyuk.2012.PRC} in the spherically symmetric approximation of nucleus.
According to Eqs.~(24), (29) in Ref.~\cite{Maydanyuk.2012.PRC}, we have:
\begin{equation}
\begin{array}{ll}
  \biggl< k_{f} \biggl| \,  e^{-i\mathbf{kr}} \displaystyle\frac{\partial}{\partial \mathbf{r}}\,
  \biggr| \,k_{i} \biggr>_\mathbf{r} =
  \sqrt{\displaystyle\frac{\pi}{2}}\:
  \displaystyle\sum\limits_{l_{\rm ph}=1}\,
    (-i)^{l_{\rm ph}}\, \sqrt{2l_{\rm ph}+1}\;
  \displaystyle\sum\limits_{\mu = \pm 1}
    \xibf_{m}\, \mu\, \times
    \Bigl[ p_{l_{\rm ph}\mu}^{M} - i\mu\: p_{l_{\rm ph}\mu}^{E} \Bigr],
\end{array}
\label{eq.2.13.2}
\end{equation}
where [see Eqs.~(38), (39) in Ref.~\cite{Maydanyuk.2012.PRC}]
\begin{equation}
\begin{array}{lcl}
\vspace{1mm}
  p_{l_{\rm ph,\mu}}^{M} & = &
    \sqrt{\displaystyle\frac{l_{i}}{2l_{i}+1}}\:
      I_{M}(l_{i},l_{f}, l_{\rm ph}, l_{i}-1, \mu) \cdot
      \Bigl\{
        J_{1}(l_{i},l_{f},l_{\rm ph}) + (l_{i}+1) \cdot J_{2}(l_{i},l_{f},l_{\rm ph})
      \Bigr\}\; - \\
\vspace{3mm}
  & - &
    \sqrt{\displaystyle\frac{l_{i}+1}{2l_{i}+1}}\:
      I_{M}(l_{i},l_{f}, l_{\rm ph}, l_{i}+1, \mu) \cdot
      \Bigl\{
        J_{1}(l_{i},l_{f},l_{\rm ph}) - l_{i} \cdot J_{2}(l_{i},l_{f},l_{\rm ph})
      \Bigr\}, \\

\vspace{1mm}
  p_{l_{\rm ph,\mu}}^{E} & = &
    \sqrt{\displaystyle\frac{l_{i}\,(l_{\rm ph}+1)}{(2l_{i}+1)(2l_{\rm ph}+1)}} \cdot
      I_{E}(l_{i},l_{f}, l_{\rm ph}, l_{i}-1, l_{\rm ph}-1, \mu) \cdot
      \Bigl\{
        J_{1}(l_{i},l_{f},l_{\rm ph}-1)\; +
        (l_{i}+1) \cdot J_{2}(l_{i},l_{f},l_{\rm ph}-1)
      \Bigr\}\; - \\
\vspace{1mm}
    & - &
    \sqrt{\displaystyle\frac{l_{i}\,l_{\rm ph}}{(2l_{i}+1)(2l_{\rm ph}+1)}} \cdot
      I_{E} (l_{i},l_{f}, l_{\rm ph}, l_{i}-1, l_{\rm ph}+1, \mu) \cdot
      \Bigl\{
        J_{1}(l_{i},l_{f},l_{\rm ph}+1)\; +
        (l_{i}+1) \cdot J_{2}(l_{i},l_{f},l_{\rm ph}+1)
      \Bigr\}\; + \\
\vspace{1mm}
  & + &
    \sqrt{\displaystyle\frac{(l_{i}+1)(l_{\rm ph}+1)}{(2l_{i}+1)(2l_{\rm ph}+1)}} \cdot
      I_{E} (l_{i},l_{f},l_{\rm ph}, l_{i}+1, l_{\rm ph}-1, \mu) \cdot
      \Bigl\{
        J_{1}(l_{i},l_{f},l_{\rm ph}-1)\; -
        l_{i} \cdot J_{2}(l_{i},l_{f},l_{\rm ph}-1)
      \Bigr\}\; - \\
  & - &
    \sqrt{\displaystyle\frac{(l_{i}+1)\,l_{\rm ph}}{(2l_{i}+1)(2l_{\rm ph}+1)}} \cdot
      I_{E} (l_{i},l_{f}, l_{\rm ph}, l_{i}+1, l_{\rm ph}+1, \mu) \cdot
      \Bigl\{
        J_{1}(l_{i},l_{f},l_{\rm ph}+1)\; -
        l_{i} \cdot J_{2}(l_{i},l_{f},l_{\rm ph}+1)
      \Bigr\},
\end{array}
\label{eq.2.13.3}
\end{equation}
and
\begin{equation}
\begin{array}{ccl}
  J_{1}(l_{i},l_{f},n) & = &
  \displaystyle\int\limits^{+\infty}_{0}
    \displaystyle\frac{dR_{i}(r, l_{i})}{dr}\: R^{*}_{f}(l_{f},r)\,
    j_{n}(k_{\rm ph}r)\; r^{2} dr, \\

  J_{2}(l_{i},l_{f},n) & = &
  \displaystyle\int\limits^{+\infty}_{0}
    R_{i}(r, l_{i})\, R^{*}_{f}(l_{f},r)\: j_{n}(k_{\rm ph}r)\; r\, dr, \\

  I_{M}\, (l_{i}, l_{f}, l_{\rm ph}, l_{1}, \mu) & = &
    \displaystyle\int
      Y_{l_{f}m_{f}}^{*}(\mathbf{n}_{\rm r})\,
      \mathbf{T}_{l_{i}\, l_{1},\, m_{i}}(\mathbf{n}_{\rm r})\,
      \mathbf{T}_{l_{\rm ph}\,l_{\rm ph},\, \mu}^{*}(\mathbf{n}_{\rm r})\; d\Omega, \\

  I_{E}\, (l_{i}, l_{f}, l_{\rm ph}, l_{1}, l_{2}, \mu) & = &
    \displaystyle\int
      Y_{l_{f}m_{f}}^{*}(\mathbf{n}_{\rm r})\,
      \mathbf{T}_{l_{i} l_{1},\, m_{i}}(\mathbf{n}_{\rm r})\,
      \mathbf{T}_{l_{\rm ph} l_{2},\, \mu}^{*}(\mathbf{n}_{\rm r})\; d\Omega.
\end{array}
\label{eq.2.13.4}
\end{equation}
Here,
$j_{n}(k_{\rm ph}r)$ is \emph{spherical Bessel function of order $n$},
$\mathbf{T}_{l_{\rm ph}l_{\rm ph}^{\prime},\mu}(\mathbf{n}_{\rm r})$ are \emph{vector spherical harmonics}.
Vectors $\xibf_{-1}$ and $\xibf_{+1}$ are (complex) vectors of circular polarization of photon emitted with opposite directions of rotation
which are related with vectors $\mathbf{e}^{\alpha}$ of polarization as
(see Ref.~\cite{Eisenberg.1973}, p.~42):
\begin{equation}
\begin{array}{ll}
  \xibf_{-1} = \displaystyle\frac{1}{\sqrt{2}} (\mathbf{e}^{(1)} - i\mathbf{e}^{(2)}), &
  \xibf_{+1} = -\displaystyle\frac{1}{\sqrt{2}} (\mathbf{e}^{(1)} + i\mathbf{e}^{(2)}).
\end{array}
\label{eq.2.13.5}
\end{equation}

Using representation~(\ref{eq.2.13.2}), the matrix element~(\ref{eq.2.12.8}) is simplified as
\begin{equation}
\begin{array}{lll}
\vspace{-0.2mm}
  p_{\rm full}^{\rm (dip)} & =  &
  Z_{\rm eff}^{\rm (dip)} \cdot
  i \hbar\, (2\pi)^{3}\,
  \displaystyle\frac{m_{\rm p}}{e}\,
  \displaystyle\frac{2\, \mu_{N}\,  m_{\rm p}}{m} \cdot
  \sqrt{\displaystyle\frac{\pi}{2}}\:
  \displaystyle\sum\limits_{l_{\rm ph}=1}\,
    (-i)^{l_{\rm ph}}\, \sqrt{2l_{\rm ph}+1}\; \times \\

  & \times &
  \biggl\{
    \displaystyle\sum\limits_{\mu = \pm 1}
      h_{m}\,
      \Bigl( \mu\, - i\, c_{0}\, \sigmabf \cdot \bigl[ \xibf_{-1} \times \xibf_{-1}^{*} \bigr] \Bigr)\;
    \bigl[ p_{l_{\rm ph}\mu}^{M} - i\mu\: p_{l_{\rm ph}\mu}^{E} \bigr]
  \biggr\},
\end{array}
\label{eq.2.14.1}
\end{equation}
where
\begin{equation}
\begin{array}{lcr}
  h_{-1} = \displaystyle\frac{1}{\sqrt{2}} (1-i), &
  h_{1}  = - \displaystyle\frac{1}{\sqrt{2}} (1+i), &
  h_{-1} + h_{1} = -i \sqrt{2}.
\end{array}
\label{eq.2.14.2}
\end{equation}
Now we take into account that two vectors $\mathbf{e}^{(1)}$ and $\mathbf{e}^{(2)}$ are vectors of polarization of photon emitted, which are perpendicular to direction of emission of this photon defined by vector $\mathbf{k}$. Modulus of vectorial multiplication $\bigl[ \mathbf{e}^{(1)} \times \mathbf{e}^{(2)} \bigr]$ equals to unity.
So, we have:
\begin{equation}
\begin{array}{lll}
  \mathbf{n}_{\rm ph} \equiv
  \displaystyle\frac{\mathbf{k}_{\rm ph}}{\bigl| \mathbf{k}_{\rm ph} \bigr|} =
  \bigl[ \mathbf{e}^{(1)} \times \mathbf{e}^{(2)} \bigr]
\end{array}
\label{eq.2.14.3}
\end{equation}
and
\begin{equation}
\begin{array}{lll}
  \Bigl[ \xibf_{-1} \times \xibf_{-1}^{*} \Bigr] & = &
  -\, \Bigl[ \xibf_{+1} \times \xibf_{+1}^{*} \Bigr] =
  -\, \bigl[ \xibf_{-1} \times \xibf_{+1} \bigr] =
  i\, \bigl[ \mathbf{e}^{1} \times \mathbf{e}^{2} \bigr] =
  i\, \mathbf{n}_{\rm ph}.
\end{array}
\label{eq.2.14.4}
\end{equation}
From here, we find property in Eq.~(\ref{eq.2.14.1}):
\begin{equation}
  \bigl[ \xibf_{m} \times \xibf_{m}^{*} \bigr] =
  -\, \mu \cdot \bigl[ \xibf_{-1} \times \xibf_{-1}^{*} \bigr] =
  -\, i\, \mu \cdot \mathbf{n}_{\rm ph},
\label{eq.2.14.5}
\end{equation}
and the matrix element (\ref{eq.2.14.1}) is simplified as
\begin{equation}
\begin{array}{lll}
  p_{\rm full}^{\rm (dip)} & = &
  Z_{\rm eff}^{\rm (dip)} \cdot
  i \hbar\, (2\pi)^{3}\,
  \displaystyle\frac{m_{\rm p}}{e}\,
  \displaystyle\frac{2\, \mu_{N}\,  m_{\rm p}}{m} \cdot
  \sqrt{\displaystyle\frac{\pi}{2}}\:
  \displaystyle\sum\limits_{l_{\rm ph}=1}\,
    (-i)^{l_{\rm ph}}\, \sqrt{2l_{\rm ph}+1}\;
  \Bigl\{
    \displaystyle\sum\limits_{\mu = \pm 1}
      h_{m}\, \bigl( \mu\, + c_{0}\, \sigmabf \cdot \mathbf{n}_{\rm ph} \bigr)\:
      \bigl[ p_{l_{\rm ph}\mu}^{M} - i\mu\: p_{l_{\rm ph}\mu}^{E} \bigr]
  \Bigr\}.
\end{array}
\label{eq.2.14.5}
\end{equation}

Analyzing such a form for the matrix element, now we introduce a new formula for fast approximated estimations of the spectra
(here, we include the largest contributions in summation):
\begin{equation}
\begin{array}{lll}
  p_{\rm full}^{\rm (dip)} & = &
  Z_{\rm eff}^{\rm (dip)} \cdot
  i \hbar\, (2\pi)^{3}\,
  \displaystyle\frac{m_{\rm p}}{e}\,
  \displaystyle\frac{2\, \mu_{N}\,  m_{\rm p}}{m} \cdot

  \bigl( 1 + \bigl|c_{0}\, \sigmabf \cdot \mathbf{n}_{\rm ph} \bigr|\, \bigr)\:

  \sqrt{\displaystyle\frac{\pi}{2}}\:
  \displaystyle\sum\limits_{l_{\rm ph}=1}\,
    (-i)^{l_{\rm ph}}\, \sqrt{2l_{\rm ph}+1}\;
  \Bigl\{
    \displaystyle\sum\limits_{\mu = \pm 1}
      h_{m}\, \bigl[ p_{l_{\rm ph}\mu}^{M} - i\mu\: p_{l_{\rm ph}\mu}^{E} \bigr]
  \Bigr\}.
\end{array}
\label{eq.2.14.6}
\end{equation}
We see appearance of a new factor
$\bigl( 1 + \bigl|c_{0}\, \sigmabf \cdot \mathbf{n}_{\rm ph} \bigr|\, \bigr)$ in this formula.
This factor characterizes explicitly influence of the magnetic moment of hyperon inside hypernucleus on matrix element of emission (and on the final bremsstrahlung spectrum).
Here, parameter $c_{0}$ is dependent on choice of hypernucleus (see Eq.~(\ref{eq.2.12.9})).

\subsection{Probability of emission of the bremsstrahlung photon
\label{sec.15}}

We define the probability of the emitted bremsstrahlung photons on the basis of the full matrix element $p_{\rm full}$ in frameworks of our previous formalism
(see Refs.~\cite{Maydanyuk_Zhang_Zou.2016.PRC,Maydanyuk_Zhang.2015.PRC,Maydanyuk.2012.PRC}, also Refs.~ and reference therein).
Here, we choose our last investigation~\cite{Maydanyuk_Zhang_Zou.2016.PRC} developed for bremsstrahlung in $\alpha$-decay (see Eq.~(22) in that paper).
But, in current research, we are interesting in the bremsstrahlung probability which is not dependent on angle $\theta_{f}$.
So, we have to integrate Eq.~(22) in Ref.~\cite{Maydanyuk_Zhang_Zou.2016.PRC} over this angle and we obtain:
\begin{equation}
\begin{array}{ccl}
  \displaystyle\frac{d\,P }{dw_{\rm ph}} & = &
  N_{0} \cdot
  \displaystyle\frac{e^{2}}{2\pi\,c^{5}}\: \displaystyle\frac{w_{\rm ph}\,E_{i}}{m_{\rm p}^{2}\,k_{i}}\: \bigl| p_{\rm full} \bigr|^{2},
\end{array}
\label{eq.2.15.2}
\end{equation}
where
$k_{i} = \displaystyle\frac{1}{\hbar}\, \sqrt{2\, \mu\, E_{i}}$ is wave number of the full nuclear system (i.e. the $\alpha$-particle and nucleus) in the initial state (i.e. in state before emission of the bremsstrahlung photon),
$E_{i}$ is energy of the full nuclear system in the initial state (it corresponds to kinetic energy of the $\alpha$-particle in the asymptotic distance from nucleus),
$\mu = m_{\alpha}\, m_{A} /(m_{\alpha} + m_{A})$ is reduced mass of system of the $\alpha$-particle and nucleus.
Here, we add an additional new factor $N_{0}$ in order to normalize calculations on experimental data.
In this research, we find the factor $N_{0}$ from the best agreement between theory and experiment
(this is a case of $\alpha$ decay of the normal nucleus \isotope[210]{Po}).
Then, we use the same found normalized factor $N_{0}$ for all other calculated spectra.
By such a way, we obtain a possibility to compare the calculated bremsstrahlung spectra for different nuclei and hypernuclei.

%



Basing on Eq.~(\ref{eq.2.14.3}),
for fast estimations in computer
Eq.~(\ref{eq.2.15.2}) for even-even nuclei with possible inclusion of $\Lambda$-hyperon can be simplified as
\begin{equation}
\begin{array}{lll}
  \displaystyle\frac{d\,P^{\rm (dip)} }{dw_{\rm ph}} =
    \Bigl\{ 1 + \bigl|c_{0}\, \sigmabf \cdot \mathbf{n}_{\rm ph} \bigr|\, \Bigr\}^{2} \cdot
    \displaystyle\frac{d\,P^{\rm (dip,\, no-\Lambda)} }{dw_{\rm ph}}, &

  \displaystyle\frac{d\,P^{\rm (dip,\, no-\Lambda)} }{dw_{\rm ph}} =
    N_{0} \cdot \displaystyle\frac{e^{2}}{2\pi\,c^{5}}\: \displaystyle\frac{w_{\rm ph}\,E_{i}}{m_{\rm p}^{2}\,k_{i}}\: \bigl| p_{\rm full}^{\rm (dip,\, no-\Lambda)} \bigr|^{2},
\end{array}
\label{eq.2.15.3}
\end{equation}
\begin{equation}
\begin{array}{lll}
\vspace{-0.2mm}
  p_{\rm full}^{\rm (dip,\, no-\Lambda)} & =  &
  Z_{\rm eff}^{\rm (dip)} \cdot
  i \hbar\, (2\pi)^{3}\,
  \displaystyle\frac{m_{\rm p}}{e}\,
  \displaystyle\frac{2\, \mu_{N}\,  m_{\rm p}}{m} \cdot
  \sqrt{\displaystyle\frac{\pi}{2}}\:
  \displaystyle\sum\limits_{l_{\rm ph}=1}\,
    (-i)^{l_{\rm ph}}\, \sqrt{2l_{\rm ph}+1} \cdot
  \displaystyle\sum\limits_{\mu = \pm 1}
    h_{m}\, \mu\, \bigl[ p_{l_{\rm ph}\mu}^{M} - i\mu\: p_{l_{\rm ph}\mu}^{E} \bigr].
\end{array}
\label{eq.2.15.4}
\end{equation}
Here, we express explicitly the bremsstrahlung probability $P^{\rm (dip,\, no-\Lambda)}$ and the matrix element $p_{\rm full}^{\rm (dip,\, no-\Lambda)}$
which belong to standard bremsstrahlung formalism without hyperons in nuclei.

\section{Discussions
\label{sec.results}}

For the first estimations of the bremsstrahlung spectra we have chosen two nuclei: \isotope[210]{Po} and \isotope[211][\Lambda]{Po}.
We explain such a choice by the following.
At present, there are experimental data for the bremsstrahlung in $\alpha$ decay for four nuclei:
\isotope[210]{Po}~\cite{D'Arrigo.1994.PHLTA,Kasagi.1997.JPHGB,Kasagi.1997.PRLTA,Boie.2007.PRL,Boie.2009.PhD},
\isotope[214]{Po}~\cite{Maydanyuk.2008.EPJA},
\isotope[226]{Ra} \cite{Giardina.2008.MPLA} and \isotope[244][\Lambda]{Cm}.
The experimental data obtained with the highest accuracy are experimental data \cite{Boie.2007.PRL,Boie.2009.PhD} for nucleus \isotope[210]{Po}.
In our approach~\cite{Maydanyuk.2006.EPJA,Maydanyuk.2008.EPJA}, we achieved the most accurate agreement with these data
(see also Refs.~\cite{Papenbrock.1998.PRLTA,Tkalya.1999.PHRVA,Jentschura.2008.PRC}).
So, we put main focus to this normal nucleus in our research.
According to Ref.~\cite{Santhosh.2018.Pramana}, the hypernucleus, which has the most close $\alpha$--nucleus interaction with the normal nucleus \isotope[210]{Po}, is \isotope[211][\Lambda]{Po}.

\begin{figure}[htbp]
\centerline{\includegraphics[width=85mm]{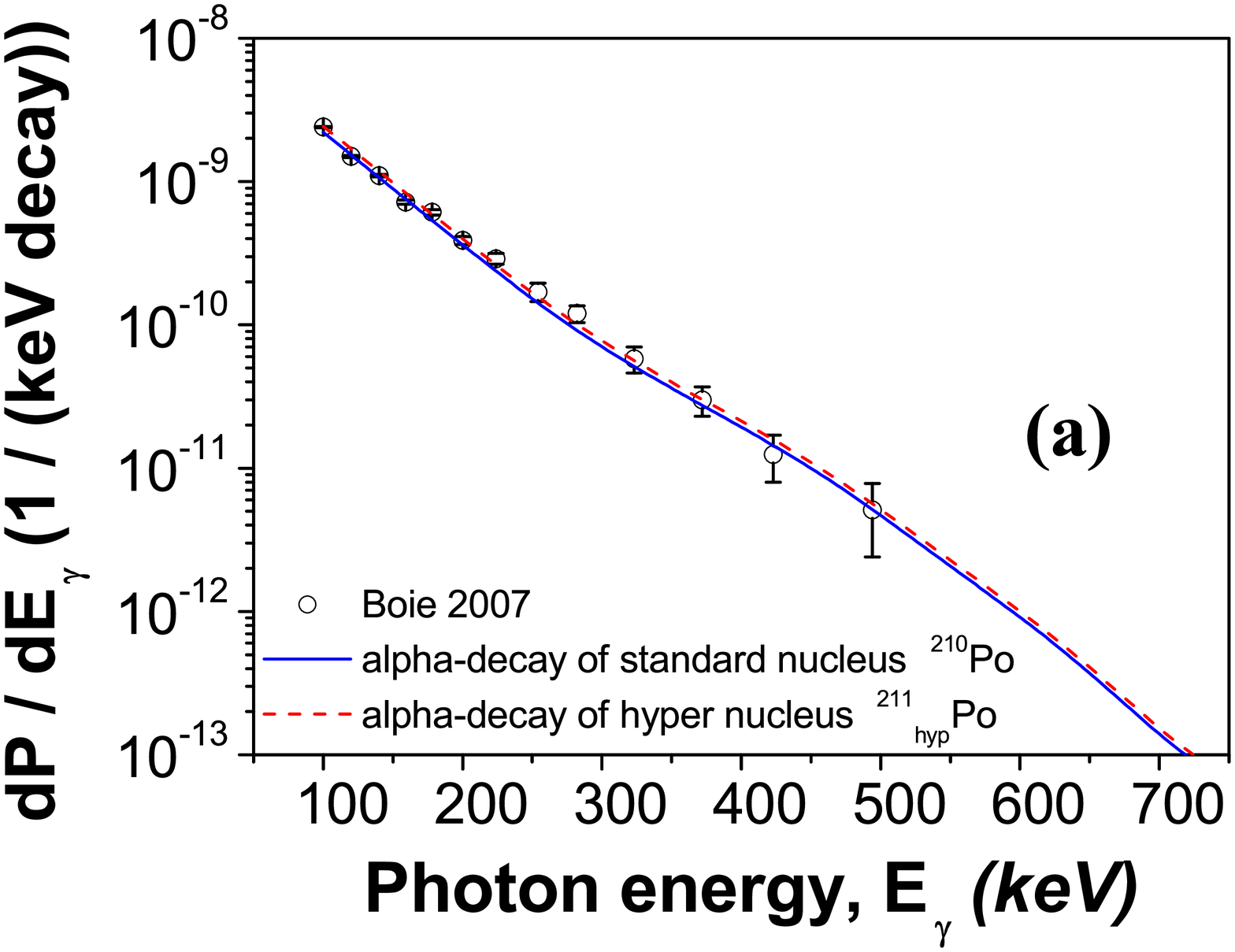}
\hspace{-1mm}\includegraphics[width=85mm]{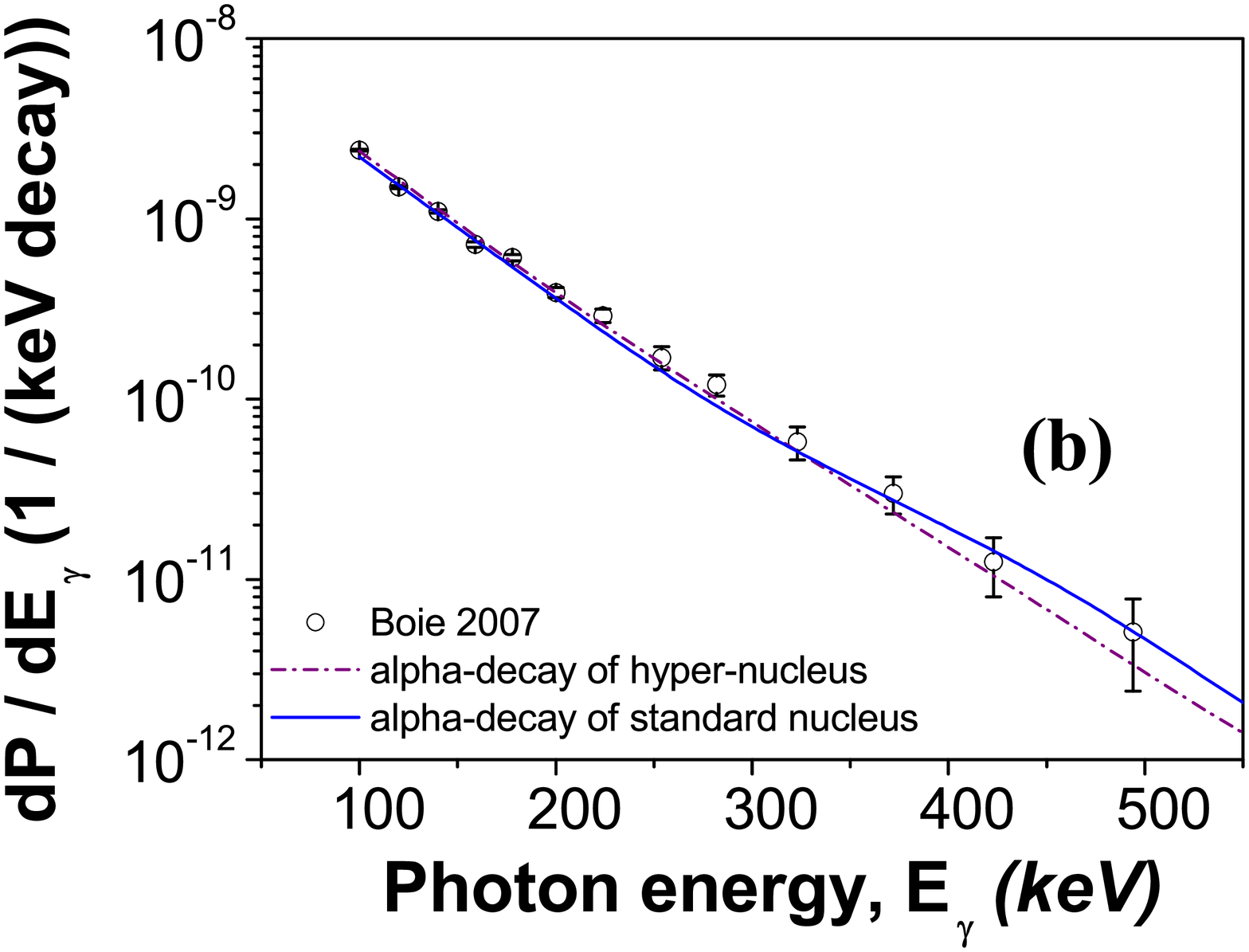}}
\vspace{-4mm}
\caption{\small (Color online)
The bremsstrahlung probabilities of photons emitted during $\alpha$ decay of the normal \isotope[210]{Po} nucleus and the \isotope[211][\Lambda]{Po} hypernucleus in comparison with experimental data
(Boie 2007:~\cite{Boie.2007.PRL,Boie.2009.PhD})
[parameters of calculations:
the bremsstrahlung probability is defined in Eq.~(\ref{eq.2.15.2}),
parameters of potentials are taken in Ref.~\cite{Santhosh.2018.Pramana}].
Here,
open circles are experimental data~\cite{Boie.2007.PRL,Boie.2009.PhD} for $\alpha$ decay of \isotope[210]{Po},
blue solid line is the calculated spectrum for $\alpha$ decay of normal nucleus \isotope[210]{Po} [(figures (a), (b)],
red dashed line is the calculated spectrum for $\alpha$ decay of hyper nucleus \isotope[211][\Lambda]{Po} with magnetic momentums of nucleons and hypeon [figure (a)],
brown dash-dotted line is the calculated spectrum for $\alpha$ decay of hyper nucleus \isotope[211][\Lambda]{Po} without magnetic momentums of nucleons and hyperon [figure (b)].
Panel (a):
One can see small difference between the spectra for the normal and hyper nuclei.
Such a difference is explained by additional contribution of emission of photons formed by the anomalous magnetic moment of hyperon inside the hyper nucleus \isotope[211][\Lambda]{Po}
(which is absent in the normal nucleus \isotope[210]{Po}).
Panel (b):
Without inclusion of the magnetic moments to calculations, difference between the spectra for \isotope[210]{Po} and \isotope[211][\Lambda]{Po} is small also.
%
\label{fig.1}}
\end{figure}

We calculated bremsstrahlung of photons emitted in $\alpha$ decay of \isotope[210]{Po} and \isotope[211][\Lambda]{Po}.
Results of such calculations in comparison with analysis of magnetic momenta of hyperon and nucleons with experimental data~\cite{Boie.2007.PRL,Boie.2009.PhD} are presented in Fig.~\ref{fig.1}~(a).
From such calculations we see that the bremsstrahlung spectrum in $\alpha$ decay of the hypernucleus \isotope[211][\Lambda]{Po} (see red dashed line in figure) is above than the bremsstrahlung spectrum in $\alpha$ decay of normal nucleus \isotope[210]{Po} (see blue solid line in figure).
Such a difference between the spectra is explained mainly by additional contribution to the full bremsstrahlung emission, which is caused by magnetic moment of hyperon inside hypernucleus.
The bremsstrahlung emission formed by such a mechanism, is of the magnetic type.
However, as we estimate this contribution is really small for all isotopes of Polonium.
Before such calculations, we estimated emission of bremsstrahlung photons without inclusion of the magnetic moments of hyperon and nucleons. In any case, potentials of interactions are different, but such difference in small also [see Fig.~\ref{fig.1}~(b)].

\begin{figure}[htbp]
\centerline{\includegraphics[width=85mm]{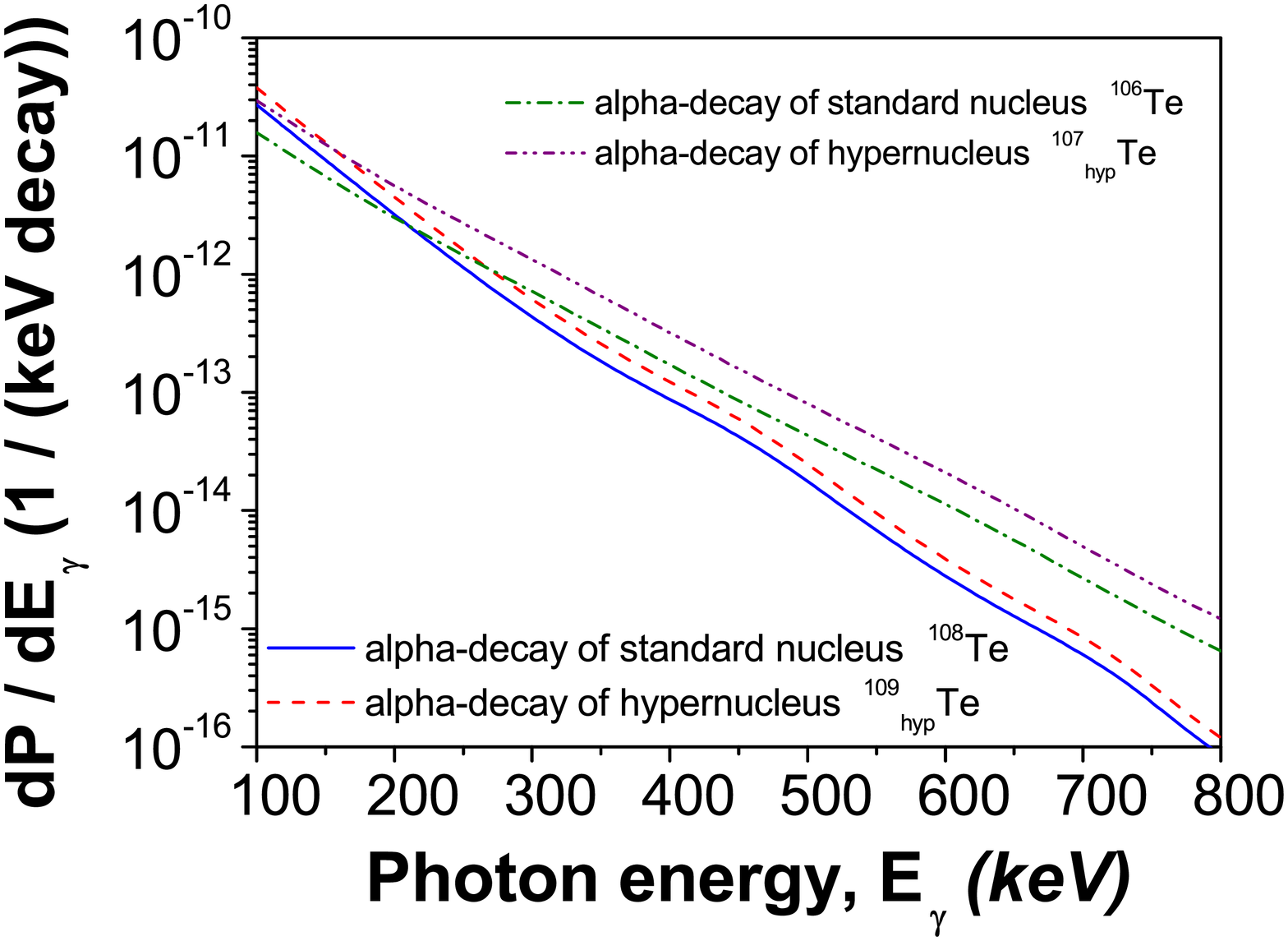}
}
\vspace{-4mm}
\caption{\small (Color online)
The bremsstrahlung probabilities of photons emitted during $\alpha$ decay of the normal nuclei \isotope[106]{Te} and \isotope[108]{Te}
in comparison with the hypernuclei \isotope[107][\Lambda]{Te} and \isotope[109][\Lambda]{Te}
[parameters of calculations:
the bremsstrahlung probability is defined in Eqs.~(\ref{eq.2.15.3})--(\ref{eq.2.15.4})
we use
$Q = 4.290$~MeV for \isotope[106]{Te} and \isotope[107][\Lambda]{Te},
$Q = 3.450$~MeV for \isotope[108]{Te} and \isotope[109][\Lambda]{Te},
$Q$-values for normal nuclei are taken from Table~I in Ref.~\cite{Duarte.2002.ADNDT}
].
Here,
blue solid line is the calculated spectrum for $\alpha$ decay of normal nucleus \isotope[108]{Te},
red dashed line is the calculated spectrum for $\alpha$ decay of hypernucleus \isotope[109][\Lambda]{Te} with magnetic momentums of nucleons and hyperon,
green dash-dotted line is the calculated spectrum for $\alpha$ decay of normal nucleus \isotope[106]{Te},
purple dash-double dotted line is the calculated spectrum for $\alpha$ decay of hypernucleus \isotope[107][\Lambda]{Te} with magnetic momentums of nucleons and hyperon.
One can see that here role of magnetic moment of hyperon inside hypernucleus in emission of bremsstrahlung photons is more essential, than in previous calculations in Fig.~\ref{fig.1} for \isotope[210]{Po} and \isotope[211][\Lambda]{Po}.
%
\label{fig.2}}
\end{figure}

Analyzing formulas (\ref{eq.2.12.10}) and (\ref{eq.2.15.3}), one can find hypernuclei, for which role of hyperon is more essential in emission of bremsstrahlung photons during $\alpha$ decay.
The simplest idea is to look for nuclei at condition of $Z_{\rm eff}^{\rm (dip)} \to 0$
(but in this paper we restrict ourselves by case of $Z_{\rm eff}^{\rm (dip)} \ne 0$).
As demonstration of this property, we estimate the bremsstrahlung spectra for the normal nuclei \isotope[106]{Te} and \isotope[108]{Te}
in comparison with the hypernuclei \isotope[107][\Lambda]{Te} and \isotope[109][\Lambda]{Te}.
Results of such calculations are presented in Fig.~\ref{fig.2}, they confirm property described above.
More high tendency of spectra for \isotope[106]{Te} and \isotope[107][\Lambda]{Te} in comparison with spectra for \isotope[108]{Te} and \isotope[109][\Lambda]{Te} is explained by higher $Q$-values for these nuclei.
One can see that emission in the $\alpha$ decay of hypernucleus has similar tendency as studied before emission in the $\alpha$ decay of normal nuclei, without existence any resonant peak in the spectrum.



\section{Conclusions and perspectives
\label{sec.conclusions}}

At first time, we investigate possibility of emission of the bremsstrahlung photons in nuclear reactions with hypernuclei.
In such an analysis we focus on interactions between $\alpha$-particles and nuclei.
In order to perform this research,
we construct a new model of the bremsstrahlung emission which accompanies interactions between $\alpha$ particles and hypernuclei.
As example for calculations and analysis, we choose $\alpha$ decay of isotopes of Polonium.
For the first estimations of the bremsstrahlung spectra we have chosen two nuclei: the normal nucleus \isotope[210]{Po} and the hypernucleus \isotope[211][\Lambda]{Po}.
%
Motivation of such a choice of reaction is the following:

\begin{enumerate}
\item
For the first estimations of bremsstrahlung in reactions with hypernuclei, we need in the most tested bremsstrahlung model and calculations.
The experimental bremsstrahlung data obtained with the highest accuracy are data~\cite{Boie.2007.PRL,Boie.2009.PhD} for $\alpha$ decay of nucleus \isotope[210]{Po}.
For this nucleus the best confirmation of the bremsstrahlung model and calculations were obtained
(see Refs.~\cite{Maydanyuk.2006.EPJA,Maydanyuk.2008.EPJA}, also calculations~\cite{Papenbrock.1998.PRLTA,Tkalya.1999.PHRVA,Jentschura.2008.PRC})).

\item
$\alpha$ Decay of hypernuclei has been already studied theoretically (see Ref.~\cite{Santhosh.2018.Pramana} and reference therein).
From such a research, interacting potentials are known, needed for our calculations.

\item
Today, heavy hypernucleus $_{\Lambda}{\rm Pb}$ has already been known experimentally (see Ref.~\cite{RewPartPhys_PDG.2018}).

\item
According to Ref.~\cite{Santhosh.2018.Pramana}, the hypernucleus, which has the most close $\alpha$--nucleus interaction with the normal nucleus \isotope[210]{Po}, is \isotope[211][\Lambda]{Po}.

\end{enumerate}


Our new contribution to the existed theory is the following:

\begin{enumerate}
\item
We generalize our previous many-nucleon bremsstrahlung model~\cite{Maydanyuk_Zhang_Zou.2016.PRC}, including new formalism for the magnetic momenta of nucleons and hyperon.

\item
We introduce the new magnetic form-factors for nucleus and $\alpha$ particle.
This characteristic can be related with anomalous magnetic momentum of hypernucleus.
It can be useful for study properties of hypernuclei.

\item
Our approach allows to study and estimate role of anomalous magnetic momenta of nucleons and hyperon in emission of bremsstrahlung photons.

\item
Our approach allows can be used for next step for study of incoherent bremsstrahlung emission in reactions with hypernuclei.

\item
We propose a new formula for fast estimations of bremsstrahlung spectra for even-even hypernuclei in computer [see Eqs.~(\ref{eq.2.15.3})--(\ref{eq.2.15.4})],
where role of magnetic moment of hyperon in hypernucleus in formation of the bremsstrahlung emission is shown explicitly.

\end{enumerate}

Our new results in study of bremsstrahlung in $\alpha$ decay of hypernuclei are the following:

\begin{enumerate}
\item
We have not found any information in literature about emission of bremsstrahlung photons during reactions with participation of hypernuclei.
We performed the first estimations of emission of the bremsstrahlung photons during $\alpha$-decay of hypernuclei,
estimating the bremsstrahlung spectra for isotopes of Polonium.

\item
Hyperon has own magnetic moment (which is different from magnetic momenta for nucleons).
By such a reason, hyperon inside hypernucleus (which is under $\alpha$-decay) forms additional bremsstrahlung emission.
This contribution of emission to the full bremsstrahlung spectrum is small, but it reinforces the full emission
(see Fig.~\ref{fig.1}, where we calculated the spectra during $\alpha$ decay of the normal nucleus \isotope[210]{Po} and the hypernucleus \isotope[211][\Lambda]{Po}).
The bremsstrahlung emission formed by such a mechanism, is of the magnetic type.

\item
The bremsstrahlung emission in the $\alpha$ decay of hypernucleus has similar tendency as studied before bremsstrahlung emission in the $\alpha$ decay of normal nuclei,
without existence any resonant peak in the spectrum (see Fig.~\ref{fig.1}).

\item
Before such calculations, we estimated emission of bremsstrahlung photons in $\alpha$ decay of \isotope[210]{Po} and \isotope[211][\Lambda]{Po} without inclusion of the magnetic moments of hyperon and nucleons to calculations.
$\alpha$-Nucleus potentials of interactions for \isotope[210]{Po} and \isotope[211][\Lambda]{Po} are different, but this difference in small \cite{Santhosh.2018.Pramana}.
As a result, we find that difference between the bremsstrahlung spectra is small, if to neglect magnetic momenta of hyperon and nucleons [see Fig.~\ref{fig.1}~(b)].

\item
We propose a way how to find hypernuclei, where role of hyperon is the most essential in emission of bremsstrahlung photons during $\alpha$ decay.
The simplest idea is to look for nuclei at condition of $Z_{\rm eff}^{\rm (dip)} \to 0$
(we restrict ourselves by case of $Z_{\rm eff}^{\rm (dip)} \ne 0$).
As example, we estimate the bremsstrahlung spectra for the normal nuclei \isotope[106]{Te} and \isotope[108]{Te}
in comparison with the hypernuclei \isotope[107][\Lambda]{Te} and \isotope[109][\Lambda]{Te} (see Fig.~\ref{fig.2}).

\end{enumerate}
%
This opens perspective to study properties of strange nuclear matter via analysis of the bremsstrahlung emission, which accompanies reactions with hypernuclei.

\section*{Acknowledgements
\label{sec.acknowledgements}}

S.~P.~Maydanyuk and Liu Xin thank the Institute of Modern Physics of Chinese Academy of Sciences for warm hospitality and support.
This work was supported by the Major State Basic Research Development Program in China (No. 2015CB856903),
the National Natural Science Foundation of China (Grant Nos. 11575254, 11447105, 11175215, 
11575060, 11375062, 11505057 and 11647306), 
the Chinese Academy of Sciences fellowships for researchers from developing countries (No. 2014FFJA0003).

\appendix
\section{Transition to coordinates of relative distances
\label{sec.app.1}}

In this Section we rewrite formalism via coordinates of relative distances.
We start from definitions (\ref{eq.2.4.1}) for coordinate of centers of masses for the $\alpha$ particle  as $\mathbf{r}_{\alpha}$, for the daughter nucleus as $\mathbf{R}_{A}$, and
for the complete system as $\mathbf{R}$:
\begin{equation}
\begin{array}{lll}
   \mathbf{r}_{\alpha} = \displaystyle\frac{1}{m_{\alpha}} \displaystyle\sum_{i=1}^{4} m_{i}\, \mathbf{r}_{\alpha i}, &
   \mathbf{R}_{A}      = \displaystyle\frac{1}{m_{A}} \displaystyle\sum_{j=1}^{A} m_{j}\, \mathbf{r}_{A j}, &
   \mathbf{R}          = \displaystyle\frac{m_{A}\mathbf{R}_{A} + m_{\alpha}\mathbf{r}_{\alpha}}{m_{A}+m_{\alpha}} =
     c_{A}\, \mathbf{R}_{A} + c_{\alpha}\, \mathbf{R}_{\alpha},
\end{array}
\label{eq.app.2.1.1}
\end{equation}
where $m_{\alpha}$ and $m_{A}$ are masses of the $\alpha$ particle and daughter nucleus, and
$c_{A} = \frac{m_{A}}{m_{A}+m_{\alpha}}$, $c_{\alpha} = \frac{m_{\alpha}}{m_{A}+m_{\alpha}}$.
Also we use definitions (\ref{eq.2.4.2}) for relative coordinate $\mathbf{r}$,
relative coordinates $\rhobf_{\alpha i}$ for nucleons of the $\alpha$-particle,
relative coordinates $\rhobf_{A j}$ for nucleons (with possible hyperon) for the daughter nucleus
as
\begin{equation}
\begin{array}{lll}
   \mathbf{r} = \mathbf{r}_{\alpha} - \mathbf{R}_{A}, &
   \rhobf_{\alpha i} = \mathbf{r}_{\alpha i} - \mathbf{r}_{\alpha}, &
   \rhobf_{A j} = \mathbf{r}_{j} - \mathbf{R}_{A}.
\end{array}
\label{eq.app.1.2}
\end{equation}
%
%
From here we find ($n=4$):
\begin{equation}
\begin{array}{lll}
  \mathbf{R} =
    \displaystyle\frac{m_{A} \mathbf{R}_{A} + m_{\alpha}\mathbf{r}_{\alpha}}{m_{A}+m_{\alpha}} =
    \displaystyle\frac{1}{m_{A}+m_{\alpha}}\;
    \Bigl\{
      \displaystyle\sum_{j=1}^{A} m_{j}\, \mathbf{r}_{A j} +
      \displaystyle\sum_{i=1}^{n} m_{i}\, \mathbf{r}_{\alpha i}
    \Bigr\}, \\

  \mathbf{r} = \mathbf{r}_{\alpha} - \mathbf{R}_{A} =
    \displaystyle\frac{1}{m_{\alpha}} \displaystyle\sum_{i=1}^{n} m_{i}\, \mathbf{r}_{\alpha i} - \displaystyle\frac{1}{m_{A}} \displaystyle\sum_{j=1}^{A} m_{j}\, \mathbf{r}_{A j}, \\

  \rhobf_{\alpha i} =
  \mathbf{r}_{\alpha i} - \mathbf{r}_{\alpha} =
  \mathbf{r}_{\alpha i} -
  \displaystyle\frac{1}{m_{\alpha}} \displaystyle\sum_{k=1}^{n=4} m_{k}\, \mathbf{r}_{\alpha k}
  \quad (i = 1 \ldots n-1), &
  \rhobf_{\alpha n} =
  -\, \displaystyle\frac{1}{m_{n}} \displaystyle\sum_{k=1}^{n-1} m_{k}\, \rhobf_{\alpha k}, \\

  \rhobf_{A j} =
  \mathbf{r}_{A j} - \mathbf{r}_{A} =
  \mathbf{r}_{A j} -
  \displaystyle\frac{1}{m_{A}} \displaystyle\sum_{k=1}^{A} m_{k}\, \mathbf{r}_{A k}
  \quad (j = 1 \ldots A-1), &
  \rhobf_{AA} =
  -\, \displaystyle\frac{1}{m_{AA}} \displaystyle\sum_{k=1}^{A-1} m_{k}\, \rhobf_{A k}.
\end{array}
\label{eq.app.1.3}
\end{equation}
Vectors $\rhobf_{\alpha n}$ and $\rhobf_{AA}$ are dependent on other $\rhobf_{\alpha 1} \ldots \rhobf_{\alpha n-1}$ and $\rhobf_{A 1}$ \ldots $\rhobf_{A A-1}$ (as we define them concerning to center of mass of studied fragment).
So, one can rewrite them as
%
\begin{equation}
\begin{array}{lll}
  \rhobf_{\alpha n} =
  \mathbf{r}_{\alpha n} - \mathbf{r}_{\alpha} =
  \mathbf{r}_{\alpha n} -
  \displaystyle\frac{1}{m_{\alpha}} \displaystyle\sum_{k=1}^{n} m_{k}\, \mathbf{r}_{\alpha k}, &

  \rhobf_{A A} =
  \mathbf{r}_{A A} - \mathbf{r}_{A} =
  \mathbf{r}_{A A} -
  \displaystyle\frac{1}{M} \displaystyle\sum_{k=1}^{A} m_{k}\, \mathbf{r}_{A k}.
\end{array}
\label{eq.app.1.4}
\end{equation}
We express old coordinates $\mathbf{r}_{\alpha}$, $\mathbf{R}_{A}$ via new coordinates $\mathbf{R}$, $\mathbf{r}$:
\[
\begin{array}{lll}
  \left\{
  \begin{array}{lll}
    \vspace{1mm}
    \mathbf{R}  = c_{A}\mathbf{R}_{A} + c_{\alpha}\, \mathbf{r}_{\alpha}, \\
    \mathbf{r} = \mathbf{r}_{\alpha} - \mathbf{R}_{A}
  \end{array}
  \right. \to

  \left\{
  \begin{array}{lll}
    \vspace{1mm}
    \mathbf{R} + c_{A}\, \mathbf{r} = c_{\alpha} \mathbf{r}_{\alpha} + c_{A} \mathbf{r}_{\alpha} = (c_{\alpha} + c_{A} )\, \mathbf{r}_{\alpha} = \mathbf{r}_{\alpha}, \\
    \mathbf{R} - c_{\alpha}\, \mathbf{r} = c_{\alpha} \mathbf{R}_{A} + c_{A} \mathbf{R}_{A} = (c_{\alpha} + c_{A})\, \mathbf{R}_{A} = \mathbf{R}_{A}
  \end{array}
  \right.
\end{array}
\]
or
\begin{equation}
\begin{array}{lll}
  \mathbf{r}_{\alpha} = \mathbf{R} + c_{A}\, \mathbf{r}, &
  \mathbf{R}_{A} = \mathbf{R} - c_{\alpha}\, \mathbf{r}.
\end{array}
\label{eq.app.1.5}
\end{equation}
Now, using (\ref{eq.app.1.2}) and (\ref{eq.app.1.3}), we rewrite old coordinates $\mathbf{r}_{\alpha i}$, $\mathbf{r}_{Aj}$ of nucleons via new coordinates $\rhobf_{\alpha i}$, :
\[
\begin{array}{lll}
  \vspace{2mm}
  \mathbf{r}_{\alpha i} = \rhobf_{\alpha i} + \mathbf{r}_{\alpha} = \rhobf_{\alpha i} + \mathbf{R} + c_{A}\, \mathbf{r}, \\
  \mathbf{r}_{Aj} = \rhobf_{A j} + \mathbf{R}_{A} = \rhobf_{A j} + \mathbf{R} - c_{\alpha}\, \mathbf{r}
\end{array}
\]
or
\begin{equation}
\begin{array}{lll}
  \mathbf{r}_{\alpha i} = \rhobf_{\alpha i} + \mathbf{R} + c_{A}\, \mathbf{r}, &
  \mathbf{r}_{Aj} = \rhobf_{A j} + \mathbf{R} - c_{\alpha}\, \mathbf{r}.
\end{array}
\label{eq.app.1.6}
\end{equation}
For numbers $i=n$ and $j=A$ it is more convenient to use [from (\ref{eq.app.1.3}) and (\ref{eq.app.1.6})]
%
\begin{equation}
\begin{array}{ll}
  \mathbf{r}_{\alpha n} = \mathbf{R} + c_{A} \mathbf{r} -
    \displaystyle\frac{1}{m_{n}} \displaystyle\sum_{k=1}^{n-1} m_{k}\, \rhobf_{\alpha k}, &
\hspace{5mm}
  \mathbf{r}_{AA} = \mathbf{R} - c_{\alpha} \mathbf{r} -
    \displaystyle\frac{1}{m_{AA}} \displaystyle\sum_{k=1}^{A-1} m_{k}\, \rhobf_{A k}.
\end{array}
\label{eq.app.1.7}
\end{equation}

From (\ref{eq.app.1.3}) we shall calculate derivatives:
\begin{equation}
\begin{array}{llll}
  \displaystyle\frac{\mathbf{dR}}{\mathbf{dr}_{\alpha i}} =
  \displaystyle\frac{m_{i}}{m_{A} + m_{\alpha}}, &

  \displaystyle\frac{\mathbf{dR}}{\mathbf{dr}_{Aj}} =
  \displaystyle\frac{m_{j}}{m_{A} + m_{\alpha}}, &

  \displaystyle\frac{\mathbf{dr}}{\mathbf{dr}_{\alpha i}} =
  \displaystyle\frac{m_{i}}{m_{\alpha}}, &

  \displaystyle\frac{\mathbf{dr}}{\mathbf{dr}_{Aj}} =
  -\, \displaystyle\frac{m_{j}}{m_{A}}. \\
\end{array}
\label{eq.app.1.8}
\end{equation}
From (\ref{eq.app.1.3}) and (\ref{eq.app.1.4}) for $\rhobf_{\alpha i}$ and $\rhobf_{A j}$ we have (at $i=1 \ldots n-1$ and $j = 1 \ldots A-1$):
\begin{equation}
\begin{array}{llllll}
  \vspace{2mm}
  \displaystyle\frac{\mathbf{d} \rhobf_{\alpha i}}{\mathbf{dr}_{\alpha i}} = \displaystyle\frac{m_{\alpha} - m_{i}}{m_{\alpha}}, &
  \displaystyle\frac{\mathbf{d} \rhobf_{\alpha i}}{\mathbf{dr}_{\alpha k\; (k \ne i,\, k \ne n=4)}} = -\,\displaystyle\frac{m_{k}}{m_{\alpha}}, &
  \displaystyle\frac{\mathbf{d} \rhobf_{\alpha i}}{\mathbf{dr}_{\alpha n}} = -\, \displaystyle\frac{m_{n}}{m_{\alpha}}, &
  \displaystyle\frac{\mathbf{d} \rhobf_{\alpha i}}{\mathbf{dr}_{A j}} = 0, \\

  \displaystyle\frac{\mathbf{d} \rhobf_{Aj}}{\mathbf{dr}_{Aj}} = \displaystyle\frac{m_{A} - m_{j}}{m_{A}}, &
  \displaystyle\frac{\mathbf{d} \rhobf_{Aj}}{\mathbf{dr}_{Ak\; (k \ne j,\, k \ne A)}} = -\,\displaystyle\frac{m_{k}}{m_{A}}, &
  \displaystyle\frac{\mathbf{d} \rhobf_{Aj}}{\mathbf{dr}_{AA}} = -\,\displaystyle\frac{m_{AA}}{m_{A}}, &
  \displaystyle\frac{\mathbf{d} \rhobf_{Aj}}{\mathbf{dr}_{\alpha i}} = 0.
\end{array}
\label{eq.app.1.9}
\end{equation}
From (\ref{eq.app.1.3}) and (\ref{eq.app.1.4}) for $\rhobf_{\alpha n}$ and $\rhobf_{AA}$ we have (at $i=1 \ldots n-1$ and $j = 1 \ldots A-1$):
\begin{equation}
\begin{array}{llllll}
&   \vspace{2mm}
  \displaystyle\frac{\mathbf{d} \rhobf_{\alpha n}}{\mathbf{dr}_{\alpha i, i \ne n}} = -\,\displaystyle\frac{m_{i}}{m_{\alpha}}, &
  \displaystyle\frac{\mathbf{d} \rhobf_{\alpha n}}{\mathbf{dr}_{\alpha n}} = 1 - \,\displaystyle\frac{m_{n}}{m_{\alpha}}, &

  \displaystyle\frac{\mathbf{d} \rhobf_{AA}}{\mathbf{dr}_{\alpha j, j \ne A}} = -\,\displaystyle\frac{m_{j}}{m_{A}}, &
  \displaystyle\frac{\mathbf{d} \rhobf_{AA}}{\mathbf{dr}_{AA}} = 1 - \,\displaystyle\frac{m_{AA}}{m_{A}}.
\end{array}
\label{eq.app.1.10}
\end{equation}

Now we shall calculate momenta connected with new independent variables $\mathbf{R}$, $\mathbf{r}$, $\rhobf_{\alpha i}$, $\rhobf_{A j}$ (at $j = 1 \ldots A-1$, $i = 1 \ldots n-1$, $n=4$ for the $\alpha$ particle).
From (\ref{eq.app.1.9}) at $i = 1 \ldots n-1$ we have:
\begin{equation}
\begin{array}{lclll}
  \vspace{1mm}
  \mathbf{p}_{\alpha i} =
    -\,i\hbar\, \displaystyle\frac{\mathbf{d}}{\mathbf{dr}_{\alpha i}} =
    -\,i\hbar\, \displaystyle\frac{\mathbf{dR}}{\mathbf{dr}_{\alpha i}}\,
      \displaystyle\frac{\mathbf{d}}{\mathbf{dR}}
    -\,i\hbar\, \displaystyle\frac{\mathbf{dr}}{\mathbf{dr}_{\alpha i}}\,
      \displaystyle\frac{\mathbf{d}}{\mathbf{dr}}
    -\,i\hbar\, \displaystyle\sum_{k=1}^{n-1}
      \displaystyle\frac{\mathbf{d}\rhobf_{\alpha k}}{\mathbf{dr}_{\alpha i}}\,
      \displaystyle\frac{\mathbf{d}}{\mathbf{d\rho}_{\alpha k}}
    -\,i\hbar\, \displaystyle\sum_{k=1}^{A-1}
      \displaystyle\frac{\mathbf{d}\rhobf_{A k}}{\mathbf{dr}_{\alpha i}}\,
      \displaystyle\frac{\mathbf{d}}{\mathbf{d\rho}_{A k}} = \\

  \vspace{1mm}
  = -\,i\hbar\, \displaystyle\frac{m_{i}}{m_{A} + m_{\alpha}}
      \displaystyle\frac{\mathbf{d}}{\mathbf{dR}}
    -\,i\hbar\, \displaystyle\frac{m_{i}}{m_{\alpha}}\,
      \displaystyle\frac{\mathbf{d}}{\mathbf{dr}}

    -\,i\hbar\, \displaystyle\sum_{k=1,\, k=i}^{n-1}
      \displaystyle\frac{\mathbf{d}\rhobf_{\alpha k}}{\mathbf{dr}_{\alpha i}}\,
      \displaystyle\frac{\mathbf{d}}{\mathbf{d\rho}_{\alpha k}}

    -\,i\hbar\, \displaystyle\sum_{k=1,\, k \ne i}^{n-1}
      \displaystyle\frac{\mathbf{d}\rhobf_{\alpha k}}{\mathbf{dr}_{\alpha i}}\,
      \displaystyle\frac{\mathbf{d}}{\mathbf{d\rho}_{\alpha k}} = \\

  \vspace{1mm}
  = -\,i\hbar\, \displaystyle\frac{m_{i}}{m_{A} + m_{\alpha}}
      \displaystyle\frac{\mathbf{d}}{\mathbf{dR}}
    -\,i\hbar\, \displaystyle\frac{m_{i}}{m_{\alpha}}\,
      \displaystyle\frac{\mathbf{d}}{\mathbf{dr}}

    -\,i\hbar\, \displaystyle\frac{m_{\alpha} - m_{i}}{m_{\alpha}}  \displaystyle\frac{\mathbf{d}}{\mathbf{d\rho}_{\alpha i}}

    +\,i\hbar\, \displaystyle\sum_{k=1, k \ne i}^{n-1}
      \displaystyle\frac{m_{i}}{m_{\alpha}}
      \displaystyle\frac{\mathbf{d}}{\mathbf{d\rho}_{\alpha k}} = \\

  = \displaystyle\frac{m_{i}}{m_{A} + m_{\alpha}}\, \mathbf{P} +
    \displaystyle\frac{m_{i}}{m_{\alpha}}\,\mathbf{p} +
    \displaystyle\frac{m_{\alpha} - m_{i}}{m_{\alpha}}\, \mathbf{\tilde{p}}_{\alpha i} -
    \displaystyle\frac{m_{i}}{m_{\alpha}}\,
      \displaystyle\sum_{k=1, k \ne i}^{n-1} \mathbf{\tilde{p}}_{\alpha k},
\end{array}
\label{eq.app.1.11}
\end{equation}
and at $i=n=4$ we have:
\begin{equation}
\begin{array}{lclll}
  \vspace{1mm}
  \mathbf{p}_{\alpha n} =
    -\,i\hbar\, \displaystyle\frac{\mathbf{d}}{\mathbf{dr}_{\alpha n}} =
    -\,i\hbar\, \displaystyle\frac{\mathbf{dR}}{\mathbf{dr}_{\alpha n}}\,
      \displaystyle\frac{\mathbf{d}}{\mathbf{dR}}
    -\,i\hbar\, \displaystyle\frac{\mathbf{dr}}{\mathbf{dr}_{\alpha n}}\,
      \displaystyle\frac{\mathbf{d}}{\mathbf{dr}}
    -\,i\hbar\, \displaystyle\sum_{k=1}^{n-1}
      \displaystyle\frac{\mathbf{d}\rhobf_{\alpha k}}{\mathbf{dr}_{\alpha n}}\,
      \displaystyle\frac{\mathbf{d}}{\mathbf{d\rho}_{\alpha k}}
    -\,i\hbar\, \displaystyle\sum_{k=1}^{A-1}
      \displaystyle\frac{\mathbf{d}\rhobf_{A k}}{\mathbf{dr}_{\alpha n}}\,
      \displaystyle\frac{\mathbf{d}}{\mathbf{d\rho}_{A k}} = \\

  \vspace{1mm}
  = -\,i\hbar\, \displaystyle\frac{m_{n}}{m_{A} + m_{\alpha}}
      \displaystyle\frac{\mathbf{d}}{\mathbf{dR}}
    -\,i\hbar\, \displaystyle\frac{m_{n}}{m_{\alpha}}\,
      \displaystyle\frac{\mathbf{d}}{\mathbf{dr}}
    +\,i\hbar\, \displaystyle\sum_{k=1}^{n-1}
      \displaystyle\frac{m_{n}}{m_{\alpha}}
      \displaystyle\frac{\mathbf{d}}{\mathbf{d\rho}_{\alpha k}} =

  \displaystyle\frac{m_{n}}{m_{A} + m_{\alpha}}\, \mathbf{P} +
    \displaystyle\frac{m_{n}}{m_{\alpha}}\,\mathbf{p} -
    \displaystyle\frac{m_{n}}{m_{\alpha}}\,
      \displaystyle\sum_{k=1}^{n-1} \mathbf{\tilde{p}}_{\alpha k}.
\end{array}
\label{eq.app.1.12}
\end{equation}
For nucleons of the nucleus we have the similar expressions, but need only change sign before momentum $\mathbf{p}$.
Summarize all final formulas:
\begin{equation}
\begin{array}{lclll}
  \vspace{1mm}
  \mathbf{p}_{\alpha i} =
  \displaystyle\frac{m_{\alpha i}}{m_{A} + m_{\alpha}}\, \mathbf{P} +
    \displaystyle\frac{m_{\alpha i}}{m_{\alpha}}\,\mathbf{p} +
    \displaystyle\frac{m_{\alpha} - m_{\alpha i}}{m_{\alpha}}\, \mathbf{\tilde{p}}_{\alpha i} -
    \displaystyle\frac{m_{\alpha i}}{m_{\alpha}}\,
      \displaystyle\sum_{k=1, k \ne i}^{n-1} \mathbf{\tilde{p}}_{\alpha k} \quad
      {\rm at }\; i = 1 \ldots n-1, \\

  \vspace{1mm}
  \mathbf{p}_{\alpha n} =
  \displaystyle\frac{m_{\alpha n}}{m_{A} + m_{\alpha}}\, \mathbf{P} +
    \displaystyle\frac{m_{\alpha n}}{m_{\alpha}}\,\mathbf{p} -
    \displaystyle\frac{m_{\alpha n}}{m_{\alpha}}\,
      \displaystyle\sum_{k=1}^{n-1} \mathbf{\tilde{p}}_{\alpha k}, \\

  \vspace{1mm}
  \mathbf{p}_{Aj} =
  \displaystyle\frac{m_{Aj}}{m_{A} + m_{\alpha}}\, \mathbf{P} -
    \displaystyle\frac{m_{Aj}}{m_{A}}\,\mathbf{p} +
    \displaystyle\frac{m_{A} - m_{Aj}}{m_{A}}\, \mathbf{\tilde{p}}_{Aj} -
    \displaystyle\frac{m_{Aj}}{m_{A}}\,
      \displaystyle\sum_{k=1, k \ne j}^{A-1} \mathbf{\tilde{p}}_{Ak} \quad {\rm at }\; j = 1 \ldots A-1, \\

  \vspace{1mm}
  \mathbf{p}_{AA} =
  \displaystyle\frac{m_{AA}}{m_{A} + m_{\alpha}}\, \mathbf{P} -
    \displaystyle\frac{m_{AA}}{m_{A}}\,\mathbf{p} -
    \displaystyle\frac{m_{AA}}{m_{A}}\,
      \displaystyle\sum_{k=1}^{A-1} \mathbf{\tilde{p}}_{Ak}.
\end{array}
\label{eq.app.1.13}
\end{equation}

\section{Operator of emission in relative coordinates
\label{sec.app.2}}


Now we will find operator of emission in new relative coordinates.
For this, we start from (\ref{eq.2.3.4}), rewriting this expression via relative momenta:
\begin{equation}
\begin{array}{lcl}
  \hat{H}_{\gamma} & = &
  -\, \sqrt{\displaystyle\frac{2\pi c^{2}}{\hbar w_{\rm ph}}}\;
  \displaystyle\sum_{i=1}^{4}
  \displaystyle\sum\limits_{\alpha=1,2}
    e^{-i\, \mathbf{k_{\rm ph}r}_{i}}\,
  \biggl\{
    \mu_{N}\, \displaystyle\frac{2 z_{i} m_{\rm p}}{m_{\alpha i}}\: \mathbf{e}^{(\alpha)} \cdot \mathbf{p}_{\alpha i} +
    i\, \mu_{i}^{\rm (an)}\, \sigmabf \cdot \Bigl( - \hbar \bigl[ \mathbf{k_{\rm ph}} \times \mathbf{e}^{(\alpha)} \bigr] + \bigl[ \mathbf{p}_{\alpha i} \times \mathbf{e}^{(\alpha)} \bigr] \Bigr)
  \biggr\}\; - \\

  & - &
  \sqrt{\displaystyle\frac{2\pi c^{2}}{\hbar w_{\rm ph}}}\;
  \displaystyle\sum_{j=1}^{A}
  \displaystyle\sum\limits_{\alpha=1,2}
    e^{-i\, \mathbf{k_{\rm ph}r}_{j}}\;
    \biggl\{
      \mu_{N}\, \displaystyle\frac{2 z_{j} m_{\rm p}}{m_{Aj}}\: \mathbf{e}^{(\alpha)} \cdot \mathbf{p}_{Aj} +
      i\, \mu_{j}^{\rm (an)}\, \sigmabf \cdot \Bigl( - \hbar \bigl[ \mathbf{k_{\rm ph}} \times \mathbf{e}^{(\alpha)} \bigr] + \bigl[ \mathbf{p}_{Aj} \times \mathbf{e}^{(\alpha)} \bigr] \Bigr)
    \biggr\}.
\end{array}
\label{eq.app.2.1}
\end{equation}
Substituting here formulas (\ref{eq.app.1.13}), we find:
\begin{equation}
\begin{array}{lcl}
  \hat{H}_{\gamma} & = &
  -\, \sqrt{\displaystyle\frac{2\pi c^{2}}{\hbar w_{\rm ph}}}\;
  \displaystyle\sum_{i=1}^{4}
  \displaystyle\sum\limits_{\alpha=1,2}
    e^{-i\, \mathbf{k_{\rm ph}r}_{i}}\,
  \biggl\{
    \mu_{N}\, \displaystyle\frac{2 z_{i} m_{\rm p}}{m_{\alpha i}}\: \mathbf{e}^{(\alpha)} \cdot
    \Bigl[
      \displaystyle\frac{m_{\alpha i}}{m_{A} + m_{\alpha}}\, \mathbf{P} +
      \displaystyle\frac{m_{\alpha i}}{m_{\alpha}}\,\mathbf{p} +
      \displaystyle\frac{m_{\alpha} - m_{\alpha i}}{m_{\alpha}}\, \mathbf{\tilde{p}}_{\alpha i} -
      \displaystyle\frac{m_{\alpha i}}{m_{\alpha}}\,
      \displaystyle\sum_{k=1, k \ne i}^{n-1} \mathbf{\tilde{p}}_{\alpha k}
    \Bigr]_{i \ne n}\; + \\

  & + &
    \mu_{N}\, \displaystyle\frac{2 z_{i} m_{\rm p}}{m_{\alpha i}}\: \mathbf{e}^{(\alpha)} \cdot
    \Bigl[
      \displaystyle\frac{m_{\alpha n}}{m_{A} + m_{\alpha}}\, \mathbf{P} +
      \displaystyle\frac{m_{\alpha n}}{m_{\alpha}}\,\mathbf{p} -
      \displaystyle\frac{m_{\alpha n}}{m_{\alpha}}\,
      \displaystyle\sum_{k=1}^{n-1} \mathbf{\tilde{p}}_{\alpha k}
    \Bigr]_{i=n} -
    i\, \hbar\, \mu_{i}^{\rm (an)}\, \sigmabf \cdot \bigl[ \mathbf{k_{\rm ph}} \times \mathbf{e}^{(\alpha)} \bigr]\; + \\

  & + &
    i\, \mu_{i}^{\rm (an)}\, \sigmabf \cdot \Bigl[
    \Bigl(
      \displaystyle\frac{m_{\alpha i}}{m_{A} + m_{\alpha}}\, \mathbf{P} +
      \displaystyle\frac{m_{\alpha i}}{m_{\alpha}}\,\mathbf{p} +
      \displaystyle\frac{m_{\alpha} - m_{\alpha i}}{m_{\alpha}}\, \mathbf{\tilde{p}}_{\alpha i} -
      \displaystyle\frac{m_{\alpha i}}{m_{\alpha}}\,
      \displaystyle\sum_{k=1, k \ne i}^{n-1} \mathbf{\tilde{p}}_{\alpha k}
    \Bigr)_{i \ne n}
    \times \mathbf{e}^{(\alpha)} \Bigr]\; + \\

\vspace{4mm}
  & + &
    i\, \mu_{i}^{\rm (an)}\, \sigmabf \cdot \Bigl[
    \Bigl(
      \displaystyle\frac{m_{\alpha n}}{m_{A} + m_{\alpha}}\, \mathbf{P} +
      \displaystyle\frac{m_{\alpha n}}{m_{\alpha}}\,\mathbf{p} -
      \displaystyle\frac{m_{\alpha n}}{m_{\alpha}}\,
      \displaystyle\sum_{k=1}^{n-1} \mathbf{\tilde{p}}_{\alpha k}
    \Bigr)_{i=n}
    \times \mathbf{e}^{(\alpha)} \Bigr]
  \biggr\}\; - \\

  & - &
  \sqrt{\displaystyle\frac{2\pi c^{2}}{\hbar w_{\rm ph}}}\;
  \displaystyle\sum_{j=1}^{A}
  \displaystyle\sum\limits_{\alpha=1,2}
    e^{-i\, \mathbf{k_{\rm ph}r}_{i}}\,
  \biggl\{
    \mu_{N}\, \displaystyle\frac{2 z_{j} m_{\rm p}}{m_{Aj}}\: \mathbf{e}^{(\alpha)} \cdot
    \Bigl[
      \displaystyle\frac{m_{Aj}}{m_{A} + m_{A}}\, \mathbf{P} -
      \displaystyle\frac{m_{Aj}}{m_{A}}\,\mathbf{p} +
      \displaystyle\frac{m_{A} - m_{Aj}}{m_{A}}\, \mathbf{\tilde{p}}_{Aj} -
      \displaystyle\frac{m_{Aj}}{m_{A}}\,
      \displaystyle\sum_{k=1, k \ne j}^{A-1} \mathbf{\tilde{p}}_{A k}
    \Bigr]_{j \ne A}\; + \\

  & + &
    \mu_{N}\, \displaystyle\frac{2 z_{j} m_{\rm p}}{m_{Aj}}\: \mathbf{e}^{(\alpha)} \cdot
    \Bigl[
      \displaystyle\frac{m_{AA}}{m_{A} + m_{\alpha}}\, \mathbf{P} -
      \displaystyle\frac{m_{AA}}{m_{A}}\,\mathbf{p} -
      \displaystyle\frac{m_{AA}}{m_{A}}\,
      \displaystyle\sum_{k=1}^{A-1} \mathbf{\tilde{p}}_{A k}
    \Bigr]_{j=A} -
    i\, \hbar\, \mu_{j}^{\rm (an)}\, \sigmabf \cdot \bigl[ \mathbf{k_{\rm ph}} \times \mathbf{e}^{(\alpha)} \bigr]\; + \\

  & + &
    i\, \mu_{j}^{\rm (an)}\, \sigmabf \cdot \Bigl[
    \Bigl(
      \displaystyle\frac{m_{Aj}}{m_{A} + m_{\alpha}}\, \mathbf{P} -
      \displaystyle\frac{m_{Aj}}{m_{A}}\,\mathbf{p} +
      \displaystyle\frac{m_{A} - m_{Aj}}{m_{A}}\, \mathbf{\tilde{p}}_{Aj} -
      \displaystyle\frac{m_{Aj}}{m_{A}}\,
      \displaystyle\sum_{k=1, k \ne i}^{n-1} \mathbf{\tilde{p}}_{\alpha k}
    \Bigr)_{j \ne A}
    \times \mathbf{e}^{(\alpha)} \Bigr]\; + \\

  & + &
    i\, \mu_{j}^{\rm (an)}\, \sigmabf \cdot \Bigl[
    \Bigl(
      \displaystyle\frac{m_{AA}}{m_{A} + m_{\alpha}}\, \mathbf{P} -
      \displaystyle\frac{m_{AA}}{m_{A}}\,\mathbf{p} -
      \displaystyle\frac{m_{AA}}{m_{A}}\,
      \displaystyle\sum_{k=1}^{n-1} \mathbf{\tilde{p}}_{A k}
    \Bigr)_{j=A}
    \times \mathbf{e}^{(\alpha)} \Bigr]
  \biggr\}.
\end{array}
\label{eq.app.2.2}
\end{equation}
In this expression, we combine terms with the similar momenta as
\begin{equation}
  \hat{H}_{\gamma} = \hat{H}_{\gamma 1}  + \hat{H}_{\gamma 2} +\hat{H}_{\gamma 3} + \hat{H}_{\gamma 4},
\label{eq.app.2.3}
\end{equation}
where
\begin{equation}
\begin{array}{lcl}
\vspace{0.5mm}
  \hat{H}_{\gamma 1} & = &
  -\, \sqrt{\displaystyle\frac{2\pi c^{2}}{\hbar w_{\rm ph}}}\;
  \displaystyle\sum_{i=1}^{4}
  \displaystyle\sum\limits_{\alpha=1,2}
    e^{-i\, \mathbf{k_{\rm ph}r}_{i}}\,
  \biggl\{
    \Bigl[
      \Bigl(\mu_{N}\, \displaystyle\frac{2 z_{i} m_{\rm p}}{m_{\alpha i}}\: \mathbf{e}^{(\alpha)} \cdot \displaystyle\frac{m_{\alpha i}}{m_{A} + m_{\alpha}} \Bigr)_{i \ne n} +
      \Bigl( \mu_{N}\, \displaystyle\frac{2 z_{i} m_{\rm p}}{m_{\alpha i}}\: \mathbf{e}^{(\alpha)} \cdot \displaystyle\frac{m_{\alpha n}}{m_{A} + m_{\alpha}} \Bigr)_{i = n}
    \Bigr]\, \mathbf{P} + \\
\vspace{0.5mm}
  & + &
    \Bigl[
      \Bigl( i\, \mu_{i}^{\rm (an)}\, \sigmabf \cdot \displaystyle\frac{m_{\alpha i}}{m_{A} + m_{\alpha}} \Bigr)_{i \ne n} +
      \Bigl( i\, \mu_{i}^{\rm (an)}\, \sigmabf \cdot \displaystyle\frac{m_{\alpha n}}{m_{A} + m_{\alpha}} \Bigr)_{i = n}
    \Bigr]\,
    \bigl[ \mathbf{P} \times \mathbf{e}^{(\alpha)} \bigr]\; + \\
\vspace{0.5mm}
  & + &
    \Bigl[
      \Bigl(\mu_{N}\, \displaystyle\frac{2 z_{i} m_{\rm p}}{m_{\alpha i}}\: \mathbf{e}^{(\alpha)} \cdot \displaystyle\frac{m_{\alpha i}}{m_{\alpha}} \Bigr)_{i \ne n} +
      \Bigl( \mu_{N}\, \displaystyle\frac{2 z_{i} m_{\rm p}}{m_{\alpha i}}\: \mathbf{e}^{(\alpha)} \cdot \displaystyle\frac{m_{\alpha n}}{m_{\alpha}} \Bigr)_{i = n}
    \Bigr]\, \mathbf{p}\; + \\
  & + &
    \Bigl[
      \Bigl( i\, \mu_{i}^{\rm (an)}\, \sigmabf \cdot \displaystyle\frac{m_{\alpha i}}{m_{\alpha}} \Bigr)_{i \ne n} +
      \Bigl( i\, \mu_{i}^{\rm (an)}\, \sigmabf \cdot \displaystyle\frac{m_{\alpha n}}{m_{\alpha}} \Bigr)_{i = n}
    \Bigr]\,
  \bigl[ \mathbf{p} \times \mathbf{e}^{(\alpha)} \bigr] \biggr\},
\end{array}
\label{eq.app.2.3.a}
\end{equation}
\begin{equation}
\begin{array}{lcl}
\vspace{0.5mm}
  \hat{H}_{\gamma 2} & = &
  -\, \sqrt{\displaystyle\frac{2\pi c^{2}}{\hbar w_{\rm ph}}}\;
  \displaystyle\sum_{i=1}^{4}
  \displaystyle\sum\limits_{\alpha=1,2}
    e^{-i\, \mathbf{k_{\rm ph}r}_{i}}\,
  \biggl\{
    \Bigl( \mu_{N}\, \displaystyle\frac{2 z_{i} m_{\rm p}}{m_{i}}\: \mathbf{e}^{(\alpha)} \cdot \displaystyle\frac{m_{\alpha} - m_{\alpha i}}{m_{\alpha}}\, \Bigr)_{i \ne n}\, \mathbf{\tilde{p}}_{\alpha i}\; + \\
\vspace{0.5mm}
  & + &
    \Bigl( - \mu_{N}\, \displaystyle\frac{2 z_{i} m_{\rm p}}{m_{i}}\: \mathbf{e}^{(\alpha)} \cdot \displaystyle\frac{m_{\alpha i}}{m_{\alpha}} \Bigr)_{i \ne n}\,
      \displaystyle\sum_{k=1, k \ne i}^{n-1} \mathbf{\tilde{p}}_{\alpha k} -
    \mu_{N}\, \displaystyle\frac{2 z_{i} m_{\rm p}}{m_{i}}\: \mathbf{e}^{(\alpha)} \cdot \Bigl[ \displaystyle\frac{m_{\alpha n}}{m_{\alpha}}\, \displaystyle\sum_{k=1}^{n-1} \mathbf{\tilde{p}}_{\alpha k} \Bigr]_{i=n} + \\
\vspace{0.5mm}
  & - &
    i\, \hbar\, \mu_{i}^{\rm (an)}\, \sigmabf \cdot \bigl[ \mathbf{k_{\rm ph}} \times \mathbf{e}^{(\alpha)} \bigr]\; + \\
  & + &
    i\, \mu_{i}^{\rm (an)}\, \sigmabf \cdot \Bigl[
    \Bigl(
      \displaystyle\frac{m_{\alpha} - m_{\alpha i}}{m_{\alpha}}\, \mathbf{\tilde{p}}_{\alpha i} -
      \displaystyle\frac{m_{\alpha i}}{m_{\alpha}}\,
      \displaystyle\sum_{k=1, k \ne i}^{n-1} \mathbf{\tilde{p}}_{\alpha k}
    \Bigr)_{i \ne n}
    \times \mathbf{e}^{(\alpha)} \Bigr] +
    i\, \mu_{i}^{\rm (an)}\, \sigmabf \cdot \Bigl[
    \Bigl(
      - \displaystyle\frac{m_{\alpha n}}{m_{\alpha}}\,
      \displaystyle\sum_{k=1}^{n-1} \mathbf{\tilde{p}}_{\alpha k}
    \Bigr)_{i=n}
    \times \mathbf{e}^{(\alpha)} \Bigr]
  \biggr\},
\end{array}
\label{eq.app.2.3.b}
\end{equation}
\begin{equation}
\begin{array}{lcl}
\vspace{0.5mm}
  \hat{H}_{\gamma 3} & = &
  -\, \sqrt{\displaystyle\frac{2\pi c^{2}}{\hbar w_{\rm ph}}}\;
  \displaystyle\sum_{j=1}^{A}
  \displaystyle\sum\limits_{\alpha=1,2}
    e^{-i\, \mathbf{k_{\rm ph}r}_{j}}\,
  \biggl\{
    \Bigl[
      \Bigl( \mu_{N}\, \displaystyle\frac{2 z_{j} m_{\rm p}}{m_{Aj}}\: \mathbf{e}^{(\alpha)} \cdot \displaystyle\frac{m_{Aj}}{m_{A} + m_{\alpha}} \Bigr)_{j \ne A} +
      \Bigl( \mu_{N}\, \displaystyle\frac{2 z_{j} m_{\rm p}}{m_{Aj}}\: \mathbf{e}^{(\alpha)} \cdot \displaystyle\frac{m_{AA}}{m_{A} + m_{\alpha}} \Bigr)_{j = A}
    \Bigr]\, \mathbf{P} + \\
\vspace{0.5mm}
  & + &
    \Bigl[
      \Bigl( i\, \mu_{j}^{\rm (an)}\, \sigmabf \cdot \displaystyle\frac{m_{Aj}}{m_{A} + m_{\alpha}} \Bigr)_{j \ne A} +
      \Bigl( i\, \mu_{j}^{\rm (an)}\, \sigmabf \cdot \displaystyle\frac{m_{AA}}{m_{A} + m_{\alpha}} \Bigr)_{j = A}
    \Bigr]\,
    \bigl[ \mathbf{P} \times \mathbf{e}^{(\alpha)} \bigr]\; - \\
\vspace{0.5mm}
  & - &
    \Bigl[
      \Bigl( \mu_{N}\, \displaystyle\frac{2 z_{j} m_{\rm p}}{m_{Aj}}\: \mathbf{e}^{(\alpha)} \cdot \displaystyle\frac{m_{Aj}}{m_{A}} \Bigr)_{j \ne A} +
      \Bigl( \mu_{N}\, \displaystyle\frac{2 z_{j} m_{\rm p}}{m_{Aj}}\: \mathbf{e}^{(\alpha)} \cdot \displaystyle\frac{m_{AA}}{m_{A}} \Bigr)_{j = A}
    \Bigr]\, \mathbf{p}\; - \\
  & - &
    \Bigl[
      \Bigl( i\, \mu_{j}^{\rm (an)}\, \sigmabf \cdot \displaystyle\frac{m_{Aj}}{m_{A}} \Bigr)_{j \ne A} +
      \Bigl( i\, \mu_{j}^{\rm (an)}\, \sigmabf \cdot \displaystyle\frac{m_{AA}}{m_{A}} \Bigr)_{j = A}
    \Bigr]\,
    \bigl[ \mathbf{p} \times \mathbf{e}^{(\alpha)} \bigr]
  \biggr\},
\end{array}
\label{eq.app.2.3.c}
\end{equation}
\begin{equation}
\begin{array}{lcl}
\vspace{0.5mm}
  \hat{H}_{\gamma 4} & = &
  -\, \sqrt{\displaystyle\frac{2\pi c^{2}}{\hbar w_{\rm ph}}}\;
  \displaystyle\sum_{j=1}^{A}
  \displaystyle\sum\limits_{\alpha=1,2}
    e^{-i\, \mathbf{k_{\rm ph}r}_{j}}\,
  \biggl\{
    \Bigl( \mu_{N}\, \displaystyle\frac{2 z_{j} m_{\rm p}}{m_{Aj}}\: \mathbf{e}^{(\alpha)} \cdot \displaystyle\frac{m_{A} - m_{Aj}}{m_{A}}\, \Bigr)_{j \ne A}\, \mathbf{\tilde{p}}_{Aj}\; + \\
\vspace{0.5mm}
  & + &
    \Bigl( - \mu_{N}\, \displaystyle\frac{2 z_{j} m_{\rm p}}{m_{Aj}}\: \mathbf{e}^{(\alpha)} \cdot \displaystyle\frac{m_{Aj}}{m_{A}} \Bigr)_{j \ne A}\,
      \displaystyle\sum_{k=1, k \ne j}^{A-1} \mathbf{\tilde{p}}_{Ak} -
    \mu_{N}\, \displaystyle\frac{2 z_{j} m_{\rm p}}{m_{Aj}}\: \mathbf{e}^{(\alpha)} \cdot \Bigl[ \displaystyle\frac{m_{AA}}{m_{A}}\, \displaystyle\sum_{k=1}^{A-1} \mathbf{\tilde{p}}_{Ak} \Bigr]_{j=A} - \\
\vspace{0.5mm}
& - &
    i\, \hbar\, \mu_{j}^{\rm (an)}\, \sigmabf \cdot \bigl[ \mathbf{k_{\rm ph}} \times \mathbf{e}^{(\alpha)} \bigr]\; + \\
& + &
    i\, \mu_{j}^{\rm (an)}\, \sigmabf \cdot \Bigl[
    \Bigl(
      \displaystyle\frac{m_{A} - m_{Aj}}{m_{A}}\, \mathbf{\tilde{p}}_{Aj} -
      \displaystyle\frac{m_{Aj}}{m_{A}}\,
      \displaystyle\sum_{k=1, k \ne j}^{A-1} \mathbf{\tilde{p}}_{Ak}
    \Bigr)_{j \ne A}
    \times \mathbf{e}^{(\alpha)} \Bigr] +
    i\, \mu_{j}^{\rm (an)}\, \sigmabf \cdot \Bigl[
    \Bigl(
      - \displaystyle\frac{m_{AA}}{m_{A}}\,
      \displaystyle\sum_{k=1}^{A-1} \mathbf{\tilde{p}}_{Ak}
    \Bigr)_{j=A}
    \times \mathbf{e}^{(\alpha)} \Bigr]
  \biggr\}.
\end{array}
\label{eq.app.2.3.d}
\end{equation}
We simplify this expression, analyzing summations over different indexes. After calculations, we obtain:
\begin{equation}
  \hat{H}_{\gamma} = \hat{H}_{P} + \hat{H}_{p} + \Delta\,\hat{H}_{\gamma} + \hat{H}_{k},
\label{eq.app.2.5}
\end{equation}
where
\begin{equation}
\begin{array}{lcl}
\vspace{0.1mm}
  \hat{H}_{P} & = &
  -\, \sqrt{\displaystyle\frac{2\pi c^{2}}{\hbar w_{\rm ph}}}\;
  \displaystyle\sum\limits_{\alpha=1,2}
  \biggl\{
    \displaystyle\sum_{i=1}^{4}
      e^{-i\, \mathbf{k_{\rm ph}r}_{i}}\,
  \biggl[
    \mu_{N}\, \displaystyle\frac{2 z_{i} m_{\rm p}}{m_{A} + m_{\alpha}}\, \mathbf{e}^{(\alpha)} \cdot \mathbf{P} +
    i\, \mu_{i}^{\rm (an)}\, \displaystyle\frac{m_{\alpha i}}{m_{A} + m_{\alpha}}\, \sigmabf \cdot \bigl[ \mathbf{P} \times \mathbf{e}^{(\alpha)} \bigr]  \biggr]\; + \\
  & + &
  \displaystyle\sum_{j=1}^{A} e^{-i\, \mathbf{k_{\rm ph}r}_{j}}\,
  \biggl[
    \mu_{N}\, \displaystyle\frac{2 z_{j} m_{\rm p}}{m_{A} + m_{\alpha}}\, \mathbf{e}^{(\alpha)} \cdot \mathbf{P} +
    i\, \mu_{j}^{\rm (an)}\, \displaystyle\frac{m_{Aj}}{m_{A} + m_{\alpha}}\, \sigmabf \cdot \bigl[ \mathbf{P} \times \mathbf{e}^{(\alpha)} \bigr]
  \biggr]
  \biggl\},
\end{array}
\label{eq.app.2.6}
\end{equation}
\begin{equation}
\begin{array}{lcl}
\vspace{0.1mm}
  \hat{H}_{p} & = &
  -\, \sqrt{\displaystyle\frac{2\pi c^{2}}{\hbar w_{\rm ph}}}\;
  \displaystyle\sum\limits_{\alpha=1,2}
  \biggl\{
    \displaystyle\sum_{i=1}^{4}
      e^{-i\, \mathbf{k_{\rm ph}r}_{i}}\,
  \biggl[
    \mu_{N}\, \displaystyle\frac{2 z_{i} m_{\rm p}}{m_{\alpha}}\, \mathbf{e}^{(\alpha)} \cdot \mathbf{p} +
    i\, \mu_{i}^{\rm (an)}\, \displaystyle\frac{m_{\alpha i}}{m_{\alpha}}\, \sigmabf \cdot \bigl[ \mathbf{p} \times \mathbf{e}^{(\alpha)} \bigr] \biggr]\; - \\

  & - &
  \displaystyle\sum_{j=1}^{A} e^{-i\, \mathbf{k_{\rm ph}r}_{j}}\,
  \biggl[
    \mu_{N}\, \displaystyle\frac{2 z_{j} m_{\rm p}}{m_{A}}\; \mathbf{e}^{(\alpha)} \cdot \mathbf{p} +
    i\, \mu_{j}^{\rm (an)}\, \displaystyle\frac{m_{Aj}}{m_{A}}\, \sigmabf \cdot \bigl[ \mathbf{p} \times \mathbf{e}^{(\alpha)} \bigr] \biggr\},
\end{array}
\label{eq.app.2.7}
\end{equation}
\begin{equation}
\begin{array}{lcl}
\vspace{0.1mm}
  & & \Delta \hat{H}_{\gamma} =
  -\, \sqrt{\displaystyle\frac{2\pi c^{2}}{\hbar w_{\rm ph}}}\;
  \displaystyle\sum\limits_{\alpha=1,2}
  \displaystyle\sum_{i=1}^{4}
    e^{-i\, \mathbf{k_{\rm ph}r}_{i}}\,
  \biggl\{
    \Bigl( \mu_{N}\, \displaystyle\frac{2 z_{i} m_{\rm p}}{m_{i}}\: \mathbf{e}^{(\alpha)} \cdot \displaystyle\frac{m_{\alpha} - m_{\alpha i}}{m_{\alpha}}\, \Bigr)_{i \ne n}\, \mathbf{\tilde{p}}_{\alpha i}\; - \\
  & - &
    \Bigl( \mu_{N}\, \displaystyle\frac{2 z_{i} m_{\rm p}}{m_{i}}\: \mathbf{e}^{(\alpha)} \cdot \displaystyle\frac{m_{\alpha i}}{m_{\alpha}} \Bigr)_{i \ne n}\,
      \displaystyle\sum_{k=1, k \ne i}^{n-1} \mathbf{\tilde{p}}_{\alpha k} -
    \mu_{N}\, \displaystyle\frac{2 z_{i} m_{\rm p}}{m_{i}}\: \mathbf{e}^{(\alpha)} \cdot \Bigl[ \displaystyle\frac{m_{\alpha n}}{m_{\alpha}}\, \displaystyle\sum_{k=1}^{n-1} \mathbf{\tilde{p}}_{\alpha k} \Bigr]_{i=n} + \\
\vspace{3.5mm}
  & + &
    i\, \mu_{i}^{\rm (an)}\, \sigmabf \cdot \Bigl[
    \Bigl(
      \displaystyle\frac{m_{\alpha} - m_{\alpha i}}{m_{\alpha}}\, \mathbf{\tilde{p}}_{\alpha i} -
      \displaystyle\frac{m_{\alpha i}}{m_{\alpha}}\,
      \displaystyle\sum_{k=1, k \ne i}^{n-1} \mathbf{\tilde{p}}_{\alpha k}
    \Bigr)_{i \ne n}
    \times \mathbf{e}^{(\alpha)} \Bigr] -
    i\, \mu_{i}^{\rm (an)}\, \displaystyle\frac{m_{\alpha n}}{m_{\alpha}}\,
      \sigmabf \cdot \Bigl[ \Bigl( \displaystyle\sum_{k=1}^{n-1} \mathbf{\tilde{p}}_{\alpha k} \Bigr)_{i=n} \times \mathbf{e}^{(\alpha)} \Bigr]
  \biggr\}\; - \\

\vspace{0.1mm}
  & - &
  \sqrt{\displaystyle\frac{2\pi c^{2}}{\hbar w_{\rm ph}}}\;
  \displaystyle\sum\limits_{\alpha=1,2}
  \displaystyle\sum_{j=1}^{A}
    e^{-i\, \mathbf{k_{\rm ph}r}_{j}}\,
  \biggl\{
    \Bigl( \mu_{N}\, \displaystyle\frac{2 z_{j} m_{\rm p}}{m_{Aj}}\: \mathbf{e}^{(\alpha)} \cdot \displaystyle\frac{m_{A} - m_{Aj}}{m_{A}}\, \Bigr)_{j \ne A}\, \mathbf{\tilde{p}}_{Aj}\; + \\
& + &
    \Bigl( - \mu_{N}\, \displaystyle\frac{2 z_{j} m_{\rm p}}{m_{Aj}}\: \mathbf{e}^{(\alpha)} \cdot \displaystyle\frac{m_{Aj}}{m_{A}} \Bigr)_{j \ne A}\,
      \displaystyle\sum_{k=1, k \ne j}^{A-1} \mathbf{\tilde{p}}_{Ak} -
    \mu_{N}\, \displaystyle\frac{2 z_{j} m_{\rm p}}{m_{Aj}}\: \mathbf{e}^{(\alpha)} \cdot \Bigl[ \displaystyle\frac{m_{AA}}{m_{A}}\, \displaystyle\sum_{k=1}^{A-1} \mathbf{\tilde{p}}_{Ak} \Bigr]_{j=A} + \\
& + &
    i\, \mu_{j}^{\rm (an)}\, \sigmabf \cdot \Bigl[
    \Bigl(
      \displaystyle\frac{m_{A} - m_{Aj}}{m_{A}}\, \mathbf{\tilde{p}}_{Aj} -
      \displaystyle\frac{m_{Aj}}{m_{A}}\,
      \displaystyle\sum_{k=1, k \ne j}^{A-1} \mathbf{\tilde{p}}_{Ak}
    \Bigr)_{j \ne A}
    \times \mathbf{e}^{(\alpha)} \Bigr] -
    i\, \mu_{j}^{\rm (an)}\, \displaystyle\frac{m_{AA}}{m_{A}}\,
      \sigmabf \cdot \Bigl[ \Bigl( \displaystyle\sum_{k=1}^{A-1} \mathbf{\tilde{p}}_{Ak} \Bigr)_{j=A} \times \mathbf{e}^{(\alpha)} \Bigr]
  \biggr\},
\end{array}
\label{eq.app.2.8}
\end{equation}
and
\begin{equation}
\begin{array}{lcl}
\vspace{0.1mm}
  \hat{H}_{k} & = &
  \sqrt{\displaystyle\frac{2\pi c^{2}}{\hbar w_{\rm ph}}}\;
  \displaystyle\sum\limits_{\alpha=1,2}
  \displaystyle\sum_{i=1}^{4}
    e^{-i\, \mathbf{k_{\rm ph}r}_{i}}\,
  \biggl\{ i\, \hbar\, \mu_{i}^{\rm (an)}\, \sigmabf \cdot \bigl[ \mathbf{k_{\rm ph}} \times \mathbf{e}^{(\alpha)} \bigr] \biggr\}\; + \\
& + &
  \sqrt{\displaystyle\frac{2\pi c^{2}}{\hbar w_{\rm ph}}}\;
  \displaystyle\sum\limits_{\alpha=1,2}
  \displaystyle\sum_{j=1}^{A}
    e^{-i\, \mathbf{k_{\rm ph}r}_{j}}\,
  \biggl\{ i\, \hbar\, \mu_{j}^{\rm (an)}\, \sigmabf \cdot \bigl[ \mathbf{k_{\rm ph}} \times \mathbf{e}^{(\alpha)} \bigr] \biggr\}.
\end{array}
\label{eq.app.2.9}
\end{equation}
Let us simplify these expressions. For $\hat{H}_{P}$ we obtain:
\begin{equation}
\begin{array}{lcl}
\vspace{0.1mm}
  \hat{H}_{P} & = &
  -\, \sqrt{\displaystyle\frac{2\pi c^{2}}{\hbar w_{\rm ph}}}\;
  \displaystyle\sum\limits_{\alpha=1,2}
  \biggl\{
    \displaystyle\sum_{i=1}^{4} e^{-i\, \mathbf{k_{\rm ph}r}_{i}}\,
      \mu_{N}\, \displaystyle\frac{2 z_{i} m_{\rm p}}{m_{A} + m_{\alpha}} +
    \displaystyle\sum_{j=1}^{A} e^{-i\, \mathbf{k_{\rm ph}r}_{j}}\,
      \mu_{N}\, \displaystyle\frac{2 z_{j} m_{\rm p}}{m_{A} + m_{\alpha}}
  \biggr\}\, \mathbf{e}^{(\alpha)} \cdot \mathbf{P} + \\
\vspace{1.0mm}
  & - &
  \sqrt{\displaystyle\frac{2\pi c^{2}}{\hbar w_{\rm ph}}}\;
    \displaystyle\sum\limits_{\alpha=1,2}
    \biggl\{
      \displaystyle\sum_{i=1}^{4} e^{-i\, \mathbf{k_{\rm ph}r}_{i}}\,
        i\, \mu_{i}^{\rm (an)}\, \displaystyle\frac{m_{\alpha i}}{m_{A} + m_{\alpha}} +

      \displaystyle\sum_{j=1}^{A} e^{-i\, \mathbf{k_{\rm ph}r}_{j}}\,
        i\, \mu_{j}^{\rm (an)}\, \displaystyle\frac{m_{Aj}}{m_{A} + m_{\alpha}}
  \biggr\}\, \sigmabf \cdot \bigl[ \mathbf{P} \times \mathbf{e}^{(\alpha)} \bigr]\; = \\

\vspace{0.1mm}
  & = &
  -\, \sqrt{\displaystyle\frac{2\pi c^{2}}{\hbar w_{\rm ph}}}\;
  \mu_{N}\, \displaystyle\frac{2 m_{\rm p}}{m_{A} + m_{\alpha}}\,
  \displaystyle\sum\limits_{\alpha=1,2}
  \biggl\{
    \displaystyle\sum_{i=1}^{4} e^{-i\, \mathbf{k_{\rm ph}r}_{i}}\, z_{i} +
    \displaystyle\sum_{j=1}^{A} e^{-i\, \mathbf{k_{\rm ph}r}_{j}}\, z_{j}
  \biggr\}\, \mathbf{e}^{(\alpha)} \cdot \mathbf{P} + \\
  & - &
  \sqrt{\displaystyle\frac{2\pi c^{2}}{\hbar w_{\rm ph}}}\;
    \displaystyle\frac{i}{m_{A} + m_{\alpha}}
    \displaystyle\sum\limits_{\alpha=1,2}
    \biggl\{
      \displaystyle\sum_{i=1}^{4} e^{-i\, \mathbf{k_{\rm ph}r}_{i}}\, \mu_{i}^{\rm (an)}\, m_{\alpha i}\, \sigmabf  +
      \displaystyle\sum_{j=1}^{A} e^{-i\, \mathbf{k_{\rm ph}r}_{j}}\, \mu_{j}^{\rm (an)}\, m_{Aj}\, \sigmabf
  \biggr\}\, \cdot \bigl[ \mathbf{P} \times \mathbf{e}^{(\alpha)} \bigr].
\end{array}
\label{eq.app.2.10}
\end{equation}
For $\hat{H}_{p}$ we obtain:
\begin{equation}
\begin{array}{lcl}
\vspace{0.1mm}
  \hat{H}_{p} & = &
  -\, \sqrt{\displaystyle\frac{2\pi c^{2}}{\hbar w_{\rm ph}}}\;
  \displaystyle\sum\limits_{\alpha=1,2}
  \biggl\{
    \displaystyle\sum_{i=1}^{4} e^{-i\, \mathbf{k_{\rm ph}r}_{i}}\, \mu_{N}\, \displaystyle\frac{2 z_{i} m_{\rm p}}{m_{\alpha}} -
    \displaystyle\sum_{j=1}^{A} e^{-i\, \mathbf{k_{\rm ph}r}_{j}}\, \mu_{N}\, \displaystyle\frac{2 z_{j} m_{\rm p}}{m_{A}}
  \biggr\}\; \mathbf{e}^{(\alpha)} \cdot \mathbf{p}\; - \\
\vspace{1.0mm}
  & - &
  \sqrt{\displaystyle\frac{2\pi c^{2}}{\hbar w_{\rm ph}}}\;
  \displaystyle\sum\limits_{\alpha=1,2}
  \biggl\{
    \displaystyle\sum_{i=1}^{4}
      e^{-i\, \mathbf{k_{\rm ph}r}_{i}}\, i\, \mu_{i}^{\rm (an)}\, \displaystyle\frac{m_{\alpha i}}{m_{\alpha}} -
    \displaystyle\sum_{j=1}^{A} e^{-i\, \mathbf{k_{\rm ph}r}_{j}}\,
      i\, \mu_{j}^{\rm (an)}\, \displaystyle\frac{m_{Aj}}{m_{A}}\, \biggr\}\;
  \sigmabf \cdot \bigl[ \mathbf{p} \times \mathbf{e}^{(\alpha)} \bigr]\; = \\

\vspace{0.1mm}
  & = &
  -\, \sqrt{\displaystyle\frac{2\pi c^{2}}{\hbar w_{\rm ph}}}\;
  2\, \mu_{N}\,  m_{\rm p}\,
  \displaystyle\sum\limits_{\alpha=1,2}
  \biggl\{
    \displaystyle\frac{1}{m_{\alpha}}\, \displaystyle\sum_{i=1}^{4} z_{i}\, e^{-i\, \mathbf{k_{\rm ph}r}_{i}}  -
    \displaystyle\frac{1}{m_{A}}\, \displaystyle\sum_{j=1}^{A} z_{j}\, e^{-i\, \mathbf{k_{\rm ph}r}_{j}}
  \biggr\}\; \mathbf{e}^{(\alpha)} \cdot \mathbf{p}\; - \\
  & - &
  i\, \sqrt{\displaystyle\frac{2\pi c^{2}}{\hbar w_{\rm ph}}}\;
  \displaystyle\sum\limits_{\alpha=1,2}
  \biggl\{
    \displaystyle\frac{1}{m_{\alpha}}
      \displaystyle\sum_{i=1}^{4}
      \mu_{i}^{\rm (an)}\, m_{\alpha i}\, e^{-i\, \mathbf{k_{\rm ph}r}_{i}}\; \sigmabf -
    \displaystyle\frac{1}{m_{A}}
      \displaystyle\sum_{j=1}^{A}
      m_{Aj}\, \mu_{j}^{\rm (an)}\, e^{-i\, \mathbf{k_{\rm ph}r}_{j}}\; \sigmabf \biggr\}\;
  \cdot \bigl[ \mathbf{p} \times \mathbf{e}^{(\alpha)} \bigr].
\end{array}
\label{eq.app.2.11}
\end{equation}
For $\hat{H}_{k}$ we obtain:
\begin{equation}
\begin{array}{lcl}
  \hat{H}_{k} =
  i\, \hbar\,
  \sqrt{\displaystyle\frac{2\pi c^{2}}{\hbar w_{\rm ph}}}\:
  \displaystyle\sum\limits_{\alpha=1,2}
  \biggl\{
    \displaystyle\sum_{i=1}^{4}
      e^{-i\, \mathbf{k_{\rm ph}r}_{i}}\, \mu_{i}^{\rm (an)}\, \sigmabf +
    \displaystyle\sum_{j=1}^{A}
      e^{-i\, \mathbf{k_{\rm ph}r}_{j}}\, \mu_{j}^{\rm (an)}\, \sigmabf
  \biggr\}
  \cdot \bigl[ \mathbf{k_{\rm ph}} \times \mathbf{e}^{(\alpha)} \bigr].
\end{array}
\label{eq.app.2.12}
\end{equation}
Now we rewrite the found solutions in relative coordinates.
Using Eqs.~(\ref{eq.app.1.6}) and (\ref{eq.app.1.7}), from Eqs.~(\ref{eq.app.2.10}) and (\ref{eq.app.2.11}) we obtain:
\begin{equation}
\begin{array}{lcl}
\vspace{-0.1mm}
  \hat{H}_{P} =
  -\, \sqrt{\displaystyle\frac{2\pi c^{2}}{\hbar w_{\rm ph}}}\;
  \mu_{N}\, \displaystyle\frac{2 m_{\rm p}}{m_{A} + m_{\alpha}}\;
  e^{-i\, \mathbf{k_{\rm ph}} \mathbf{R}}
  \displaystyle\sum\limits_{\alpha=1,2}
  \biggl\{
    e^{-i\, c_{A}\, \mathbf{k_{\rm ph}} \mathbf{r}}
      \displaystyle\sum_{i=1}^{4} z_{i}\, e^{-i\, \mathbf{k_{\rm ph}} \rhobf_{\alpha i}} +
    e^{i\, c_{\alpha}\, \mathbf{k_{\rm ph}} \mathbf{r} }
      \displaystyle\sum_{j=1}^{A} z_{j}\, e^{-i\, \mathbf{k_{\rm ph}} \rhobf_{Aj}}
  \biggr\}\, \mathbf{e}^{(\alpha)} \cdot \mathbf{P} + \\
\vspace{1.0mm}
  -\;
  \sqrt{\displaystyle\frac{2\pi c^{2}}{\hbar w_{\rm ph}}}\;
    \displaystyle\frac{i}{m_{A} + m_{\alpha}}\;
    e^{-i\, \mathbf{k_{\rm ph}} \mathbf{R}}\,
    \displaystyle\sum\limits_{\alpha=1,2}
    \biggl\{
      e^{-i\, c_{A}\, \mathbf{k_{\rm ph}} \mathbf{r}}\,
        \displaystyle\sum_{i=1}^{4} \mu_{i}^{\rm (an)}\, m_{\alpha i}\, e^{-i\, \mathbf{k_{\rm ph}} \rhobf_{\alpha i}}\, \sigmabf  +
      e^{i\, c_{\alpha}\, \mathbf{k_{\rm ph}} \mathbf{r}}\,
        \displaystyle\sum_{j=1}^{A} \mu_{j}^{\rm (an)}\, m_{Aj}\, e^{-i\, \mathbf{k_{\rm ph}} \rhobf_{Aj}}\, \sigmabf
  \biggr\}\, \times \\
  \times\; \bigl[ \mathbf{P} \times \mathbf{e}^{(\alpha)} \bigr],
\end{array}
\label{eq.app.2.13}
\end{equation}
\begin{equation}
\begin{array}{lll}
\vspace{-0.1mm}
  \hat{H}_{p} =
  -\, \sqrt{\displaystyle\frac{2\pi c^{2}}{\hbar w_{\rm ph}}}\;
  2\, \mu_{N}\,  m_{\rm p}\,
  e^{-i\, \mathbf{k_{\rm ph}} \mathbf{R}}
  \displaystyle\sum\limits_{\alpha=1,2}
  \biggl\{
    e^{-i\, c_{A} \mathbf{k_{\rm ph}} \mathbf{r}}\, \displaystyle\frac{1}{m_{\alpha}}\,
      \displaystyle\sum_{i=1}^{4} z_{i}\, e^{-i\, \mathbf{k_{\rm ph}} \rhobf_{\alpha i}} -
    e^{i\, c_{\alpha} \mathbf{k_{\rm ph}} \mathbf{r}}\,  \displaystyle\frac{1}{m_{A}}\,
      \displaystyle\sum_{j=1}^{A} z_{j}\, e^{-i\, \mathbf{k_{\rm ph}} \rhobf_{Aj}}
  \biggr\}\; \mathbf{e}^{(\alpha)} \cdot \mathbf{p}\; - \\
\vspace{1.0mm}
  -\;
  i\, \sqrt{\displaystyle\frac{2\pi c^{2}}{\hbar w_{\rm ph}}}\;
  e^{-i\, \mathbf{k_{\rm ph}} \mathbf{R}}
  \displaystyle\sum\limits_{\alpha=1,2}
  \biggl\{
    e^{-i\, c_{A} \mathbf{k_{\rm ph}} \mathbf{r}} \displaystyle\frac{1}{m_{\alpha}}\,
    \displaystyle\sum_{i=1}^{4}
      \mu_{i}^{\rm (an)}\, m_{\alpha i}\;
      e^{-i\, \mathbf{k_{\rm ph}} \rhobf_{\alpha i}}\, \sigmabf -
    e^{i\, c_{\alpha} \mathbf{k_{\rm ph}} \mathbf{r}} \displaystyle\frac{1}{m_{A}}
    \displaystyle\sum_{j=1}^{A}
      \mu_{j}^{\rm (an)}\, m_{Aj}\;
      e^{-i\, \mathbf{k_{\rm ph}} \rhobf_{Aj}}\, \sigmabf \biggr\}\; \times \\
  \times\; \bigl[ \mathbf{p} \times \mathbf{e}^{(\alpha)} \bigr].
\end{array}
\label{eq.app.2.14}
\end{equation}
\begin{equation}
\begin{array}{lcl}
  \hat{H}_{k} & = &
  i\, \hbar\,
  \sqrt{\displaystyle\frac{2\pi c^{2}}{\hbar w_{\rm ph}}}\:
  e^{-i\, \mathbf{k_{\rm ph}} \mathbf{R}}\,
  \displaystyle\sum\limits_{\alpha=1,2}
  \biggl\{
    e^{-i\, c_{A}\, \mathbf{k_{\rm ph}} \mathbf{r}}\,
    \displaystyle\sum_{i=1}^{4}
      \mu_{i}^{\rm (an)}\, e^{-i\, \mathbf{k_{\rm ph}} \rhobf_{\alpha i}}\, \sigmabf +
    e^{i\, c_{\alpha}\, \mathbf{k_{\rm ph}} \mathbf{r}}\,
    \displaystyle\sum_{j=1}^{A}
      \mu_{j}^{\rm (an)}\, e^{-i\, \mathbf{k_{\rm ph}} \rhobf_{Aj}}\, \sigmabf
  \biggr\}
  \cdot \bigl[ \mathbf{k_{\rm ph}} \times \mathbf{e}^{(\alpha)} \bigr].
\end{array}
\label{eq.app.2.15}
\end{equation}

\section{Electric and magnetic form-factors
\label{sec.app.3}}

We substitute explicit formulation (\ref{eq.2.8.1}) for wave function $F (\mathbf{r}, \beta_{A}, \beta_{\alpha})$ to the obtained matrix element (\ref{eq.2.8.7}):
\begin{equation}
\begin{array}{lll}
\vspace{-0.1mm}
  & M_{2} = -\, (2\pi)^{3} \delta (\mathbf{K}_{f} - \mathbf{k}_{\rm ph}) \cdot
  \displaystyle\sum\limits_{\alpha=1,2}
  \biggl\langle
    \Phi_{\rm \alpha - nucl, f} (\mathbf{r}) \cdot
    \psi_{\rm nucl, f} (\beta_{A}) \cdot
    \psi_{\alpha, f} (\beta_{\alpha})\,
  \biggl|\, \times \\

\vspace{-0.1mm}
  \times &
  2\, \mu_{N}\,  m_{\rm p}\,
  \Bigl\{
    e^{-i\, c_{A} \mathbf{k_{\rm ph}} \mathbf{r}}\, \displaystyle\frac{1}{m_{\alpha}}\,
      \displaystyle\sum_{i=1}^{4} z_{i}\, e^{-i\, \mathbf{k_{\rm ph}} \rhobf_{\alpha i}} -
    e^{i\, c_{\alpha} \mathbf{k_{\rm ph}} \mathbf{r}}\,  \displaystyle\frac{1}{m_{A}}\,
      \displaystyle\sum_{j=1}^{A} z_{j}\, e^{-i\, \mathbf{k_{\rm ph}} \rhobf_{Aj}}
  \Bigr\}\; \mathbf{e}^{(\alpha)} \cdot \mathbf{p}\; + \\
\vspace{-0.1mm}
  + &
  i\,
  \Bigl\{
    e^{-i\, c_{A} \mathbf{k_{\rm ph}} \mathbf{r}} \displaystyle\frac{1}{m_{\alpha}}\,
    \displaystyle\sum_{i=1}^{4}
      \mu_{i}^{\rm (an)}\, m_{\alpha i}\;
      e^{-i\, \mathbf{k_{\rm ph}} \rhobf_{\alpha i}}\, \sigmabf -
    e^{i\, c_{\alpha} \mathbf{k_{\rm ph}} \mathbf{r}} \displaystyle\frac{1}{m_{A}}
    \displaystyle\sum_{j=1}^{A}
      \mu_{j}^{\rm (an)}\, m_{Aj}\;
      e^{-i\, \mathbf{k_{\rm ph}} \rhobf_{Aj}}\, \sigmabf \Bigr\}
    \cdot \bigl[ \mathbf{p} \times \mathbf{e}^{(\alpha)} \bigr]\, \times \\
  \times &
  \biggr|\,
    \Phi_{\rm \alpha - nucl, i} (\mathbf{r}) \cdot
    \psi_{\rm nucl, i} (\beta_{A}) \cdot
    \psi_{\alpha, i} (\beta_{\alpha})
  \biggr\rangle.
\end{array}
\label{eq.app.3.1}
\end{equation}
We rewrite integration over variable $\mathbf{r}$ explicitly:
\begin{equation}
\begin{array}{lll}
\vspace{-0.1mm}
  & M_{2} = -\, (2\pi)^{3} \delta (\mathbf{K}_{f} - \mathbf{k}_{\rm ph}) \cdot
  \displaystyle\sum\limits_{\alpha=1,2}
  \displaystyle\int\limits_{}^{}
    \Phi_{\rm \alpha - nucl, f}^{*} (\mathbf{r}) \cdot
  \Bigl\langle
    \psi_{\rm nucl, f} (\beta_{A}) \cdot
    \psi_{\alpha, f} (\beta_{\alpha})\,
  \Bigl|\, \times \\

\vspace{-0.1mm}
  \times &
  2\, \mu_{N}\,  m_{\rm p}\,
  \Bigl\{
    e^{-i\, c_{A} \mathbf{k_{\rm ph}} \mathbf{r}}\, \displaystyle\frac{1}{m_{\alpha}}\,
      \displaystyle\sum_{i=1}^{4} z_{i}\, e^{-i\, \mathbf{k_{\rm ph}} \rhobf_{\alpha i}} -
    e^{i\, c_{\alpha} \mathbf{k_{\rm ph}} \mathbf{r}}\,  \displaystyle\frac{1}{m_{A}}\,
      \displaystyle\sum_{j=1}^{A} z_{j}\, e^{-i\, \mathbf{k_{\rm ph}} \rhobf_{Aj}}
  \Bigr\}\; \mathbf{e}^{(\alpha)} \cdot \mathbf{p}\; + \\
\vspace{-0.1mm}
  + &
  i\,
  \Bigl\{
    e^{-i\, c_{A} \mathbf{k_{\rm ph}} \mathbf{r}} \displaystyle\frac{1}{m_{\alpha}}\,
    \displaystyle\sum_{i=1}^{4}
      \mu_{i}^{\rm (an)}\, m_{\alpha i}\;
      e^{-i\, \mathbf{k_{\rm ph}} \rhobf_{\alpha i}}\, \sigmabf -
    e^{i\, c_{\alpha} \mathbf{k_{\rm ph}} \mathbf{r}} \displaystyle\frac{1}{m_{A}}
    \displaystyle\sum_{j=1}^{A}
      \mu_{j}^{\rm (an)}\, m_{Aj}\;
      e^{-i\, \mathbf{k_{\rm ph}} \rhobf_{Aj}}\, \sigmabf \Bigr\}
    \cdot \bigl[ \mathbf{p} \times \mathbf{e}^{(\alpha)} \bigr]\, \times \\
  \times &
  \Bigr|\,
    \psi_{\rm nucl, i} (\beta_{A}) \cdot
    \psi_{\alpha, i} (\beta_{\alpha})
  \Bigr\rangle \cdot
  \Phi_{\rm \alpha - nucl, i} (\mathbf{r})\; \mathbf{dr}.
\end{array}
\label{eq.app.3.2}
\end{equation}
We calculate this equation further as
\begin{equation}
\begin{array}{lll}
\vspace{-0.1mm}
  & M_{2} = -\, (2\pi)^{3} \delta (\mathbf{K}_{f} - \mathbf{k}_{\rm ph}) \cdot
  \displaystyle\sum\limits_{\alpha=1,2}
  \displaystyle\int\limits_{}^{}
    \Phi_{\rm \alpha - nucl, f}^{*} (\mathbf{r}) \cdot \biggl\{ \\

\vspace{-0.1mm}
  \times &
  2\, \mu_{N}\,  m_{\rm p}\,
  \Bigl\langle
    \psi_{\rm nucl, f} (\beta_{A}) \cdot
    \psi_{\alpha, f} (\beta_{\alpha})\,
  \Bigl|\,
    \Bigl\{
      e^{-i\, c_{A} \mathbf{k_{\rm ph}} \mathbf{r}}\, \displaystyle\frac{1}{m_{\alpha}}\, \displaystyle\sum_{i=1}^{4} z_{i}\, e^{-i\, \mathbf{k_{\rm ph}} \rhobf_{\alpha i}} -
      e^{i\, c_{\alpha} \mathbf{k_{\rm ph}} \mathbf{r}}\,  \displaystyle\frac{1}{m_{A}}\, \displaystyle\sum_{j=1}^{A} z_{j}\, e^{-i\, \mathbf{k_{\rm ph}} \rhobf_{Aj}}
    \Bigr\}\; \times \\
\vspace{0.5mm}
  \times &
  \Bigr|
    \psi_{\rm nucl, i} (\beta_{A}) \cdot
    \psi_{\alpha, i} (\beta_{\alpha})
  \Bigr\rangle \cdot
  \mathbf{e}^{(\alpha)} \mathbf{p}\; + \\
\vspace{-0.1mm}
  + &
  i\,
  \Bigl\langle
    \psi_{\rm nucl, f} (\beta_{A}) \cdot
    \psi_{\alpha, f} (\beta_{\alpha})\,
  \Bigl|\,
    \Bigl\{
      e^{-i\, c_{A} \mathbf{k_{\rm ph}} \mathbf{r}} \displaystyle\frac{1}{m_{\alpha}}\,
      \displaystyle\sum_{i=1}^{4}
        \mu_{i}^{\rm (an)}\, m_{\alpha i}\; e^{-i\, \mathbf{k_{\rm ph}} \rhobf_{\alpha i}}\, \sigmabf -
      e^{i\, c_{\alpha} \mathbf{k_{\rm ph}} \mathbf{r}} \displaystyle\frac{1}{m_{A}}
      \displaystyle\sum_{j=1}^{A}
        \mu_{j}^{\rm (an)}\, m_{Aj}\; e^{-i\, \mathbf{k_{\rm ph}} \rhobf_{Aj}}\, \sigmabf \Bigr\}\; \times \\
  \times &
  \Bigr|\,
    \psi_{\rm nucl, i} (\beta_{A}) \cdot
    \psi_{\alpha, i} (\beta_{\alpha})
  \Bigr\rangle
  \bigl[ \mathbf{p} \times \mathbf{e}^{(\alpha)} \bigr]\,
  \biggr\} \cdot
  \Phi_{\rm \alpha - nucl, i} (\mathbf{r})\; \mathbf{dr}
\end{array}
\label{eq.app.3.3}
\end{equation}
or
\begin{equation}
\begin{array}{lll}
\vspace{-0.1mm}
  & M_{2} = -\, (2\pi)^{3} \delta (\mathbf{K}_{f} - \mathbf{k}_{\rm ph}) \cdot
  \displaystyle\sum\limits_{\alpha=1,2}
  \displaystyle\int\limits_{}^{}
    \Phi_{\rm \alpha - nucl, f}^{*} (\mathbf{r})\; \times \\
\vspace{-0.1mm}
  \times &
  \biggl\{
  2\, \mu_{N}\,  m_{\rm p}\,
  \Bigl\{
    e^{-i\, c_{A} \mathbf{k_{\rm ph}} \mathbf{r}}\, \displaystyle\frac{1}{m_{\alpha}}\,
    \Bigl\langle \psi_{\rm nucl, f} (\beta_{A}) \cdot \psi_{\alpha, f} (\beta_{\alpha})\, \Bigl|\,
      \displaystyle\sum_{i=1}^{4} z_{i}\, e^{-i\, \mathbf{k_{\rm ph}} \rhobf_{\alpha i}}
    \Bigr| \psi_{\rm nucl, i} (\beta_{A}) \cdot \psi_{\alpha, i} (\beta_{\alpha}) \Bigr\rangle\; - \\

\vspace{0.5mm}
  - &
    e^{i\, c_{\alpha} \mathbf{k_{\rm ph}} \mathbf{r}}\,  \displaystyle\frac{1}{m_{A}}\,
    \Bigl\langle \psi_{\rm nucl, f} (\beta_{A}) \cdot \psi_{\alpha, f} (\beta_{\alpha})\, \Bigl|\,
      \displaystyle\sum_{j=1}^{A} z_{j}\, e^{-i\, \mathbf{k_{\rm ph}} \rhobf_{Aj}}
    \Bigr| \psi_{\rm nucl, i} (\beta_{A}) \cdot \psi_{\alpha, i} (\beta_{\alpha}) \Bigr\rangle
  \Bigr\} \cdot
  \mathbf{e}^{(\alpha)} \mathbf{p}\; + \\

\vspace{-0.1mm}
  + &
  i\,
  \Bigl\{
    e^{-i\, c_{A} \mathbf{k_{\rm ph}} \mathbf{r}} \displaystyle\frac{1}{m_{\alpha}}\,
    \Bigl\langle \psi_{\rm nucl, f} (\beta_{A}) \cdot \psi_{\alpha, f} (\beta_{\alpha})\, \Bigl|\,
      \displaystyle\sum_{i=1}^{4}
        \mu_{i}^{\rm (an)}\, m_{\alpha i}\; e^{-i\, \mathbf{k_{\rm ph}} \rhobf_{\alpha i}}\, \sigmabf
    \Bigr| \psi_{\rm nucl, i} (\beta_{A}) \cdot \psi_{\alpha, i} (\beta_{\alpha}) \Bigr\rangle\; - \\

\vspace{-0.1mm}
  - &
    e^{i\, c_{\alpha} \mathbf{k_{\rm ph}} \mathbf{r}} \displaystyle\frac{1}{m_{A}}
    \Bigl\langle \psi_{\rm nucl, f} (\beta_{A}) \cdot \psi_{\alpha, f} (\beta_{\alpha})\, \Bigl|\,
      \displaystyle\sum_{j=1}^{A}
        \mu_{j}^{\rm (an)}\, m_{Aj}\; e^{-i\, \mathbf{k_{\rm ph}} \rhobf_{Aj}}\, \sigmabf
    \Bigr| \psi_{\rm nucl, i} (\beta_{A}) \cdot \psi_{\alpha, i} (\beta_{\alpha}) \Bigr\rangle
  \Bigr\}\;
  \bigl[ \mathbf{p} \times \mathbf{e}^{(\alpha)} \bigr]\,
  \biggr\}\;
  \times \\
  \times &
  \Phi_{\rm \alpha - nucl, i} (\mathbf{r})\; \mathbf{dr}.
\end{array}
\label{eq.app.3.4}
\end{equation}
Here, we take into account that function $\psi_{\alpha, s} (\beta_{\alpha})$ is dependent of variables $\rhobf_{\alpha n}$ (i.e. it is not dependent on variables $\rhobf_{A m}$),
as the function $\psi_{\rm nucl, s} (\beta_{A})$ is dependent on variables $\rhobf_{A m}$ (i.e. it is not dependent on variables $\rhobf_{\alpha n}$).
On such a basis, we rewrite Eq.~(\ref{eq.app.3.4}) as
\begin{equation}
\begin{array}{lll}
\vspace{-0.1mm}
  & M_{2} = -\, (2\pi)^{3} \delta (\mathbf{K}_{f} - \mathbf{k}_{\rm ph}) \cdot
  \displaystyle\sum\limits_{\alpha=1,2}
  \displaystyle\int\limits_{}^{}
    \Phi_{\rm \alpha - nucl, f}^{*} (\mathbf{r})\; \times \\
\vspace{-0.1mm}
  \times &
  \biggl\{
  2\, \mu_{N}\,  m_{\rm p}\,
  \biggl[
    e^{-i\, c_{A} \mathbf{k_{\rm ph}} \mathbf{r}}\, \displaystyle\frac{1}{m_{\alpha}}\,
    \biggl\langle \psi_{\alpha, f} (\beta_{\alpha})\, \biggl|\,
      \displaystyle\sum_{i=1}^{4} z_{i}\, e^{-i\, \mathbf{k_{\rm ph}} \rhobf_{\alpha i}}
    \biggr| \psi_{\alpha, i} (\beta_{\alpha}) \biggr\rangle \cdot
    \Bigl\langle \psi_{\rm nucl, f} (\beta_{A}) \Bigr|\, \psi_{\rm nucl, i} (\beta_{A}) \Bigr\rangle\; - \\

\vspace{0.5mm}
  - &
    e^{i\, c_{\alpha} \mathbf{k_{\rm ph}} \mathbf{r}}\,  \displaystyle\frac{1}{m_{A}}\,
    \biggl\langle \psi_{\rm nucl, f} (\beta_{A})\, \biggl|\,
      \displaystyle\sum_{j=1}^{A} z_{j}\, e^{-i\, \mathbf{k_{\rm ph}} \rhobf_{Aj}}
    \biggr| \psi_{\rm nucl, i} (\beta_{A}) \biggr\rangle \cdot
    \Bigl\langle \psi_{\alpha, f} (\beta_{\alpha})\, \Bigl|\, \psi_{\alpha, i} (\beta_{\alpha}) \Bigr\rangle\:
  \biggr] \cdot
  \mathbf{e}^{(\alpha)} \mathbf{p}\; + \\

\vspace{-0.1mm}
  + &
  i\,
  \biggl[
    e^{-i\, c_{A} \mathbf{k_{\rm ph}} \mathbf{r}} \displaystyle\frac{1}{m_{\alpha}}\,
    \biggl\langle \psi_{\alpha, f} (\beta_{\alpha})\, \biggl|\,
      \displaystyle\sum_{i=1}^{4}
        \mu_{i}^{\rm (an)}\, m_{\alpha i}\; e^{-i\, \mathbf{k_{\rm ph}} \rhobf_{\alpha i}}\, \sigmabf
    \biggr| \psi_{\alpha, i} (\beta_{\alpha}) \biggr\rangle \cdot
    \Bigl\langle \psi_{\rm nucl, f} (\beta_{A}) \Bigr|\, \psi_{\rm nucl, i} (\beta_{A}) \Bigr\rangle\; - \\

\vspace{-0.1mm}
  - &
    e^{i\, c_{\alpha} \mathbf{k_{\rm ph}} \mathbf{r}} \displaystyle\frac{1}{m_{A}}
    \biggl\langle \psi_{\rm nucl, f} (\beta_{A})\, \biggl|\,
      \displaystyle\sum_{j=1}^{A}
        \mu_{j}^{\rm (an)}\, m_{Aj}\; e^{-i\, \mathbf{k_{\rm ph}} \rhobf_{Aj}}\, \sigmabf
    \biggr| \psi_{\rm nucl, i} (\beta_{A}) \biggr\rangle \cdot
    \Bigl\langle \psi_{\alpha, f} (\beta_{\alpha})\, \Bigl|\, \psi_{\alpha, i} (\beta_{\alpha}) \Bigr\rangle\:
  \biggr]\:
  \bigl[ \mathbf{p} \times \mathbf{e}^{(\alpha)} \bigr]\,
  \biggr\}\;
  \times \\
  \times &
  \Phi_{\rm \alpha - nucl, i} (\mathbf{r})\; \mathbf{dr}.
\end{array}
\label{eq.app.3.5}
\end{equation}
We take into account normalization condition for wave functions as
\begin{equation}
\begin{array}{lll}
  \Bigl\langle \psi_{\rm nucl, f} (\beta_{A}) \Bigr|\, \psi_{\rm nucl, i} (\beta_{A}) \Bigr\rangle = 1, &
  \Bigl\langle \psi_{\alpha, f} (\beta_{\alpha})\, \Bigl|\, \psi_{\alpha, i} (\beta_{\alpha}) \Bigr\rangle = 1,
\end{array}
\label{eq.app.3.6}
\end{equation}
and Eq.~(\ref{eq.app.3.5}) is transformed to
\begin{equation}
\begin{array}{lll}
\vspace{-0.1mm}
  & M_{2} = -\, (2\pi)^{3} \delta (\mathbf{K}_{f} - \mathbf{k}_{\rm ph}) \cdot
  \displaystyle\sum\limits_{\alpha=1,2}
  \displaystyle\int\limits_{}^{}
    \Phi_{\rm \alpha - nucl, f}^{*} (\mathbf{r})\; \times \\
\vspace{-0.1mm}
  \times &
  \biggl\{
  2\, \mu_{N}\,  m_{\rm p}\,
  \biggl[
    e^{-i\, c_{A} \mathbf{k_{\rm ph}} \mathbf{r}}\, \displaystyle\frac{1}{m_{\alpha}}\,
    \Bigl\langle \psi_{\alpha, f} (\beta_{\alpha})\, \Bigl|\,
      \displaystyle\sum_{i=1}^{4} z_{i}\, e^{-i\, \mathbf{k_{\rm ph}} \rhobf_{\alpha i}}
    \Bigr| \psi_{\alpha, i} (\beta_{\alpha}) \Bigr\rangle\; - \\
\vspace{0.5mm}
  - &
    e^{i\, c_{\alpha} \mathbf{k_{\rm ph}} \mathbf{r}}\,  \displaystyle\frac{1}{m_{A}}\,
    \Bigl\langle \psi_{\rm nucl, f} (\beta_{A})\, \Bigl|\,
      \displaystyle\sum_{j=1}^{A} z_{j}\, e^{-i\, \mathbf{k_{\rm ph}} \rhobf_{Aj}}
    \Bigr| \psi_{\rm nucl, i} (\beta_{A}) \Bigr\rangle
  \biggr]\,
  \mathbf{e}^{(\alpha)} \mathbf{p}\; + \\

\vspace{-0.1mm}
  + &
  i\,
  \biggl[
    e^{-i\, c_{A} \mathbf{k_{\rm ph}} \mathbf{r}} \displaystyle\frac{1}{m_{\alpha}}\,
    \Bigl\langle \psi_{\alpha, f} (\beta_{\alpha})\, \Bigl|\,
      \displaystyle\sum_{i=1}^{4}
        \mu_{i}^{\rm (an)}\, m_{\alpha i}\; e^{-i\, \mathbf{k_{\rm ph}} \rhobf_{\alpha i}}\, \sigmabf
    \Bigr| \psi_{\alpha, i} (\beta_{\alpha}) \Bigr\rangle\; - \\
\vspace{-0.1mm}
  - &
    e^{i\, c_{\alpha} \mathbf{k_{\rm ph}} \mathbf{r}} \displaystyle\frac{1}{m_{A}}
    \Bigl\langle \psi_{\rm nucl, f} (\beta_{A})\, \Bigl|\,
      \displaystyle\sum_{j=1}^{A}
        \mu_{j}^{\rm (an)}\, m_{Aj}\; e^{-i\, \mathbf{k_{\rm ph}} \rhobf_{Aj}}\, \sigmabf
    \Bigr| \psi_{\rm nucl, i} (\beta_{A}) \Bigr\rangle\,
  \biggr]\,
  \bigl[ \mathbf{p} \times \mathbf{e}^{(\alpha)} \bigr]\,
  \biggr\}\;
  \times \\
  \times &
  \Phi_{\rm \alpha - nucl, i} (\mathbf{r})\; \mathbf{dr}.
\end{array}
\label{eq.app.3.7}
\end{equation}

Now we introduce new definitions of \emph{electric and magnetic form-factors} of the $\alpha$-particle and nucleus as
\begin{equation}
\begin{array}{lll}
\vspace{1mm}
  F_{\alpha,\, {\rm el}} = &
    \displaystyle\sum\limits_{n=1}^{4}
    \biggl\langle \psi_{\alpha, f} (\beta_{\alpha})\, \biggl|\,
      z_{n}\, e^{-i \mathbf{k}_{\rm ph} \rhobf_{\alpha n} }
    \biggr|\,  \psi_{\alpha, i} (\beta_{\alpha}) \biggr\rangle , \\

\vspace{1mm}
  F_{A,\, {\rm el}} = &
    \displaystyle\sum\limits_{m=1}^{A}
    \biggl\langle \psi_{\rm nucl, f} (\beta_{A}) \biggl|\,
      z_{m}\, e^{-i \mathbf{k}_{\rm ph} \rhobf_{A m} }
    \biggr|\, \psi_{\rm nucl, i} (\beta_{A}) \biggr\rangle , \\

\vspace{1mm}
  \mathbf{F}_{\alpha,\, {\rm mag}} = &
    \displaystyle\sum_{i=1}^{4}
    \Bigl\langle \psi_{\alpha, f} (\beta_{\alpha})\, \Bigl|\,
      \mu_{i}^{\rm (an)}\, m_{\alpha i}\; e^{-i\, \mathbf{k_{\rm ph}} \rhobf_{\alpha i}}\, \sigmabf
    \Bigr| \psi_{\alpha, i} (\beta_{\alpha}) \Bigr\rangle, \\

  \mathbf{F}_{A,\, {\rm mag}} = &
    \displaystyle\sum_{j=1}^{A}
    \Bigl\langle \psi_{\rm nucl, f} (\beta_{A})\, \Bigl|\,
        \mu_{j}^{\rm (an)}\, m_{Aj}\; e^{-i\, \mathbf{k_{\rm ph}} \rhobf_{Aj}}\, \sigmabf
    \Bigr| \psi_{\rm nucl, i} (\beta_{A}) \Bigr\rangle\,
\end{array}
\label{eq.app.3.8}
\end{equation}
and we take into account this definition for relative momentum as
\begin{equation}
  \mathbf{p} = - i \hbar\, \mathbf{\displaystyle\frac{d}{dr}}.
\label{eq.app.3.9}
\end{equation}
Then Eq.~(\ref{eq.app.3.7}) can be rewritten as
\begin{equation}
\begin{array}{lll}
\vspace{-0.1mm}
  M_{2} & = &
  i \hbar\, (2\pi)^{3} \delta (\mathbf{K}_{f} - \mathbf{k}_{\rm ph}) \cdot
  \displaystyle\sum\limits_{\alpha=1,2}
  \displaystyle\int\limits_{}^{}
    \Phi_{\rm \alpha - nucl, f}^{*} (\mathbf{r})\; \times \\
\vspace{0.5mm}
  & \times &
  \biggl\{
  2\, \mu_{N}\,  m_{\rm p}\,
  \Bigl[
    e^{-i\, c_{A} \mathbf{k_{\rm ph}} \mathbf{r}}\, \displaystyle\frac{1}{m_{\alpha}}\, F_{\alpha,\, {\rm el}} -
    e^{i\, c_{\alpha} \mathbf{k_{\rm ph}} \mathbf{r}}\,  \displaystyle\frac{1}{m_{A}}\, F_{A,\, {\rm el}}
  \Bigr]\,
  \mathbf{e}^{(\alpha)}\, \mathbf{\displaystyle\frac{d}{dr}}\; + \\
  & + &
  i\,
  \Bigl[
    e^{-i\, c_{A} \mathbf{k_{\rm ph}} \mathbf{r}} \displaystyle\frac{1}{m_{\alpha}}\, \mathbf{F}_{\alpha,\, {\rm mag}} -
    e^{i\, c_{\alpha} \mathbf{k_{\rm ph}} \mathbf{r}} \displaystyle\frac{1}{m_{A}}\, \mathbf{F}_{A,\, {\rm mag}}
  \Bigr]\,
  \Bigl[ \mathbf{\displaystyle\frac{d}{dr}} \times \mathbf{e}^{(\alpha)} \Bigr]\,
  \biggr\} \cdot
  \Phi_{\rm \alpha - nucl, i} (\mathbf{r})\; \mathbf{dr}.
\end{array}
\label{eq.app.3.10}
\end{equation}



\end{document}